\documentclass{aa} 
\usepackage{graphicx}
\usepackage{txfonts}
\usepackage{xcolor}
\usepackage{textgreek}
\begin{document} 

        \hyphenation{MontAGN}
        \title{High angular resolution polarimetric imaging\\ of the nucleus of NGC~1068:}
        \titlerunning{HAR polarisation in NGC~1068: location of polarising mechanisms}
        
        \subtitle{Disentangling the polarising mechanisms}
        
        \author{L. Grosset\inst{1,2}\thanks{\email{lucas.grosset@obspm.fr}} 
                \and 
                        D. Rouan\inst{1} 
                \and
                        F. Marin\inst{3} 
                \and 
                         D. Gratadour\inst{1,4}
                \and
                 		 E. Lagadec\inst{5}
				\and
                       	 S. Hunziker\inst{6}
                \and
                         M. Montargès\inst{1,7}
                \and
                		 Y. Magnard\inst{8}
                \and
                		 M. Carle\inst{9}
                \and
                		 J. Pragt\inst{10}
                \and
                		 C. Petit\inst{11}
                        }

   \institute{
                  LESIA, Observatoire de Paris, PSL Research University, CNRS, Sorbonne Universit\'es, UPMC Univ. Paris 06, Univ. Paris Diderot, Sorbonne Paris Cit\'e, 5 place Jules Janssen, 92190 Meudon, France
          \and
                  SOFIA Science Center, USRA, NASA Ames Research Center, Moffett Field, CA 94035, USA
          \and
          		  Université de Strasbourg, CNRS, Observatoire Astronomique de Strasbourg, UMR 7550, F-67000 Strasbourg, France
          \and
           		  Research School of Astronomy and Astrophysics, Australian National University, Canberra, ACT 2611, Australia	 
          \and 
          		  Université Côte d'Azur, Observatoire de la Côte d'Azur, CNRS, Laboratoire Lagrange, Bd de l'Observatoire, CS 34229, 06304 Nice cedex 4, France
          \and
          		  ETH Zurich, Institute for Particle Physics and Astrophysics,Wolfgang-Pauli-Strasse 27, CH-8093 Zurich, Switzerland
          \and
          		  Institute of Astronomy, KU Leuven, Celestijnenlaan 200D B2401, 3001 Leuven, Belgium
          \and
          		  Univ. Grenoble Alpes, CNRS, IPAG, F-38000 Grenoble, France
          \and
                  Aix Marseille Universit\'e, CNRS, CNES,  LAM, Marseille, France 
          \and
                  NOVA Optical Infrared Instrumentation Group, Oude Hoogeveensedijk 4, 7991 PD Dwingeloo, The Netherlands
          \and
                  DOTA, ONERA, Université Paris Saclay, F-91123, Palaiseau France 
                  }

   \date{Received xx, 2020; accepted xx, 202x}

 
  \abstract
   {Polarisation is a decisive method to study the inner region of active galactic nuclei (AGNs) since it is not affected by contrast issues similarly to how classical imaging is. When coupled to high angular resolution (HAR), polarisation can help to disentangle the location of the various polarising mechanisms and then give an insight on the physics taking place on the core of AGNs.}
   {We obtained a new data set of HAR polarimetric images of the archetypal Seyfert~2 nucleus of NGC~1068 observed with SPHERE/VLT and we aim in this paper at presenting the polarisation maps and at spatially separating the location of the polarising mechanisms, thus deriving constraints on the organisation of the dust material in the inner region of this AGN.}
   {With four new narrow bands images between the visible and the near infrared combined to older broad band observations, we studied the wavelength dependency of the polarisation properties from 0.7 to 2.2 \textmu m of three selected regions within the inner 2'' surrounding the central engine. We then compared these measurements to radiative transfer simulations of scattering and dichroic absorption processes, using the Monte-Carlo code MontAGN.}
   {We establish a detailed table of the relative importance of the polarising mechanism as a function of the aperture and of the wavelength. We are able to separate the dominant polarising mechanisms in the three regions of the ionisation cone, the extended envelop of the torus and the very central bright source of the AGN. Thus, we estimate the contribution of the different polarisation mechanisms to the observed polarisation flux in these regions. Dichroic absorption is estimated to be responsible for about 99~\% of the polarised flux coming from the photo-centre. However, this contribution would be restricted to this location only, double scattering process being the most important contributor to polarisation in the equatorial plane of the AGN and single scattering being dominant in the polar outflow bi-cone.}
   {Despite that results are in good agreement with larger apertures measurements, the variety of situations with different mechanisms at play highlights the importance of spatial resolution for the interpretation of polarisation measurements. We also refine the estimation of the integrated optical depth in the visible of the obscuring structure to a range of 20 to 100, constraining the geometry of the inner region of this AGN.}


   \maketitle
%


\section{Introduction}


Thanks to the continuous progress of instrumentation, our understanding of the organisation of the centre of active galactic nuclei (AGNs) has been growing fast in the last few years.

In particular, we are now resolving at an unprecedented parsec size spatial scale the core of nearby AGNs thanks to the development of high angular resolution (HAR) techniques, in particular interferometry and extreme adaptive optics (AO), with the help of polarisation. Recent ALMA observations of two of the closest and brightest AGNs, NGC~1068 \citep{Garcia-Burillo2016,Garcia-Burillo2019,Imanishi2018,Imanishi2020,Lopez-Rodriguez2019,Impellizzeri2019} and Circinus \citep{Izumi2018}, resolved at different frequencies the molecular material surrounding the central engine (CE), whose associated dust is thought to block the light emitted when observing the nucleus edge on, as proposed years ago by \cite{Antonucci1985,Antonucci1993}. Using the GRAVITY interferometer (at VLTI), the \cite{Sturm2018} was able to constrain the shape of the broad line region of nearby quasar 3C 273 to a rotating thick disk and identified the sublimation region in NGC~1068 \citep{Pfuhl2020}.
Finally, this view is completed thanks to the high contrast imaging coupled to polarimetry, as shown for instance by \cite{Packham1997,Packham2007,Lopez-Rodriguez2015} with different instruments in the near to mid infrared range \citep{Gratadour2015,Grosset2018}. 


Indeed, polarimetric imaging is 
less affected by contrast issues because polarisation provides more measurable intrinsic parameters of the light than just its intensity, namely the two parameters describing the orientation of the linear polarisation and for some instruments the circular polarisation. Therefore, by looking at the polarisation of the incoming light, we can get access to information on the material responsible for the polarisation, either by scattering, absorption or emission, even in the vicinity of bright sources. Polarisation has revealed itself to be very efficient to study media close to a very bright source typical of AGNs environments, as demonstrated by \cite{Antonucci1985}, using polarised light to detect broad lines in the spectrum of NGC~1068.


The model of the obscuring material surrounding the CE has been undergoing evolutions, from simple torus shape in the early 90's \citep{Antonucci1993} to the current complex, clumpy and dynamical environments such as those modelled by \cite{Izumi2018}. Identifying the mechanism responsible for the polarisation in AGNs brings information on the nature and characteristics of the scatterers. Several phenomena can be responsible for polarisation, principally scattering on dust grains or on electrons \citep{Antonucci1985} or dichroic absorption or emission by elongated dust grains \citep{Efstathiou1997,Lopez-Rodriguez2015}.

As concern the mechanisms giving rise to polarisation, single scattering is currently observed in the polarisation of circum-stellar environments, as for example by \cite{Kervella2015} with SPHERE/ZIMPOL (VLT, ESO's Paranal observatory). More complex signatures have also been invoked in young stellar objects, as the double scattering process presented by \cite{Bastien1990} and simulated by \cite{Murakawa2010}. This mechanism is also thought to be present in the outer envelop of the obscuring material of AGNs \citep{Grosset2018} to account for the observed polarisation in the 20~$\times$~60~pc central region \citep{Gratadour2015}.

Dichroism is expected to induce polarisation closer to the very central core of AGN. For elongated grains to emit or absorb photons with a preferential polarisation orientation, it is required to have a mechanism responsible for aligning these grains on a large enough scale \citep{Efstathiou1997}, typically on few parsecs size in the case of AGNs \citep{Lopez-Rodriguez2015}. Magnetic field has been a promising candidate for aligning dust grains, and the measured polarisation could then be used to constrain magnetic field properties, as achieved recently by \cite{Lopez-Rodriguez2015}.

There has been in the last years a growing number of simulation works to better understand the Physics of AGNs, through polarisation properties: \cite{Nenkova2002,Alonso2011,Marin2015,Lopez-Rodriguez2015,Grosset2018} for example. The importance of polarisation in the understanding of AGNs is also revealed by the large quantity of polarimetric measurements on the AGN of NGC~1068, the archetypal Seyfert~2 galaxy (D~$\approx$~14~Mpc), at all wavelengths on the last 50 years, compiled by \cite{Marin2018}.

On this AGN, \cite{Antonucci1985} first invoked scattering as the mechanism responsible for the polarisation, as it would scatter above the equatorial plane toward the observer the light emitted in the broad line region, hidden by the torus. Discussions about the nature of the scatterers, between electrons and dust grains, have been ongoing, \cite{Antonucci1993} favouring electrons due to the observed independence of the polarisation with wavelength. Polarimetric observations by \cite{Packham1997} in the near infrared (NIR) confirmed the location of scatterers in the polar region of NGC~1068 and it is now expected that both electrons and dust grains are populating this region. \cite{Packham2007} and \cite{Lopez-Rodriguez2016} indeed detected polarisation in the narrow line region around 10~\textmu m, most likely produced by dichroic emission by dust, while \cite{Marin2015} combined both electrons and dust grains populations in their simulations to study the polar outflows. Scattering on dust grains has now been observed on NGC~1068 at medium spatial scales \citep[at about 100 to 200~pc]{Gratadour2015}.


\vspace{0.5cm}
We obtained new polarimetric HAR images of NGC~1068 thanks to SPHERE (installed at the Nasmyth focus of the UT3 - Melipal - of the VLT, at ESO's Paranal observatory) in three narrow bands (NB) in the NIR and one in the visible. In this paper, we present the data sets and discuss the evolution of the polarisation in the different observed structures in the inner region as a function of wavelength, in order to constrain the polarising mechanisms. We detail in section 2 these new observations, and the obtained polarimetric maps. We then present in section 3 the polarisation as a function of wavelength in different selected regions of the AGN. In section 4, we detail how dichroism was introduced in the numerical simulations conducted using our code MontAGN. We compare in section 5 the observations to the simulations. We then discuss in section 6 this comparison and the new insight brought by these new observations before concluding.

\section{Observations and data processing}

\subsection{Observations summary}

This study is based on three data sets of the central region of the Seyfert 2 galaxy NGC 1068, obtained in polarimetric mode at HAR with the instrument SPHERE \citep{Beuzit2008} at VLT, ESO Paranal Observatory.

The first data are those of the SPHERE science verification (SV) observing program, obtained in H and Ks bands with the sub-module IRDIS \citep{Dohlen2008,Langlois2014}, published and analysed in \cite{Gratadour2015} and \cite{Grosset2018}. The second set was observed between the 11$^{\textrm{th}}$ and the 14$^{\textrm{th}}$ of September 2016, also with the IRDIS system, bringing three additional NBs polarimetric images in the NIR, namely Continuum H, Continuum K1 and Continuum K2 (hereafter Cnt~H, Cnt~K1 and Cnt~K2 respectively). Filters properties can be found in Table~\ref{table:obs}. As for the first data set, the bright nucleus was used as a guide source for the SPHERE extreme adaptive optics system SAXO \citep{Fusco2006}, giving a correction quality comparable to what was achieved in the SV observation (SAXO could not be used at full capacity in both observations due to flux limitation). Seeing was comparable to previous run, ranging between 0.7 and 1.2'' (a complete summary is available on Table~\ref{table:obs}). Investigations on the radial profile of the central peak in all five NIR bands by \cite{Rouan2019} show no clear modification of the PSF (see their figure 4) that could indicate change in the resolution and the achieved angular resolution is estimated to 60~mas (i.e. 4~pc at 15~Mpc).

NGC~1068 was also observed with ZIMPOL, another SPHERE sub-system \citep{Schmid2018} in the 12$^{\textrm{th}}$ of October 2016, in the frame of the Other Science SPHERE guaranteed time observation (GTO), adding the narrow band NR (Narrow Red, in the visible at 645.9~nm, see Table~\ref{table:obs} for filter characteristics) to the available polarimetric data. Similarly to the IRDIS observations, SAXO was used but, due to the lack of photons, no information about the achieved angular resolution could be retrieved. SPARTA, the SPHERE system aimed at estimating the AO efficiency is indeed known to suffer bias in the case of faint targets, and these informations are thus not reliable, as described in \cite{Milli2017arXiv}, highlighted in their figure 2. The seeing was comparable to those of IRDIS and we can expect the AO correction to be similar, with lower achieved resolution due to the wavelength-dependency on the AO correction. The achieved resolution can be estimated around 100--150~mas based on the smallest structure detectable on the ZIMPOL image. A summary of the observations is presented in Table \ref{table:obs}.

\begin{table*}[ht!]
\caption[SPHERE observation log.]{Log of the SPHERE observations (IRDIS and ZIMPOL)}
\label{table:obs} 
\centering
\begin{tabular}{|l||c|c|c||c|c||c|}
\hline\hline 
Filter & Date & Obs time & Sky time & \textlambda$_0$ & $\Delta_\textrm{\textlambda}$ & Seeing \\
 & (UT) & (min) & (min) & (\textmu m) & (\textmu m) & ('') \\
(1) & (2) & (3) & (4) & (5) & (6) & (7) \\
\hline
H & 10-11/12/2014 & 17.07 & 6.40 & 1.625 & 0.290 & 0.91 to 1.22 \\
Ks & 10-11/12/2014 & 11.33 & 2.00 & 2.182 & 0.300 & 0.73 to 0.78 \\
Cnt H & 12-14/09/2016 & 85.33 & 32.00 & 1.573 & 0.023 & 0.87 to 1.07 \\
Cnt K1 & 12-14/09/2016 & 85.33 & 42.67 & 2.091 & 0.034 & 0.82 to 1.13 \\
Cnt K2 & 12-14/09/2016 & 106.67 & 42.27 & 2.266 & 0.032 & 0.75 to 0.98 \\
NR & 12/10/2016 & 12.25 & - & 0.6459 & 0.0567 & $\approx 0.89$ \\
\hline
\end{tabular}
\tablefoot{ 
(1) Filter name according to the SPHERE user manual. (2) Date of the observation (UT) (3) Observation integration duration, combining all the HWP and dithering positions (4) Sky observation integration time, taken with one preferential HWP position (see text)  (5) Filter central wavelength (in \textmu m) (6) Filter full width at half maximum (in \textmu m) (7) Seeing (in arcseconds) as measured by the observatory, corrected for airmass. This is different from the achieved angular resolution, which is around 60~mas in the NIR and 100--150~mas in the red, as described in text.}
\end{table*}

When observing in NIR, because of the sky emission becoming important, it is usual to match the sky observation integration time to the on-target integration time. However, because of the specific data processing of dual polarimetric imaging, we do not need sky acquisitions for each Half-Wave Plate (HWP) position in polarimetric measurements, as displayed by Table \ref{table:obs}. More details about the polarimetric sky subtraction strategy will be developed in an upcoming paper.

\subsection{Data reduction}

\begin{figure*}[ht!]
 \centering
 \includegraphics[width=0.45\textwidth,clip]{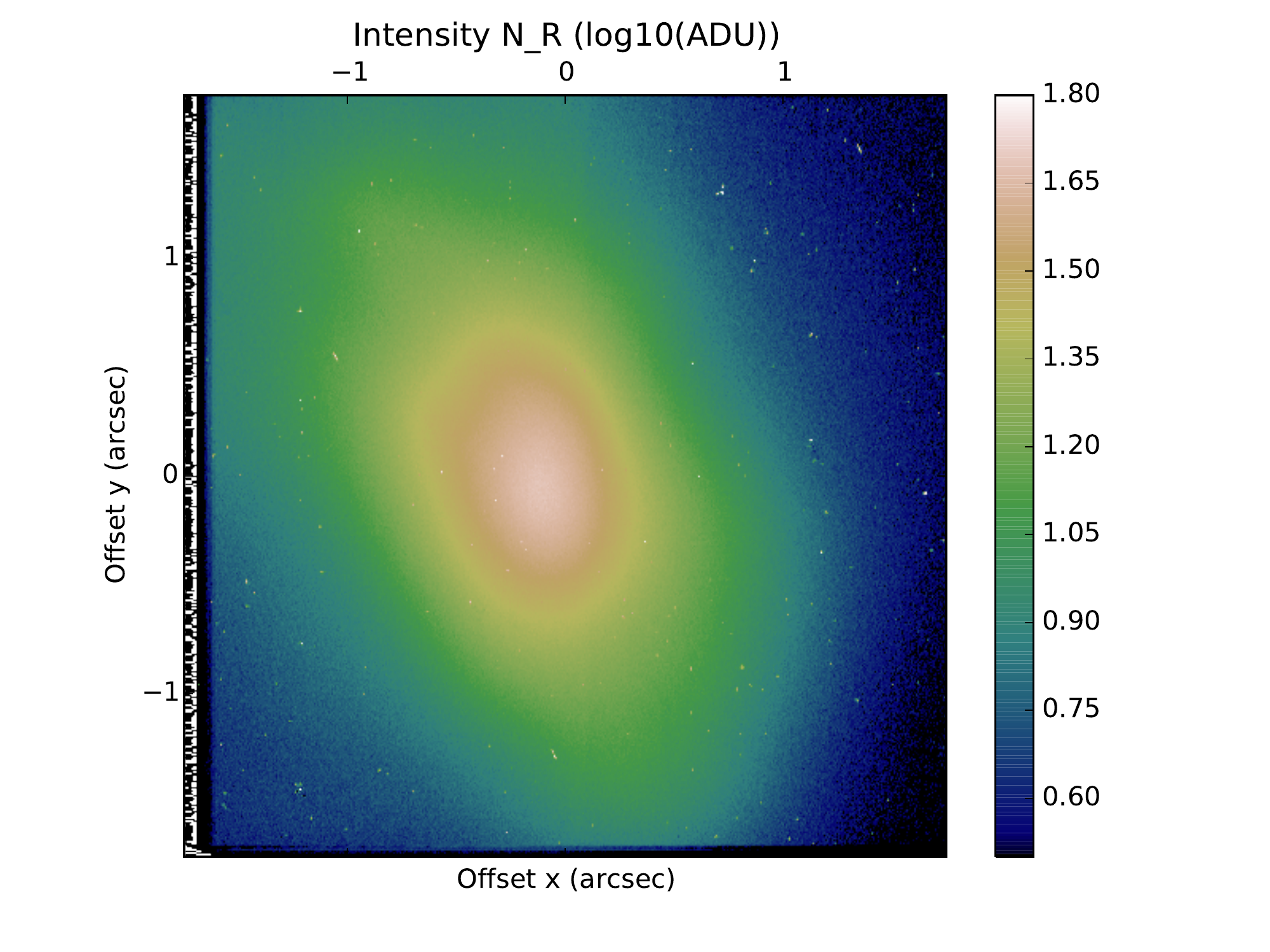}
 \includegraphics[width=0.45\textwidth,clip]{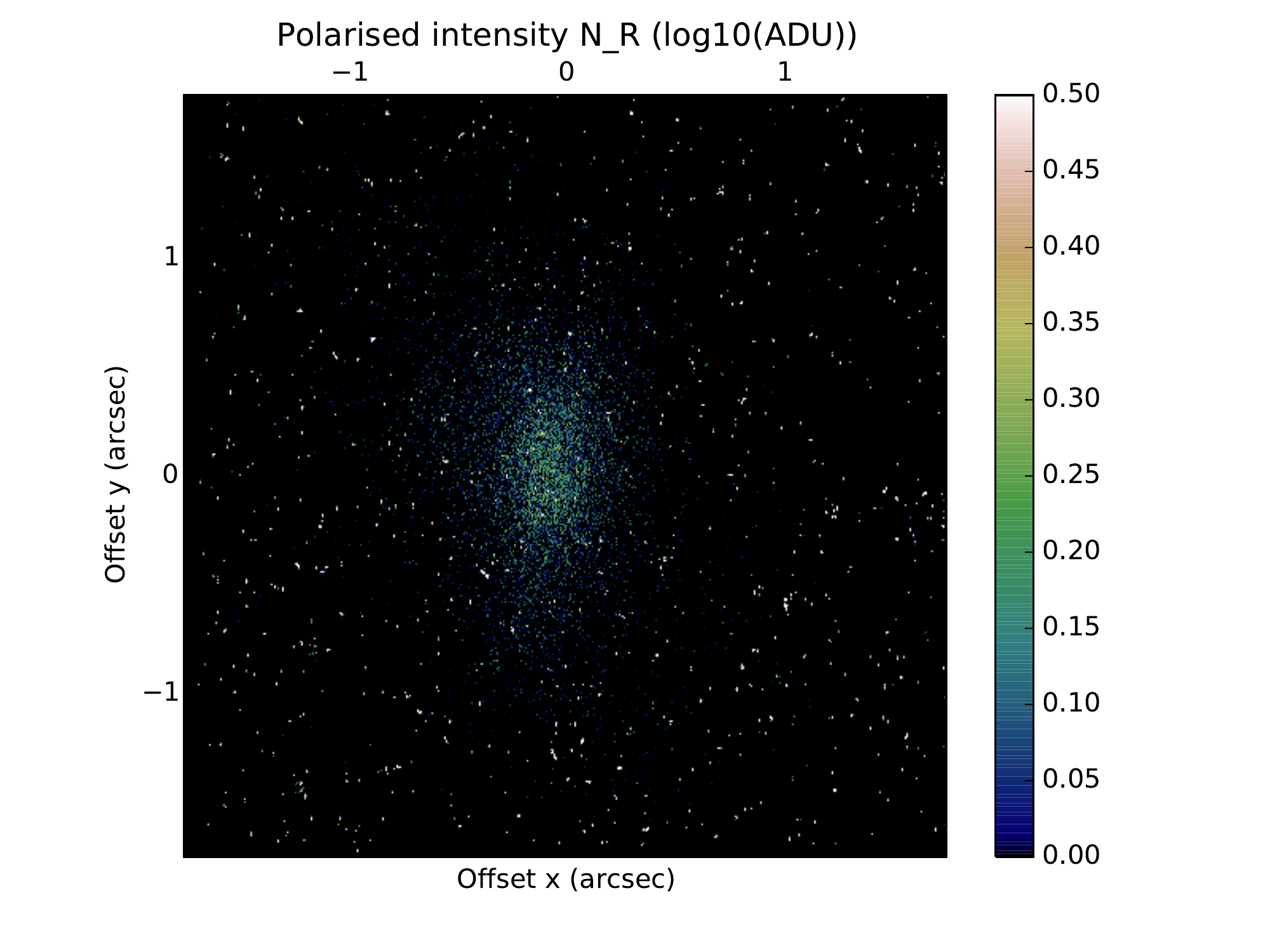}
 
 \includegraphics[width=0.45\textwidth,clip]{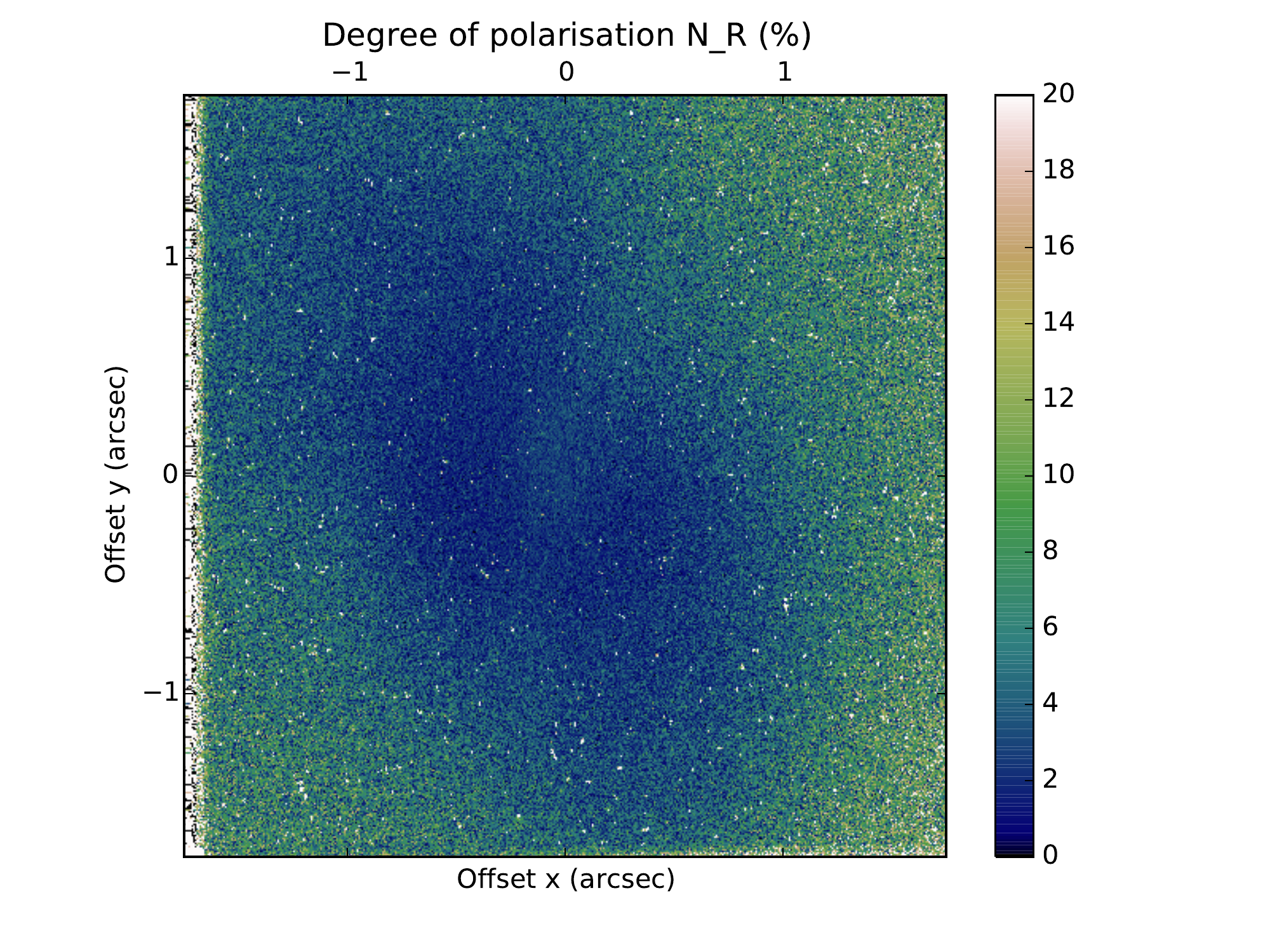}
 \includegraphics[width=0.45\textwidth,clip]{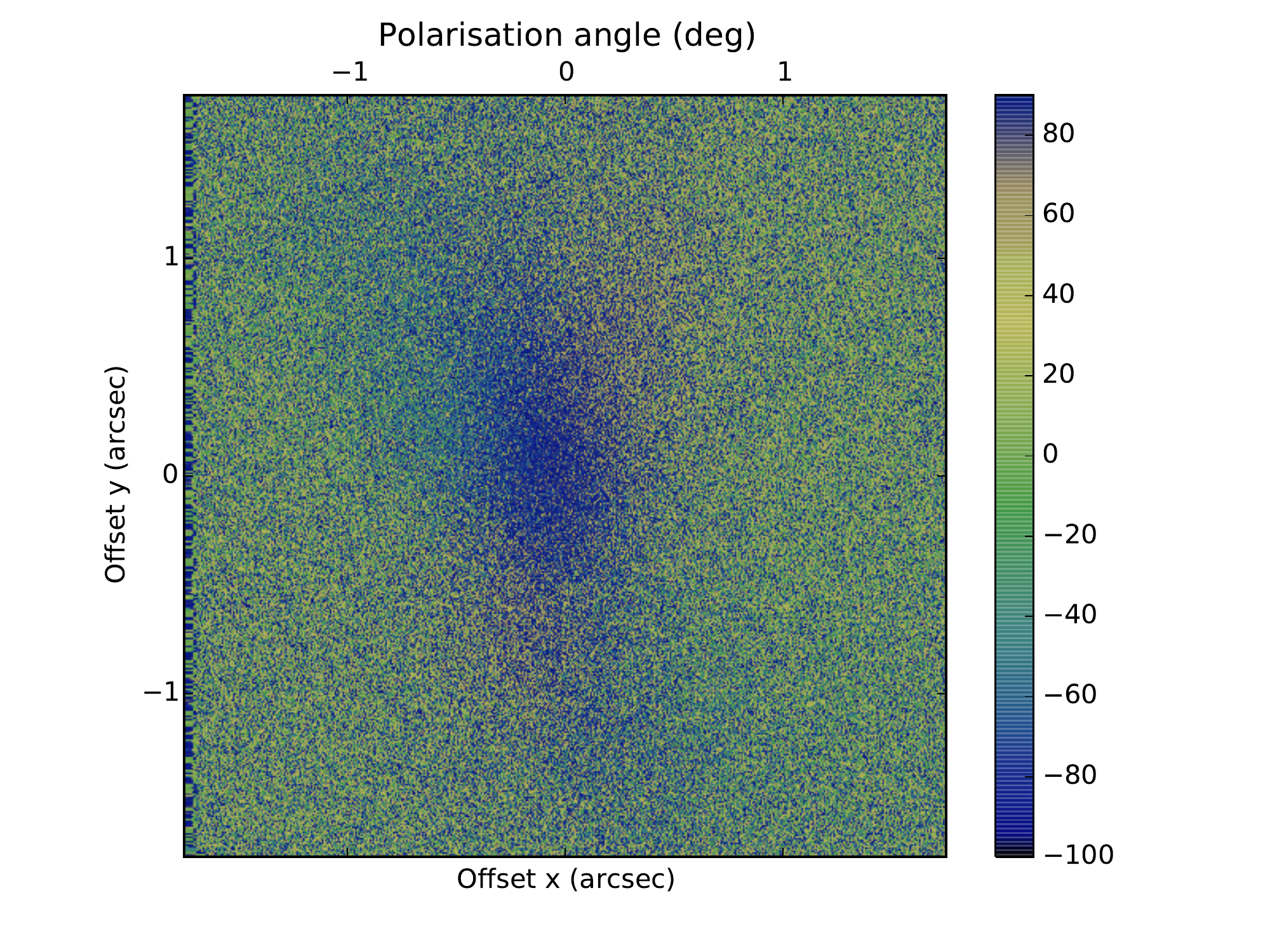}
  \caption{Maps of intensity (in $\log_{10}$(ADU/s), upper left panel), polarised intensity (in $\log_{10}$(ADU/s), upper right panel), linear degree of polarisation (in \%, bottom left panel) and linear angle of polarisation (in degrees, bottom right panel) in NR (645.9~nm) with ZIMPOL. North is up, and East is to the left.}
  \label{fig:NR}
\end{figure*}

ZIMPOL polarimetric maps (0.0036'' per pixel, \citealt{Schmid2018}) were created using the official SPHERE-ZIMPOL pipeline. Maps of total and polarised intensities, degree of linear polarisation and angle of linear polarisation are displayed in Figure~\ref{fig:NR}.

All IRDIS polarimetric images (from both observing runs) were reduced with the same dedicated pipeline including sky subtraction, flat field correction, realignment after a median filter of size $3\times3$ pixels$^2$. The 8 final images were then realigned precisely (IRDIS records two images with perpendicular polarisation at the same time), before being combined using the inverse matrix method\footnote{We also used double differences and double ratios methods, with very similar outputs (see \cite{Tinbergen1996} for more detail about these methods).}, as described in Appendix \ref{App:matrinvers}, to produce the intensity and polarisation Q and U maps. As detailed in Section~\ref{sec:K2} and Appendix~\ref{App:derot}, Cnt~K2 (2266~nm) images were selected to avoid depolarisation due to SPHERE derotator.

From these obtained maps, we finally transform them onto polarised intensity $Ip_{lin}$, linear polarisation degree $P_{lin}$ and linear polarisation position angle $\theta_{lin}$ maps (Polarisation position angle reference follows \cite{IAU1973} recommendations, starting from North and counting positive from North to East) following 

\begin{equation}
\label{eq:P}
P_{lin} = \frac{\sqrt{Q^2 + U^2}}{I}~\textrm{,}
\end{equation}
\begin{equation}
\label{eq:Ip}
Ip_{lin} = P_{lin}\times I
\end{equation}
and
\begin{equation}
\label{eq:theta}
\theta_{lin} = \frac{1}{2}~\mathrm{atan2}(U,Q)
\end{equation}
respectively.

\begin{figure*}[ht!]
 \centering
 \includegraphics[width=0.45\textwidth,clip]{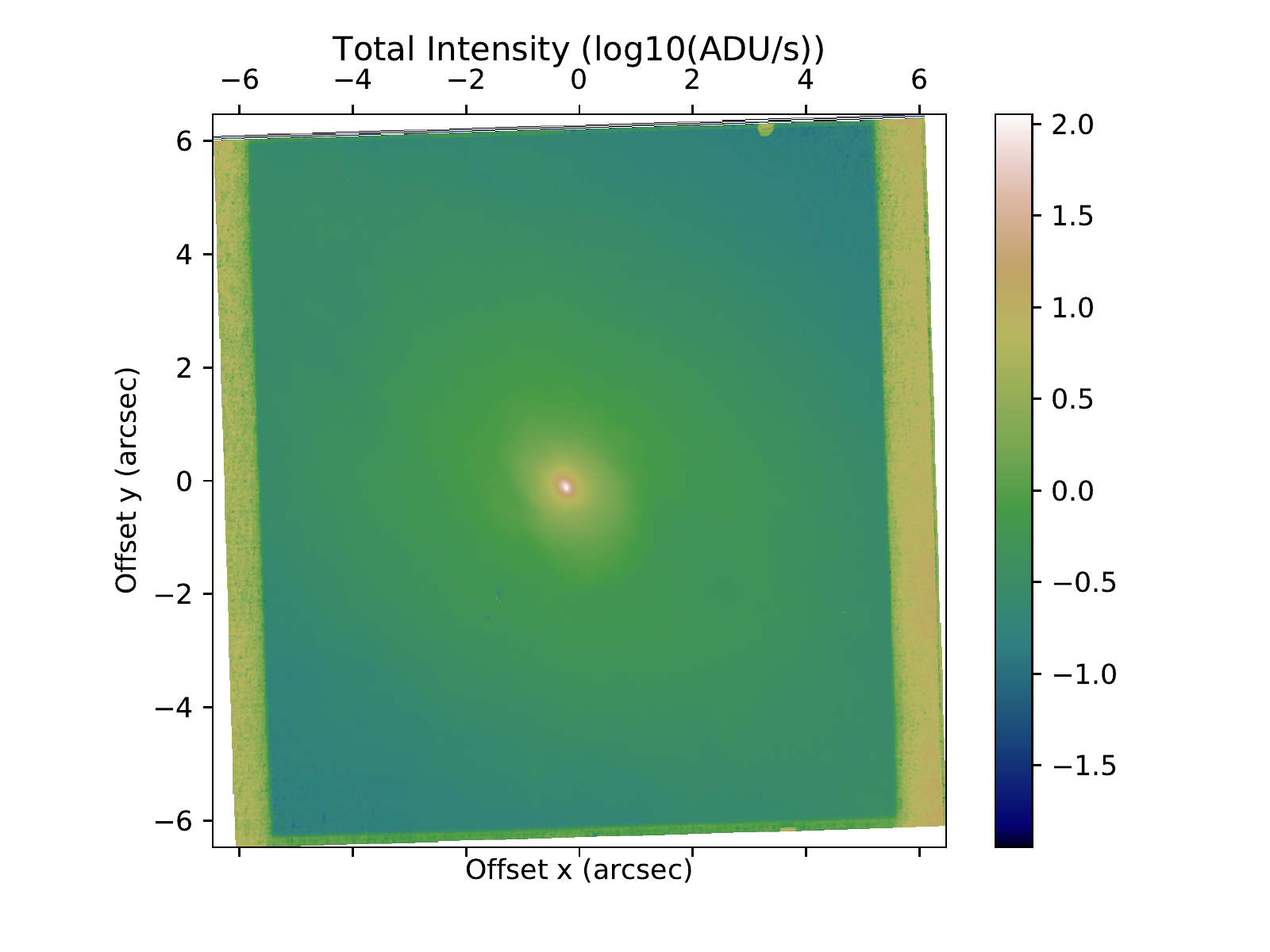}
 \includegraphics[width=0.45\textwidth,clip]{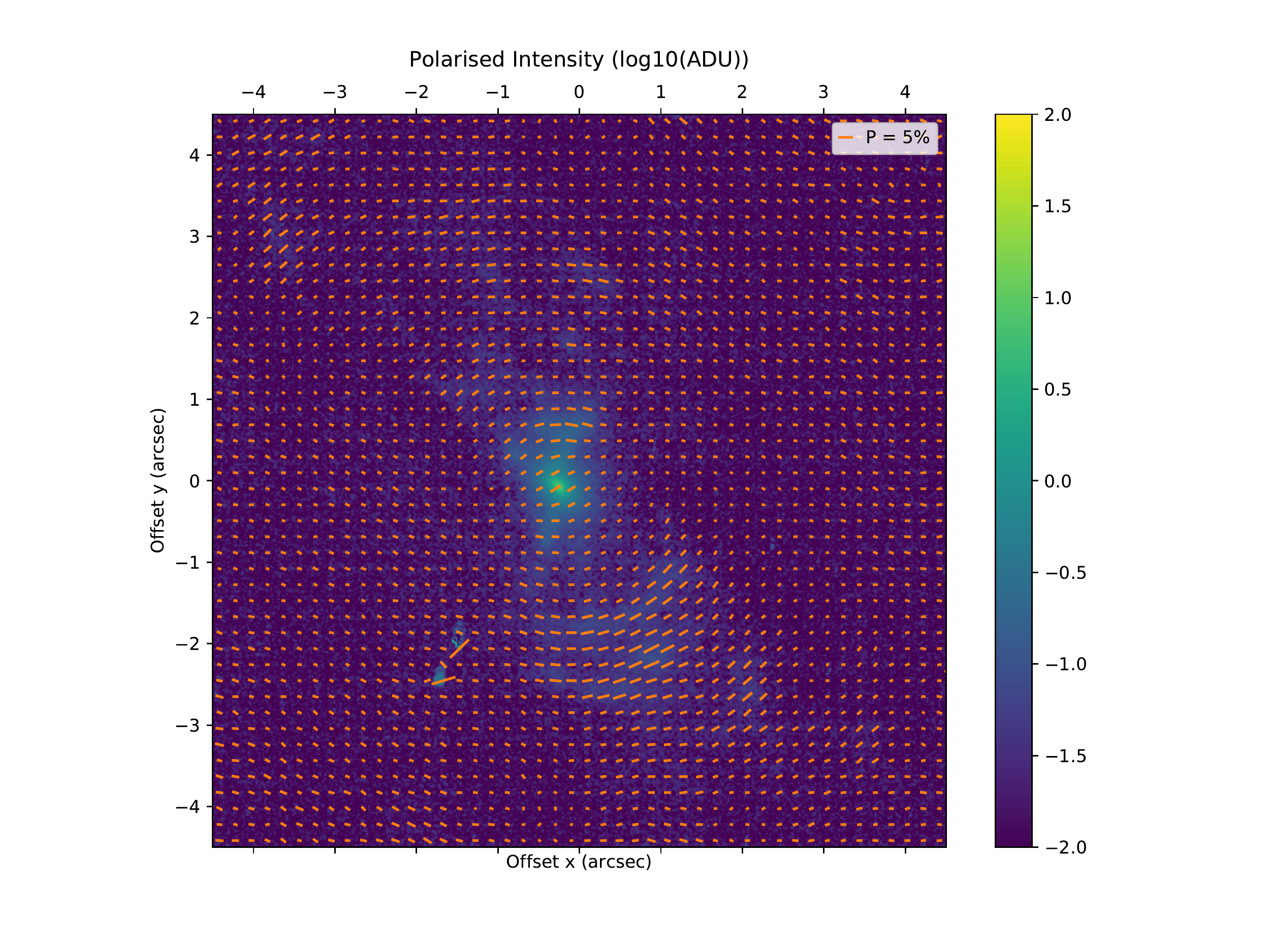}
  
 \includegraphics[width=0.45\textwidth,clip]{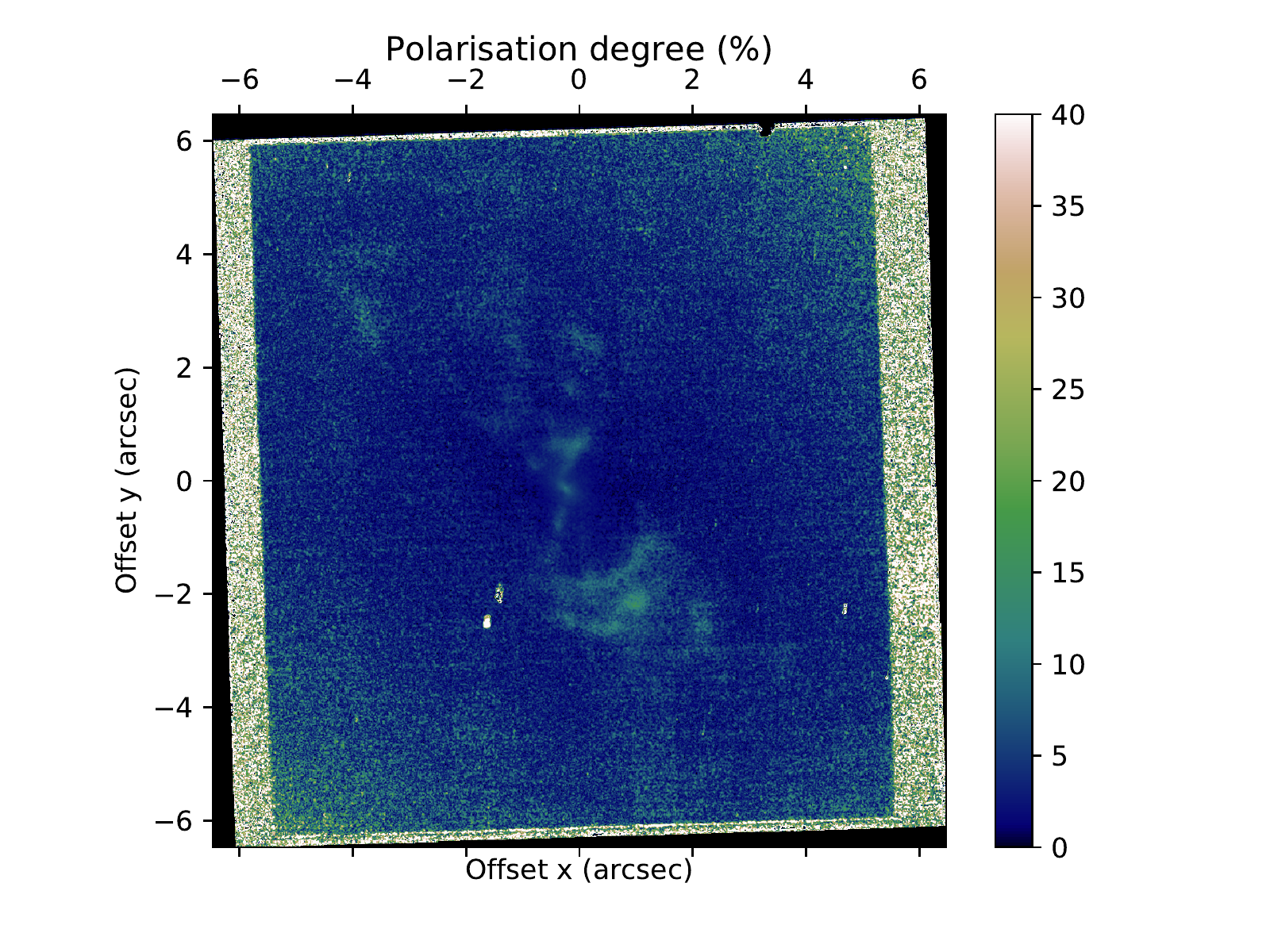}
 \includegraphics[width=0.45\textwidth,clip]{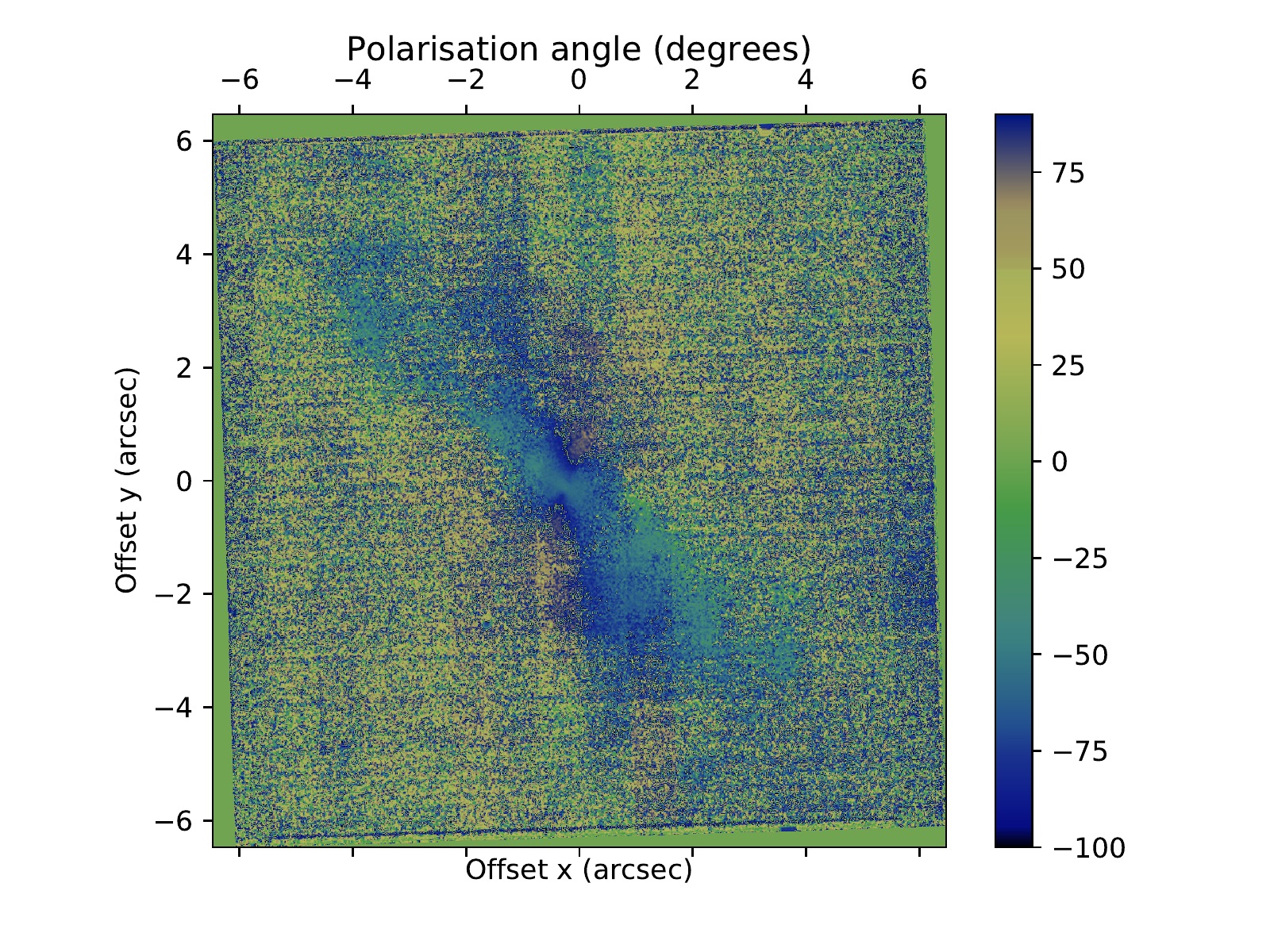}
  \caption{Maps of intensity (in $\log_{10}$(ADU/s), upper left panel), polarised intensity (in $\log_{10}$(ADU/s), upper right panel), linear degree of polarisation (in \%, bottom left panel) and linear angle of polarisation (in degrees, bottom right panel) in Cnt~H (1573~nm). Polarisation vectors have been over-plotted to the Polarised intensity maps of upper right panel, with a length relative to the local polarisation degree. A reference length for a 5~\% vector is shown. North is up, and East is to the left.}
  \label{fig:NB1}
\end{figure*}

\begin{figure*}[ht!]
 \centering
  \includegraphics[width=0.45\textwidth,clip]{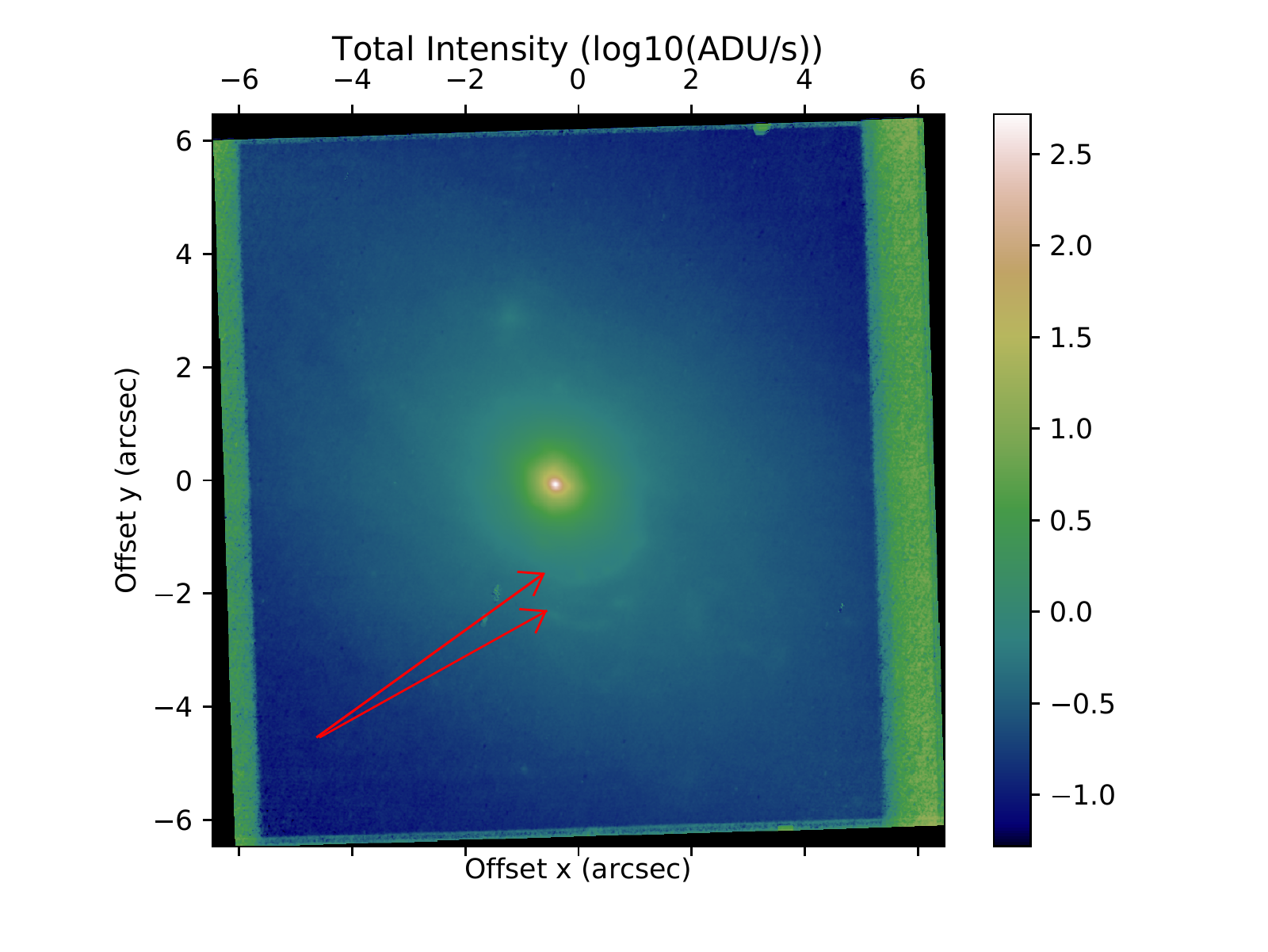}
  \includegraphics[width=0.45\textwidth,clip]{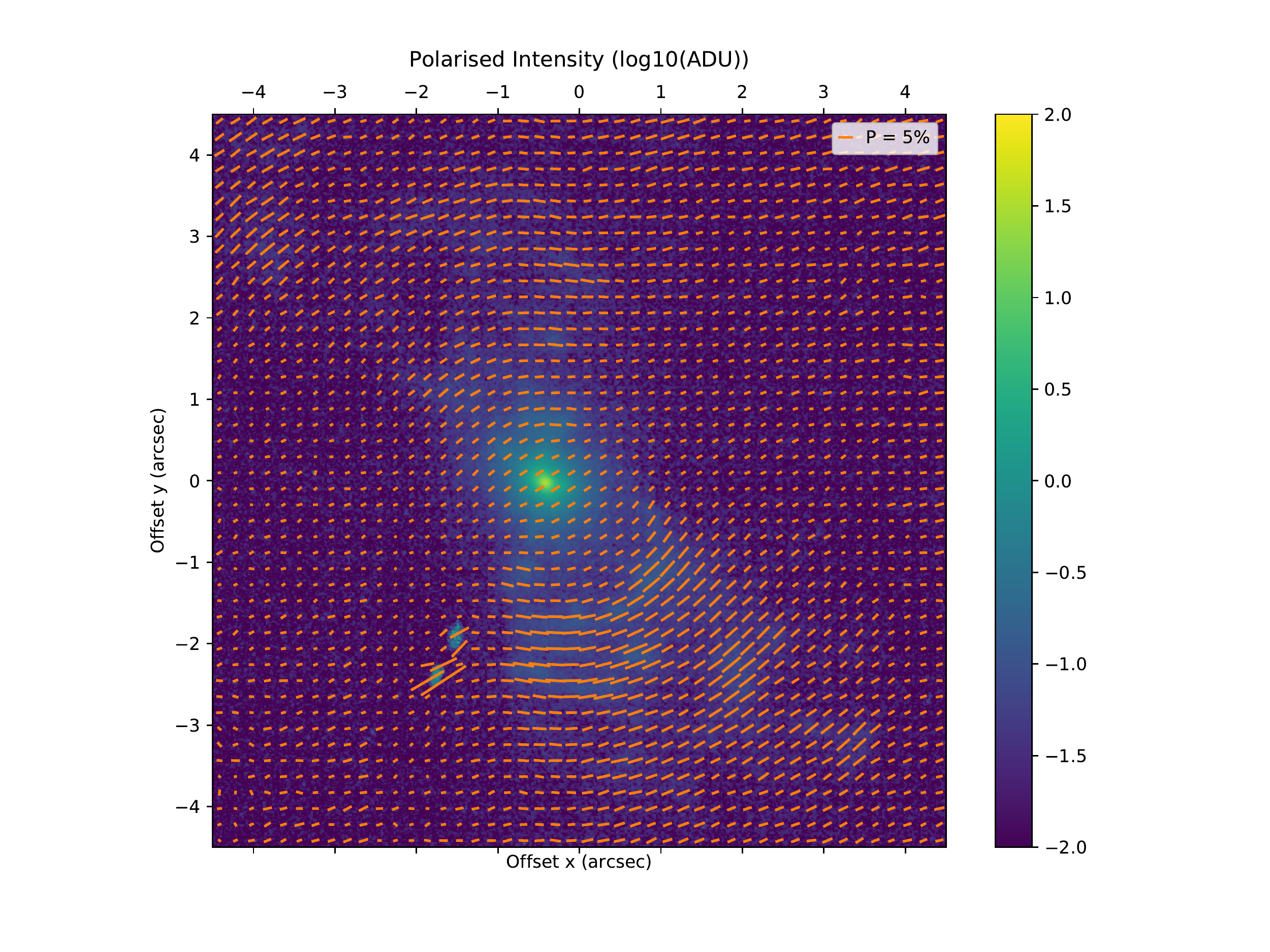}
  
  \includegraphics[width=0.45\textwidth,clip]{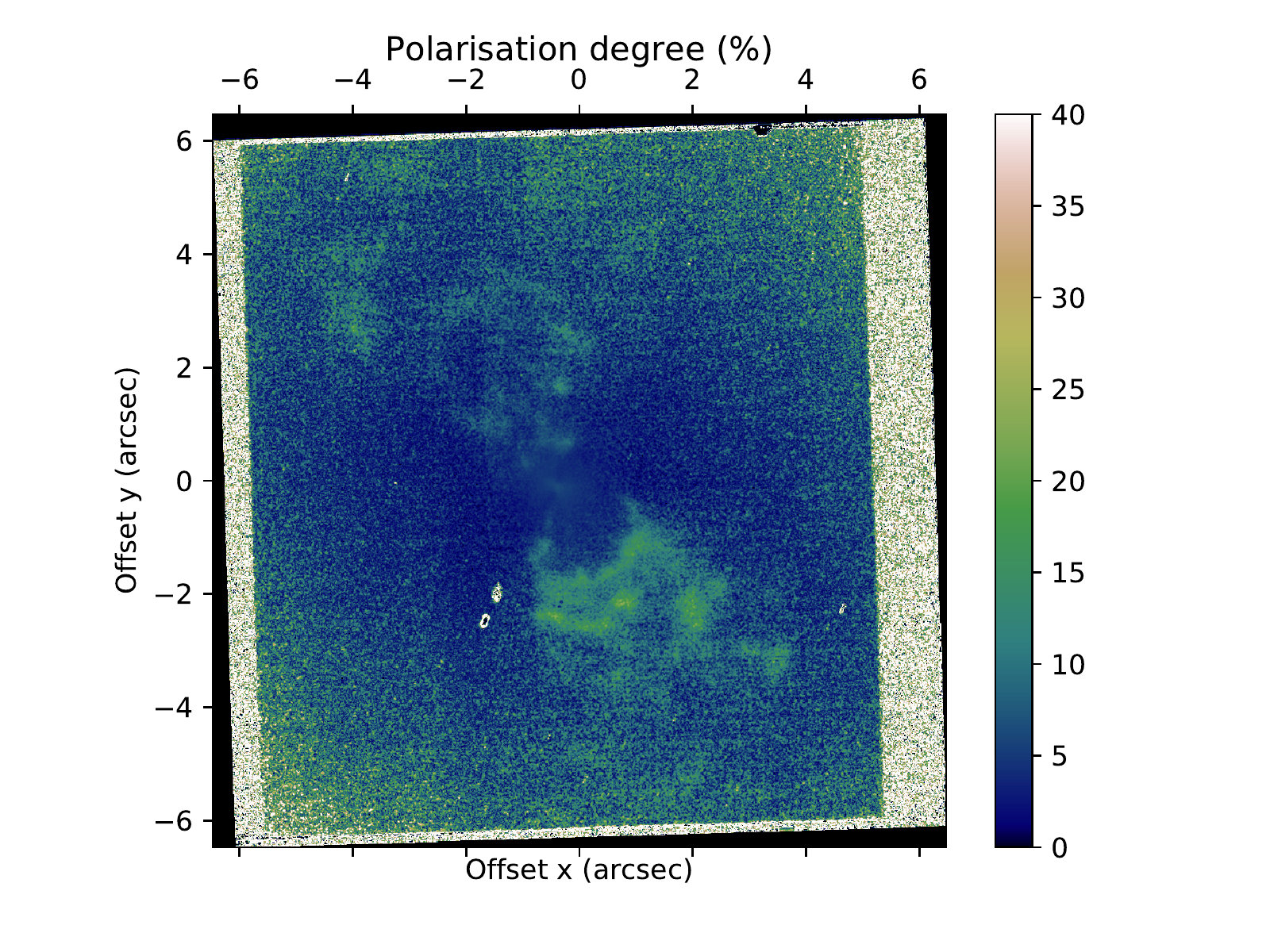}
  \includegraphics[width=0.45\textwidth,clip]{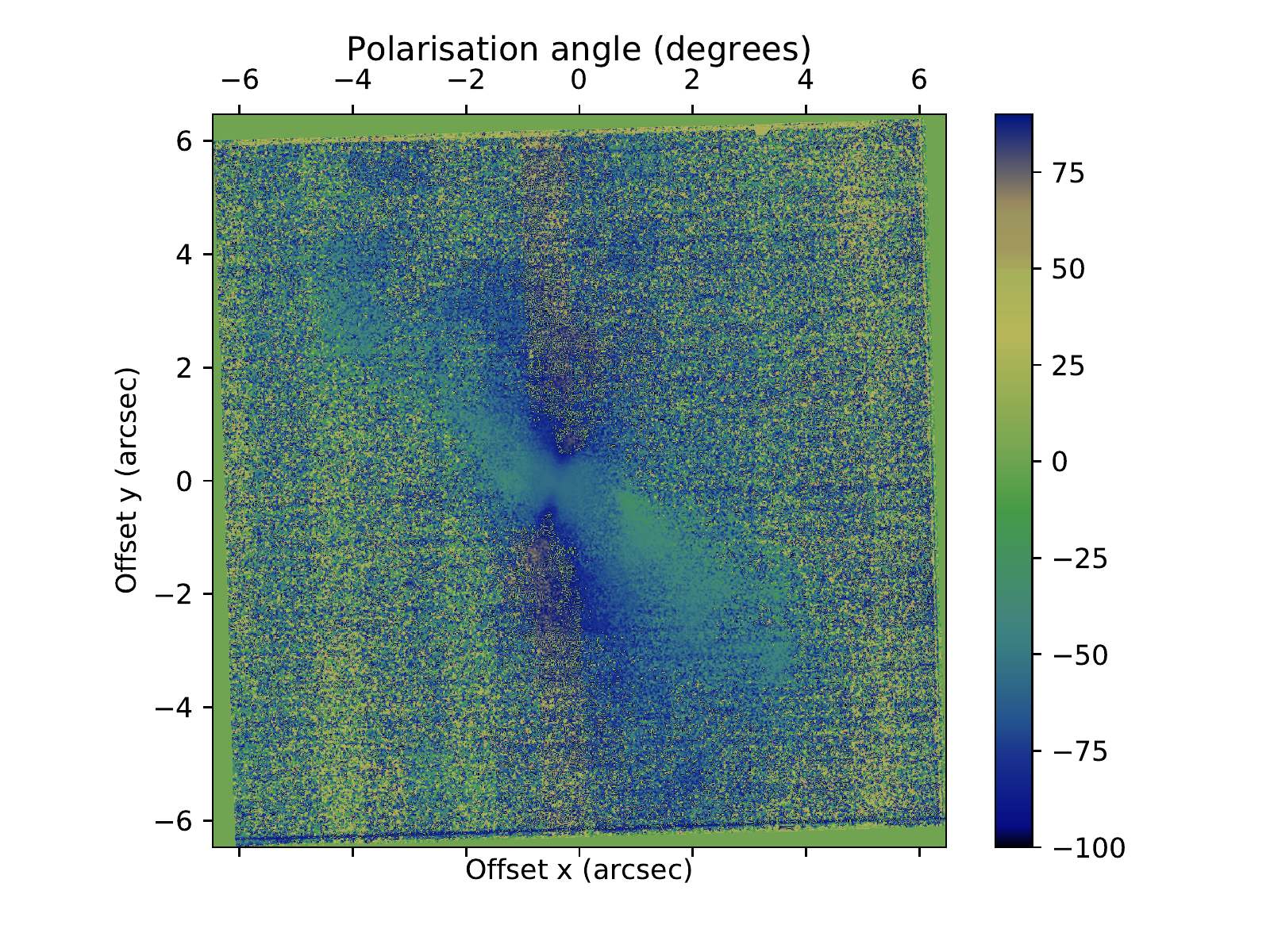}
  \caption{Same as Fig.~\ref{fig:NB1} for Cnt~K1 (2091~nm). Red arrows indicate the position of the two South-western arcs.}
  \label{fig:NB2}
\end{figure*}

\begin{figure*}[ht!]
 \centering
  \includegraphics[width=0.45\textwidth,clip]{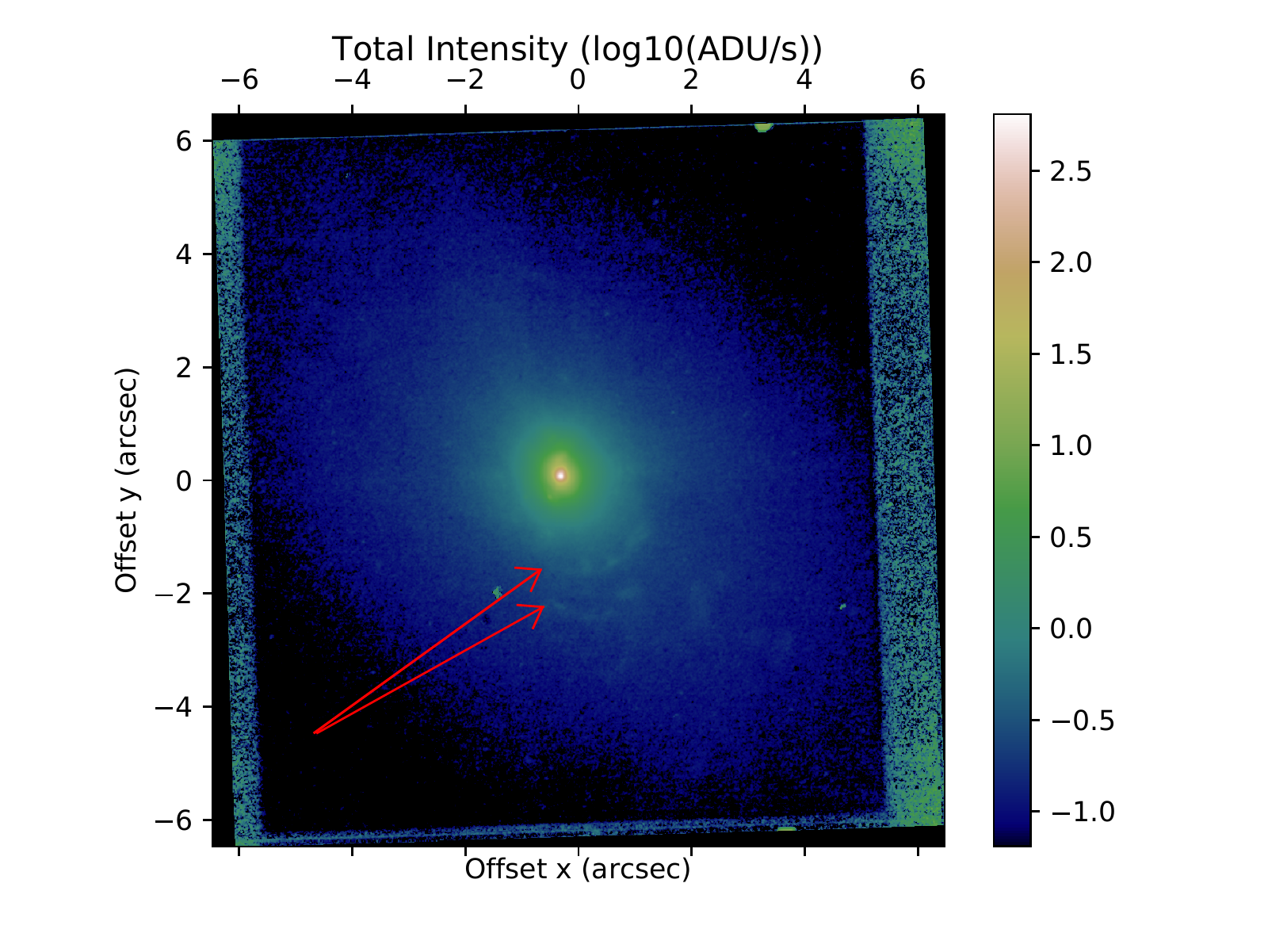}
  \includegraphics[width=0.45\textwidth,clip]{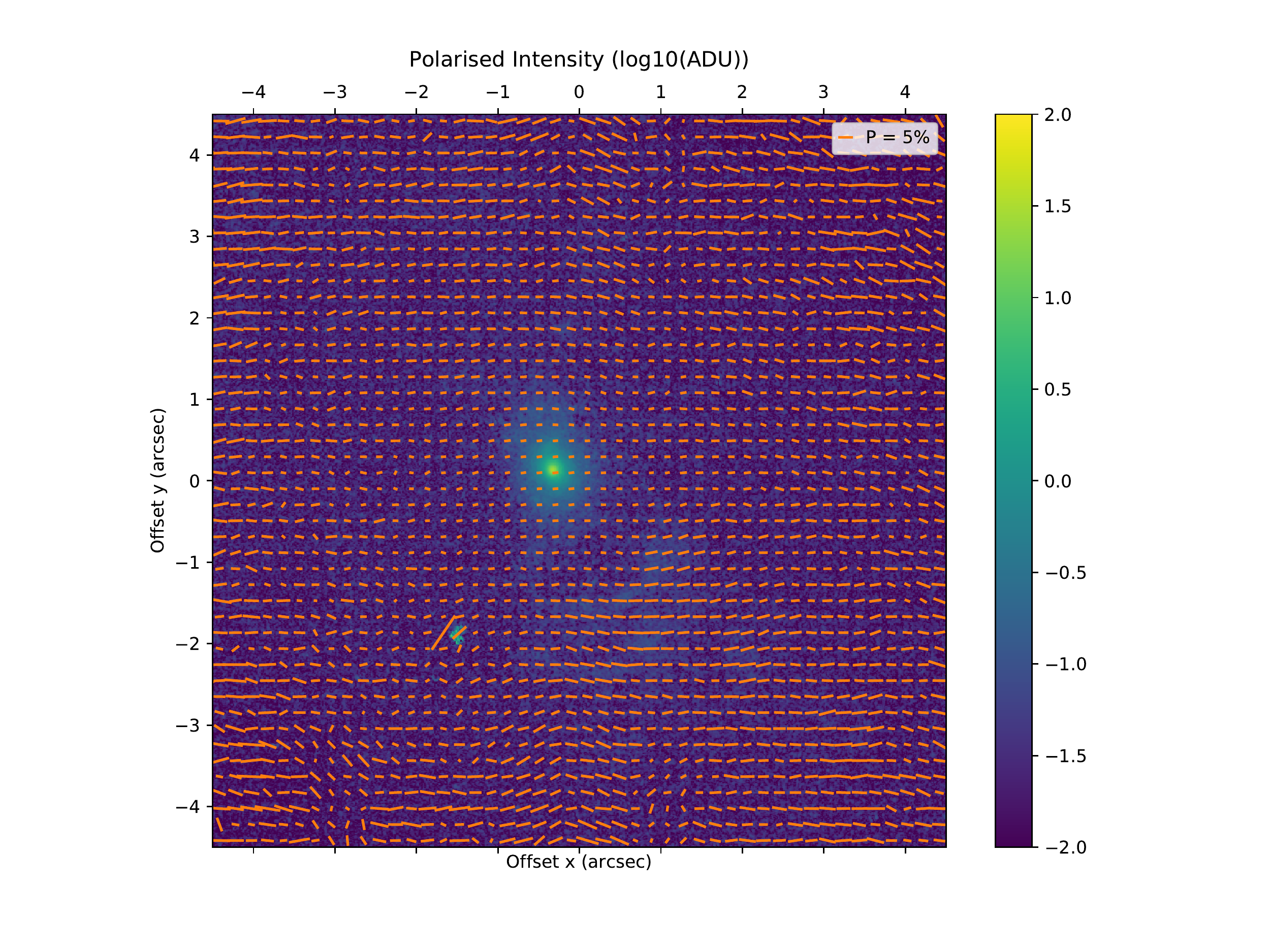}
  
  \includegraphics[width=0.45\textwidth,clip]{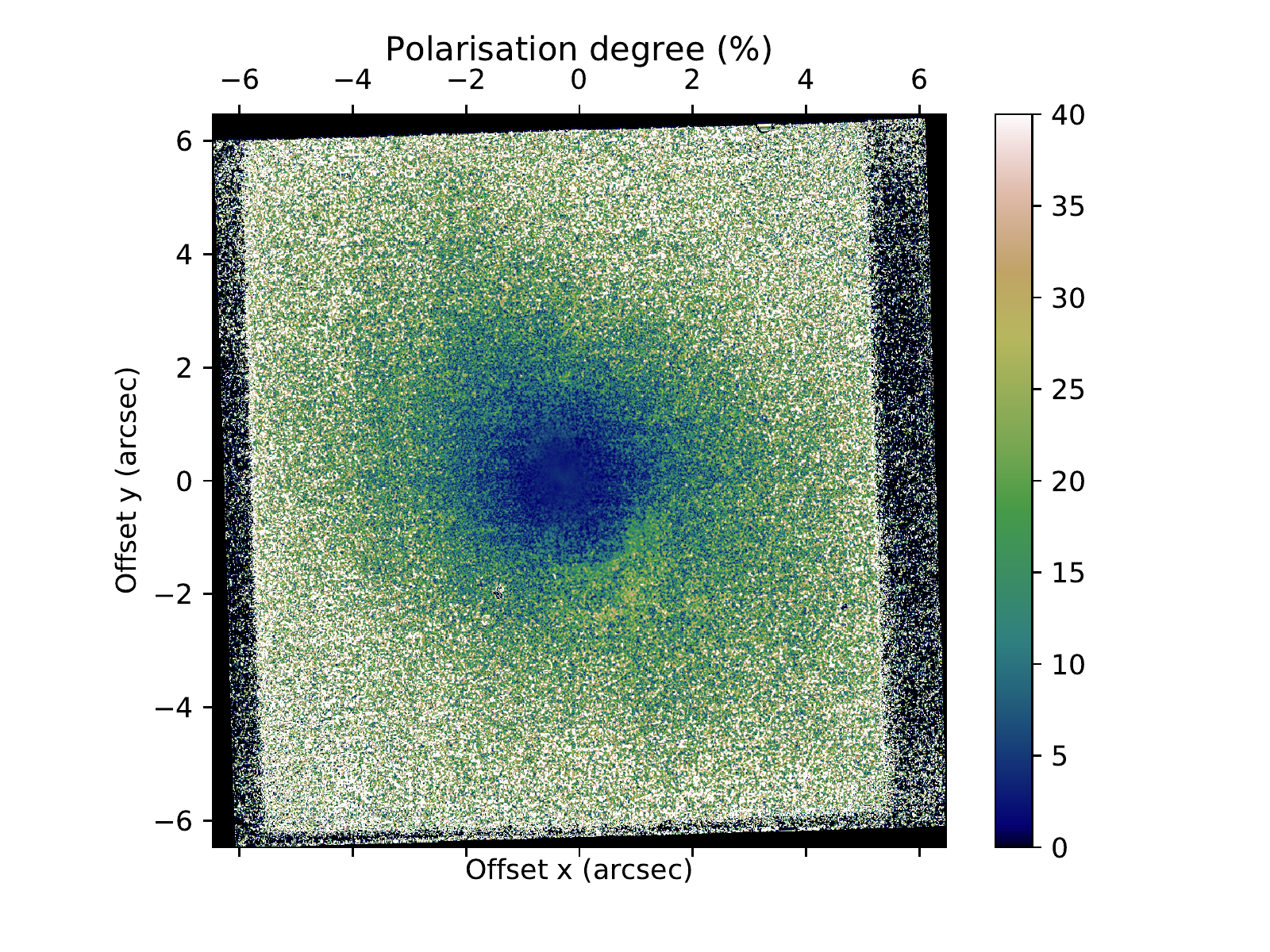}
  \includegraphics[width=0.45\textwidth,clip]{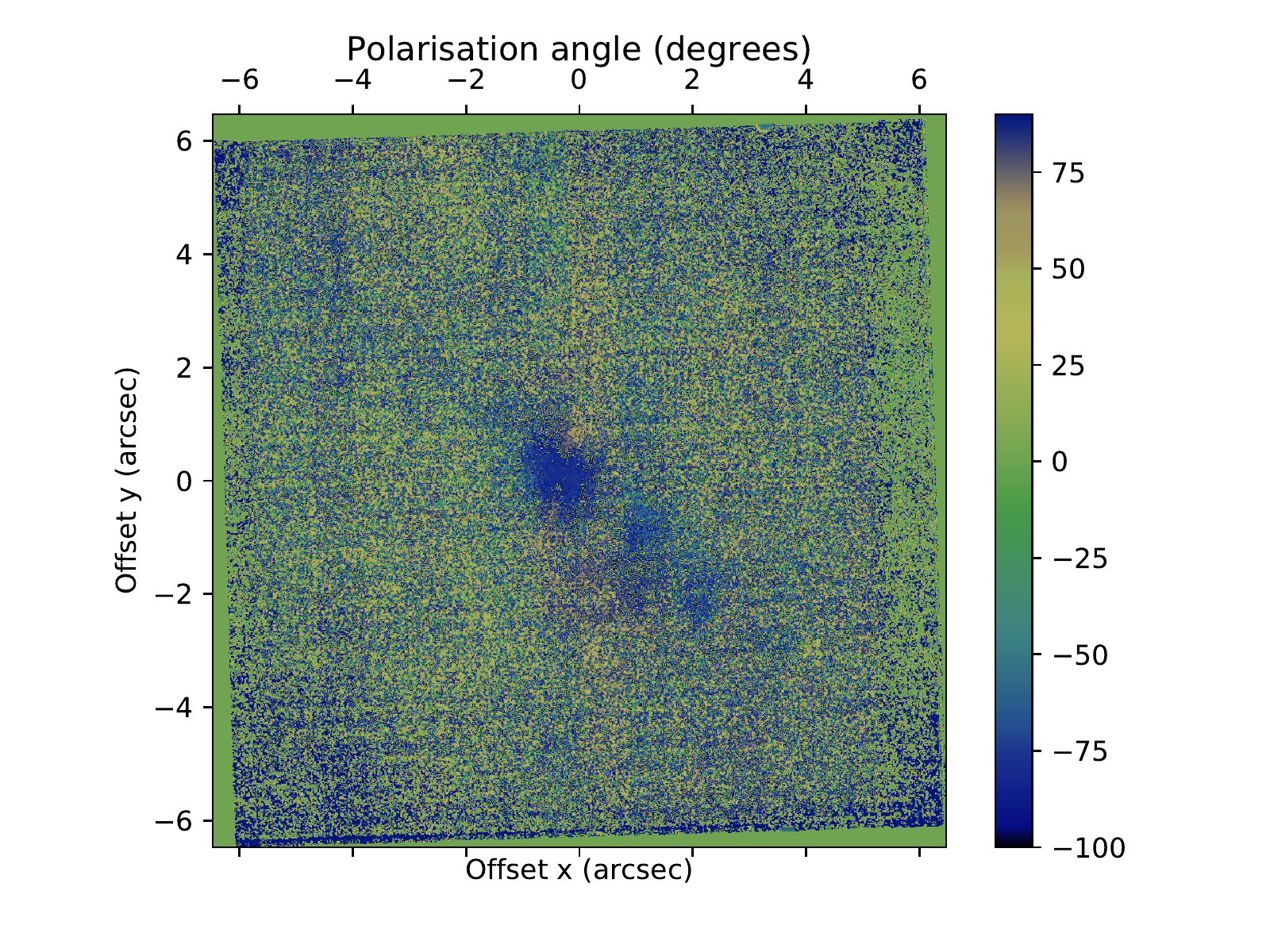}
  \caption{Same as Fig.~\ref{fig:NB1} for Cnt~K2 (2266~nm). Red arrows indicate the position of the two South-western arcs.}
 \label{fig:NB3}
\end{figure*}

Maps were then corrected from distortion and true North orientation. Images were shrinked along the vertical axis by a factor 1.006 and rotated using the orientation of the true North as measured by \citealt{Maire2016} of $-1.75^\circ$. 
Final NB maps, for IRDIS, with pixels of 0.01225'' (more precisely evolving between 0.012251'' and 0.012265'' per pixel, from H to Ks band, according to \citealt{Maire2016}) for the total and polarised intensities, the degree of linear polarisation, and the linear angle of polarisation are shown in Figures~\ref{fig:NB1}, \ref{fig:NB2} and \ref{fig:NB3}. 

\section{Polarisation in selected areas}

As can be seen in Figures~\ref{fig:NB1}, \ref{fig:NB2} and \ref{fig:NB3}, the Cnt~H (1573~nm) and Cnt~K1 (2091~nm) polarimetric maps harbour the same structures as already identified by \cite{Gratadour2015} on broad H and Ks bands (1625 and 2182~nm respectively). Namely, we identify a highly polarised (5 to 10~\%) central source surrounded by a well defined polarised hourglass shaped region tracing the double ionisation cones and two low polarisation regions on the perpendicular direction to the hourglass axis at the level on the waist (North-West and South-East), interpreted as a signature of the obscuring material \citep{Gratadour2015,Grosset2018}. The Cnt~K2 (2266~nm) maps share some common points with the previous two wavelengths, the polarised central source is also present, although less polarised, and the low polarisation region have a similar extension. However they also show some differences, the hourglass shaped structure being identified on the polarised intensity map but harder to distinguish on the polarisation degree and angle maps and the polarisation angle maps showing rather different behaviour. These peculiarities will be discussed in Section \ref{sec:K2}.

With a smaller field of view (FOV) of only $3.6'' \times 3.6''$, the ZIMPOL NR (645.9~nm) image is to be compared with the central region of the NIR maps (FOV of $11'' \times 12.5''$). 
In the polarisation degree map, we can recognise the elongated North-South structure seen in the inner 1'' of the NIR maps, but no other structure can be identified.

\vspace{0.5cm}
To investigate the properties of the different identified structures with respect to the wavelength, we extracted the global polarimetric signal from several peculiar regions in the four NBs and the two broad bands (BB). We will focus on three important regions, the two south-western arcs laying between 1.5 and 2.5'' from the photo-centre, within the ionisation cone, the very central region at the photo-centre and the low polarisation double region surrounding the central source on the South-Est and North-West directions, between 0.5 and 1.5'' from the centre.

\subsection{Polarisation uncertainty}

One major effect to be taken into account when measuring polarisation within aperture is the impact of the aperture's size on the measured polarisation. It is expected (see Table 2 of \citealt{Lopez-Rodriguez2016} and \citealt{Marin2018} for polarisation dependency on the aperture size in NGC~1068) that the larger the aperture is the smaller the polarisation degree due to both increasing dilution by non polarised light and the inclusion of differently polarised light within the aperture. This later effect is especially important here since we do see clear differences at small scale between regions in the inner 2'' of the AGN, in particular between the very central region, the low polarisation region and the South-western arcs. One major advantage of HAR in polaro-imaging is indeed to better resolve the polarised emission.

As discussed in \cite{Tinbergen1996}, it is difficult to evaluate the quality of polarimetric maps. Signal to Noise Ratio (SNR) for example is not well defined since a higher intensity will not correspond to higher degree of polarisation. Furthermore, when the degree of polarisation is small, neither the polarisation angle nor the polarisation degree follow Gaussian distributions. 

We thus used two different methods, detailed in Appendix~\ref{App:polar_sig} to evaluate the uncertainty on our polarisation measurements. In our apertures, this uncertainty is generally ranging between 15 and 20~\% (see error bars of Figure~\ref{fig:SA} and Figure~\ref{fig:centre}).

\subsection{South-western arcs}

\begin{figure}[ht!]
 \centering
 \includegraphics[width=0.45\textwidth,clip]{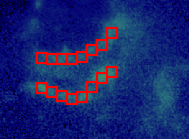}
  \caption{Position of the apertures along the two arcs, South-West from the central source, on top of a polarisation degree map in H (1.6 $\mu$m).}
  \label{fig:SA_ap}
\end{figure}

\begin{figure*}[ht!]
 \centering
 \includegraphics[width=0.45\textwidth,clip]{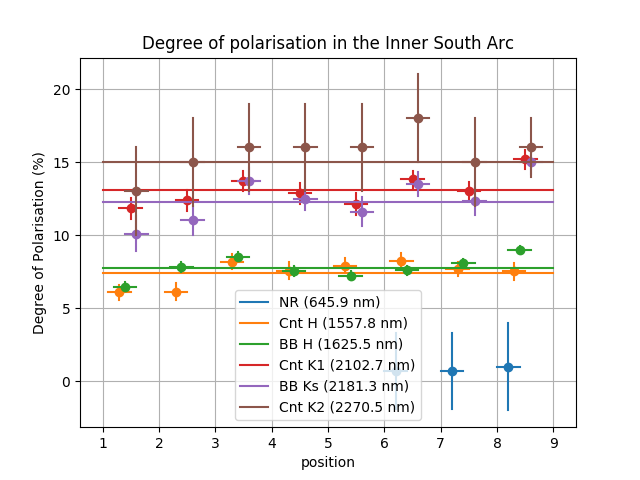}
 \includegraphics[width=0.45\textwidth,clip]{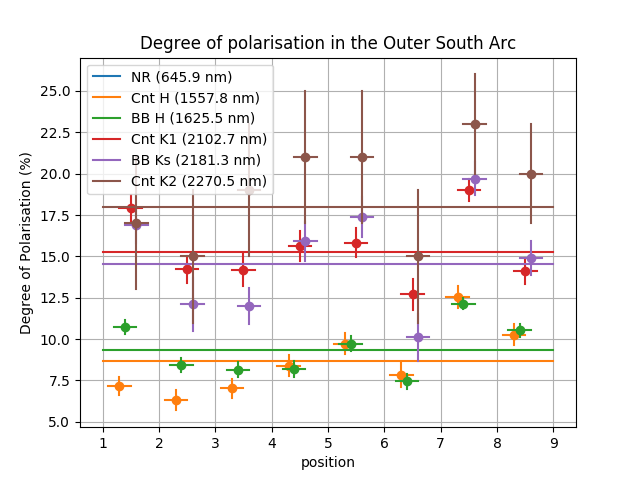}
 \includegraphics[width=0.45\textwidth,clip]{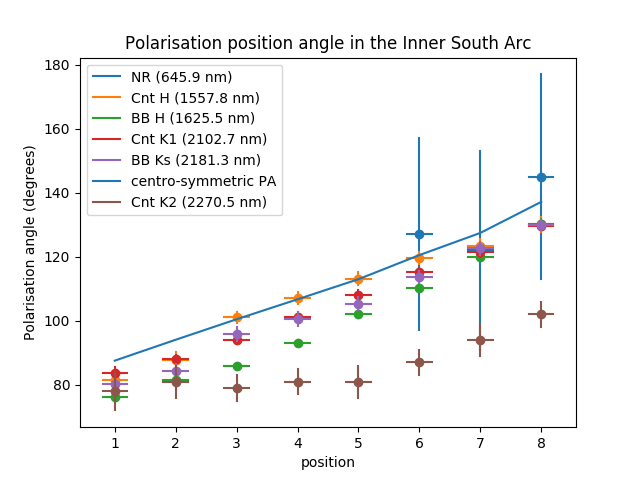}
 \includegraphics[width=0.45\textwidth,clip]{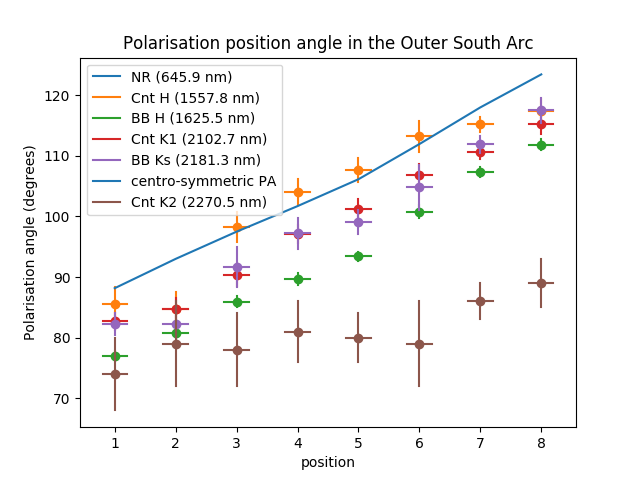}
  \caption{Degree (in \%, first row) and angle (in degrees, second row) of polarisation as a function of position in the arc, from East to West, for the inner South-western arc (first column) and the outer South-western arc (second column), for all the observing bands (colour coded). Average value of the polarisation degree in a given band is displayed by the respective colour horizontal lines. In plots of the second row, blue line corresponds to the expected polarisation position angle for these apertures if due to single scattering of light coming from the photo-centre of the AGN.} 
  \label{fig:SA}
\end{figure*}

The two South-western arcs, identifiable around 2'' South-West of the central source, are detected in all the NIR polarisation maps and directly on the intensity map of Cnt~K1 and Cnt~K2 (2091 and 2266~nm respectively). We extracted the polarisation in 8 regions, of size $0.2''\times 0.2''$\footnote{The size of the aperture was selected to be larger than any achieved resolution and to contain sufficient flux for significance, while keeping the information as local as possible}, from East to West along each of the arcs. A zoom on the two arcs and the apertures used here are displayed in Figure~\ref{fig:SA_ap}. Aperture  measurements are based on extraction of signal on the I, Q and U maps, translated onto final polarimetric parameters using eq. \ref{eq:P} and \ref{eq:theta}. Error bars are based on the ``pseudo-noise'' approach of eq. \ref{eq:sigma_pol}. The measured polarisation degree and polarisation position angle for all filters are displayed in Figure~\ref{fig:SA}.

\begin{figure}[ht!]
 \centering
 \includegraphics[width=0.48\textwidth,clip]{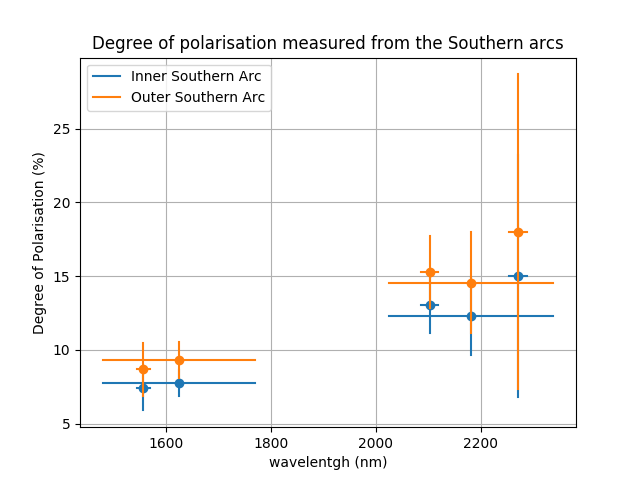}
  \caption{Degree of polarisation (in \%), averaged for all 8 apertures in both arcs (colour coded) as a function of wavelength.} 
  \label{fig:P_SA}
\end{figure}

The first row of Figure~\ref{fig:SA} shows that the degree of linear polarisation (displayed versus a position offset in this figure) in the South-western structures tends to increase as a function of wavelength in all apertures. This is illustrated in Figure~\ref{fig:P_SA}, that represents the evolution of the averaged degree of polarisation as a function of wavelength in both arcs. We note that they evolve on a very similar linear increase, evidence for an unique mechanism being responsible for the polarisation in the two arcs. Furthermore, the degree of polarisation, always higher in the outer arc, might suggest a scattering geometry configuration closer to 90$^\circ$ for this arc.

The polarisation position angle, displayed in the second row of Figure~\ref{fig:SA}, is consistent with the expected variation for a centro-symmetric pattern, which is shown as the blue curve, computed as the orthogonal direction to the photo-centre for every aperture position. Cnt~K2 (2266~nm) is the most divergent from this pattern, and this case will be discussed in Section~\ref{sec:K2}. We note however that there is a trend, stronger for the outer South-western arc, for the measured polarisation position angles to be offset by up to 20$^\circ$, below the expected angles. Possible reasons for this offset are multiples and are difficult to disentangle. An offset between the emission centre of the scattered light and the reference centre of the image (the photo-centre in NIR) would produce an offset, but this offset should not depend on wavelength. A better explanation would be the addition of slightly different polarisation orientation all along the extension of the arcs on the line of sight (LOS). Inclination of these arcs with respect to the plane of the sky would also introduce an offset in the polarisation since the polarisation angle would not be 90$^\circ$ in this case.


\subsection{Central region}
\label{sec:centre}

We analysed polarisation in the inner 2'' of the AGN in different apertures in order to properly interpret this complex region, where multiple polarisation mechanisms such as dichroic absorption by aligned elongated grains, Thomson scattering on electrons and Mie scattering on dust grains, may be in competition.

\begin{figure}[ht!]
 \centering
 \includegraphics[width=0.48\textwidth,clip]{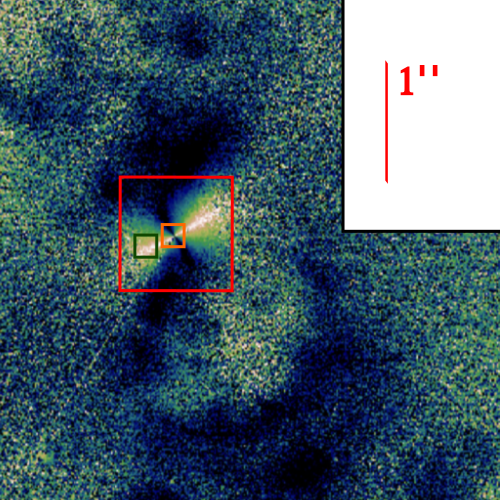}
   \caption{Position and size of the apertures used for polarimetric measurements, superimposed to the differential polarisation angle map of BB H (1625~nm, from \cite{Gratadour2015}). The orange aperture corresponds to the ``very centre'', the green to the ``off-centre'' and the red to the ``central region'' as described in text.}
   \label{fig:centre_pos}
 \end{figure}

\begin{figure*}[ht!]
 \centering
 \includegraphics[width=0.48\textwidth,clip]{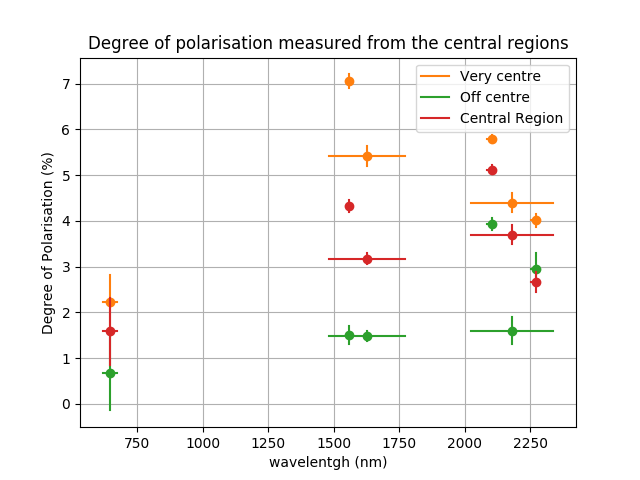}
 \includegraphics[width=0.48\textwidth,clip]{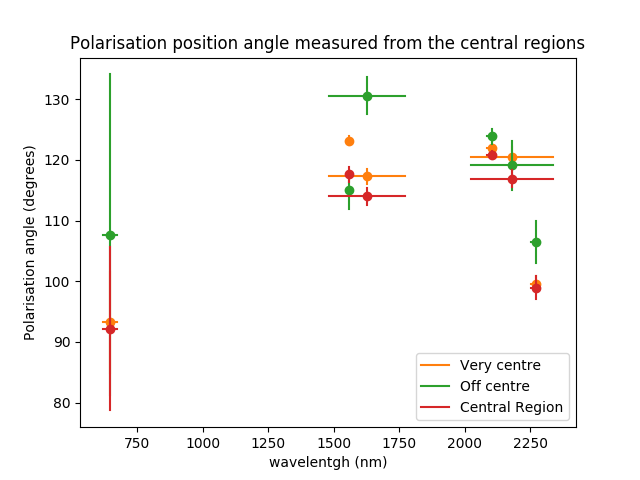}
  \caption{Degree (in \%, left panel) and angle (in degrees, right panel) of polarisation as a function of wavelength (in nm) for the three apertures in the central region of NGC~1068. ``very centre'' refers to the very inner 0.2'', ``off-centre'' to the region with an offset on the torus extension direction and ``central region'' to the large aperture integrating all the flux. See text of Section~\ref{sec:centre} for more detail about these apertures.}
  \label{fig:centre}
\end{figure*}

In particular, we analysed the very central 0.2'' (hereafter ``very centre''), encompassing the bright photo-centre, that shows higher polarisation degree than in its immediate surrounding and that dominates the polarised flux. We also extracted polarisation from the very low polarisation regions surrounding this photo-centre, in the South-East and North-West directions, on the expected extension of the obscuring material (hereafter ``off centre''). We finally considered a large aperture of 1'' size (hereafter the ``central region'') in order to compare this study to other polarimetric investigations references of this region at larger scale. These apertures are identified in Figure~\ref{fig:centre_pos} and the measured polarisation degree and polarisation position angle, as a function of wavelength, are displayed in Figure~\ref{fig:centre}.

We also noted that, because the centre is very bright, the wings of the PSF of the central source could affect the innermost measurements. Indeed, because this central source is polarised, part of the measured polarisation could be contaminated by the higher polarisation from the centre, especially in the low polarisation regions. The PSF in the K (2182~nm) band for SPHERE, contributes up to 10~\% of the flux at $\approx$~0.12'' from the centre, as shown with the same data by \cite{Rouan2019}, for example by their Figure~3, comparing the K profile to the PSF profile at K. This contribution of the central PSF then drops at larger distances. In particular, this implies that the polarisation degree contamination is smaller than 0.7~\% in K in the ``off centre'' aperture, which ranges between $\approx$~0.2 and 0.4'' from the centre. As this PSF contribution is even smaller in H (1625~nm), we thus ensured that it does not affect significantly our measurements.

Figure~\ref{fig:centre} reveals very different trends for the three apertures. The large aperture ``central region'' has an almost linearly increasing polarisation degree from 1.5 to 4~\% between 0.7~\textmu m to 2.2~\textmu m (left panel), while the ``very centre'' and the ``off centre'' regions evolve in a less monotonic way. As expected, the ``off centre'' aperture displays the lower polarisation degrees, $<2~\%$, with increases only at 2.0~\textmu m to about 2-3~\%. At the opposite, the ``very centre'' has a higher polarisation degree, starting at about 2~\% in R, increasing to a maximum of 5-7~\% at $\approx 1.6$~\textmu m and then decreasing to $\approx 4$~\% at 2.2~\textmu m.

As regards the polarisation position angle, it does not seem to undergo noticeable evolution, for all apertures and wavelengths. Measurements range between 90 and 130 degrees, encompassing the expected polarisation orientation in the central region in the NIR of $\approx 120^\circ$ as measured by previous studies \citep{Packham2007,Gratadour2015}.

\subsection{The case of Cnt~K2}
\label{sec:K2}

In most of the polarimetric maps and measurements in this work, Cnt~K2 (2266~nm) results diverge from the other NIR ones. In particular, it exhibits a rather different behaviour compared to the Cnt~K1 and Ks filters, which are however at close wavelengths (2091 and 2182~nm respectively). The Cnt~K2 intensity map features a narrower central peak, leading to a lower flux in the surrounding of the very central peak.  This limits the precision of polarimetric measurements since non significant differences between Q and U maps at low SNR are creating random polarisation levels and explains at some extend these differences. Indeed, these random polarisation levels outside the 1'' inner region pollutes the true polarised signal making its extraction more uncertain. We also note that the polarisation in the most central region is slightly lower in this band, even on the area not affected by this problem. More problematic is the polarisation position angle, which reveals very different patterns from other bands, within regions however shaped very similarly. Both the polarisation position angle and degree are modified by a selection of the frames according to the SPHERE derotator position angle (see Appendix~\ref{App:derot} for more details), and we thus investigate the effect of this on our polarimetric measurements in Appendix~\ref{App:CntK2}.

Since both the binned polarisation degree and polarised intensity maps of Figure~\ref{fig:cntK2} reveal the features observed in the other NIR maps (especially the same polarisation degree of $\approx$~15~\% in the South-western arcs), the difference in intensity is likely to be the most impacting effect on polarisation degree for this NB. Using the selection for raw images only leads to a lower SNR. Thus, our polarisation degree measurements should not be contaminated at a significant level by systematics coming from derotator depolarisation issues. However, binning of polarisation position angle map (Figure~\ref{fig:cntK2}) does not lead to an improvement and an impact of the derotator position angle on the measured angle is very likely there. We will thus not go further in our investigations for this polarisation position angle map.

\section{Simulation of dichroism}

\cite{Grosset2018} succeeded in reproducing using the radiative transfer code MontAGN the constant polarisation position angle and low polarisation degree over the central region of NGC~1068 (about $20 \times 60$~pc). This analysis was based on a double scattering mechanism on spherical dust grains: photons undergo two scattering events, one in the ionisation cone where they are scattered back toward the equatorial plane where they are scattered a second time in the direction of the observer. These results are still in good agreement with the present paper's measurements. Especially, the overall polarisation position angle in the inner region and the low polarisation degree found in the South-East / North-West extension surrounding the very centre, are compatible with double scattering and is therefore strengthening this interpretation. However, the relatively high  ($\approx 15~\%$) polarisation degree of the photo-centre itself could not be explained by these simulations.

\subsection{Framework}

Here we extend this simulation work through simulations of polarimetric signal produced by dichroism. We used MontAGN \citep{Grosset2018} to reproduce absorption of photons along a straight path through a region containing aligned elongated dust grains. As we aim at reproducing a signal observed between 0.5 and 2.3~\textmu m, we focused on dichroic absorption. Indeed, at these wavelengths absorption is the dominant mechanism \citep{Efstathiou1997} since temperature is expected to be high enough for a significant dust emission below 2.3~\textmu m (ie. above $\approx$~400-500~K) only in the very inner region ($\le 1$~pc) of the obscuring material and could thus be included in the central source. Indeed, this emission would still have to suffer the dichroic absorption by the rest of the colder dust structure surrounding it and will have little impact on the final polarisation, despite being possibly intrinsically polarised. If this effect is present, it would lower the dominant polarisation orientation flux, and thus slightly decrease the dilution fraction\footnote{a factor of 2 in flux, unlikely large, would only lead to a lowering of the dilution fraction by up to $\approx$~5\% because of the flux ratio of $\ge$100 between the two polarisation orientation observed in this simulation work.} required to match the observed polarisation (see Section~\ref{sec:dilution} for more details about dilution).

As MontAGN does not allow yet the user to align the dust grains orientation along a preferential direction, it is impossible to simulate dichroic absorption in the scattering framework of \cite{Grosset2018}. We thus focused on dichroism that we investigated by simulating separately four polarisation orientation components (Q$^+$, U$^+$, Q$^-$ and U$^-$). For each of them, we set the dust properties to match the dust population encountered by this particular polarisation orientation. We then combined the output files to simulate the polarimetric properties of the resulting signal, generated by dichroic absorption only.

\subsection{Producing dichroism signal}

Following this method, we simulated polarimetric signals produced by dichroic absorption for a range of optical depth ($\tau_V$~$\in$~$ \left[2.5,200\right]$) and for two different grain axis ratios. We used a ratio of major axis to minor axis of 1.5 (hereafter ``r1:1.5'') and a ratio of 2 (hereafter ``r1:2''). In both cases, we used a MRN \citep{Mathis1977} size distribution with a power law of -3.5, ranging between 5 and 250~nm, with variation depending on the axis ratios. Thus, for the first population, we used distribution ranging between 5 and 125~nm for the minor axis and between 7.5 and 188~nm for the major axis. For the ``r1:2'' population, we kept the same ratio for minor axis (between 5 and 125~nm) and set the major axis distribution between 10 and 250~nm. We show in this paper results for silicates dust grains, as simulations with graphites gave similar results. In a first step, photons packets were emitted using a flat spectrum\footnote{we used a flat spectrum since we are more interested in transmissivity and since polarisation does not depend on the absolute intensity} between 0.5 and 3~\textmu m and dust densities were set to match the chosen optical depth at 0.5~\textmu m for each dust mixture integration.

Because elongated grains preferentially absorb photons polarised along their major axis, photons with different polarisation orientations will not encounter the same optical depth along their path through the same dust structure. For example, for our ``r1:1.5'' model we get values of $\tau_V$ of 43.2, 6.8 and 25.0 along the Q$^+$, Q$^-$ and U$^\pm$ polarisation orientations respectively for an averaged $\tau_V=25.0$.

\begin{figure*}[ht!]
 \centering
 \includegraphics[width=0.48\textwidth,clip]{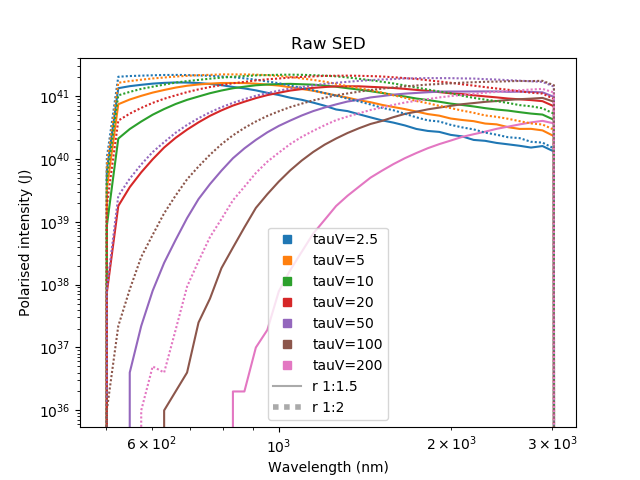}
 \includegraphics[width=0.48\textwidth,clip]{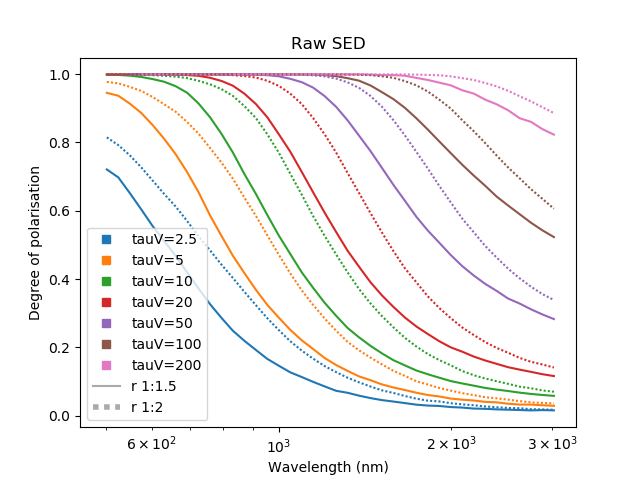}
  \caption{Spectra of polarised intensity (first panel, with arbitrary intensity unit) and degree of polarisation (second panel) from dichroism simulated with MontAGN.}
  \label{fig:dichro_raw}
\end{figure*}

This difference of optical depth will translate into different final fluxes for each of the polarisation orientations, thus creating a polarised signal. Simulated raw polarisation degree and polarised intensity spectra are displayed in Figure~\ref{fig:dichro_raw}.

Polarisation is very important at short wavelengths as shown by right panel of Figure~\ref{fig:dichro_raw}. Indeed, at short wavelengths, only one of the polarisation orientation propagates significantly through the dust structure because of its optical depth (low fluxes, as shown by left panel), creating a 100~\% polarised signal. At longer wavelength, the decrease of optical depth allow a fraction of the orthogonal polarisation to escape the obscuring material, and we thus observe a decrease in the polarised intensity and in the polarisation degree, especially for low optical depth models (where p~$\approx 0$, right panel). 

\subsection{Dilution}
\label{sec:dilution}

\begin{figure*}[ht!]
 \centering
 \includegraphics[width=0.48\textwidth,clip]{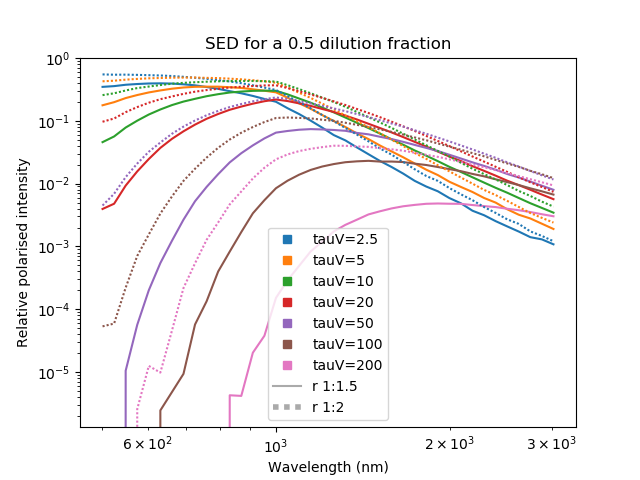}
 \includegraphics[width=0.48\textwidth,clip]{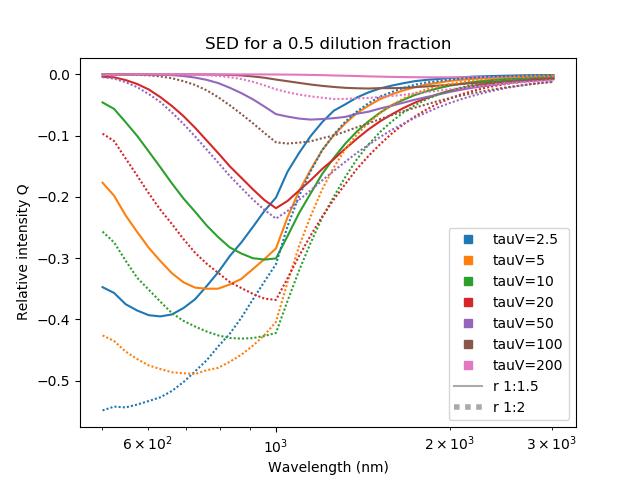}
  \caption{Spectra of relative polarised intensity (first panel) and normalised polarisation Q intensity (second panel) for an AGN emission from MontAGN simulations. Polarised intensity is normalised to the maximum of the dichroic flux.}
  \label{fig:QIp}
\end{figure*}

To compare our models with observations, we need to take into account additional effects. First, the flat spectrum used for emission needs to be adapted to our AGN case and we multiplied the output flux, after normalisation, by a typical emission spectrum. We used a simple source SED (like \citealt{Siebenmorgen2015} for example) based on \cite{Rowan-Robinson1995} approximation for AGNs:
\begin{equation}
\lambda F_\lambda \propto \lambda^{-0.5} \mathrm{~if~\lambda < 1~ \mu m;}\\
\lambda F_\lambda \propto \lambda^{-3} \mathrm{~else.}
\end{equation}
Polarised intensity and polarisation Q corrected spectra are shown in Figure~\ref{fig:QIp}, for $\tau_V$ in the range 2.5 - 200 and for the two grain axis ratios. The Q polarisation spectrum (right panel) is the main polarisation indicator since we aligned the grain axis along the $\pm$Q axis (Photons with a $\pm$U polarisation orientation will encounter a 50/50 grain axis ratio mixture), and indeed, values of U do not exceed 0.0015 (0.15~\% of the maximum of the dichroic flux). Absolute value of polarisation Q evolution shows two phases. It first increases\footnote{Q has negative values since here polarisation is produced horizontally with our model} when the -Q photons, corresponding to the lowest optical depth, escape first and decrease when +Q photons become also able to escape. The polarised intensity graph (left panel), translates the same trends when weighted by total intensity.

In a second step, we need to properly account for dilution by all other emission sources contributing to the flux measured within the apertures. This is very critical since the high polarisation degree values on the left panel of Figure~\ref{fig:dichro_raw} (at wavelength $\leq 1$~\textmu m) corresponds to very low fluxes (left panel) and will therefore be reduced much more than the lower polarisation degree obtained at higher fluxes (at $\lambda \geq 1$~\textmu m).

\begin{figure*}[ht!]
 \centering
 \includegraphics[width=0.48\textwidth,clip]{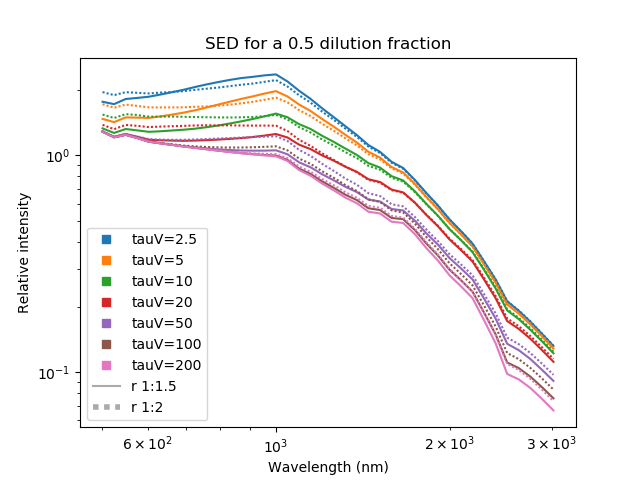}
 \includegraphics[width=0.48\textwidth,clip]{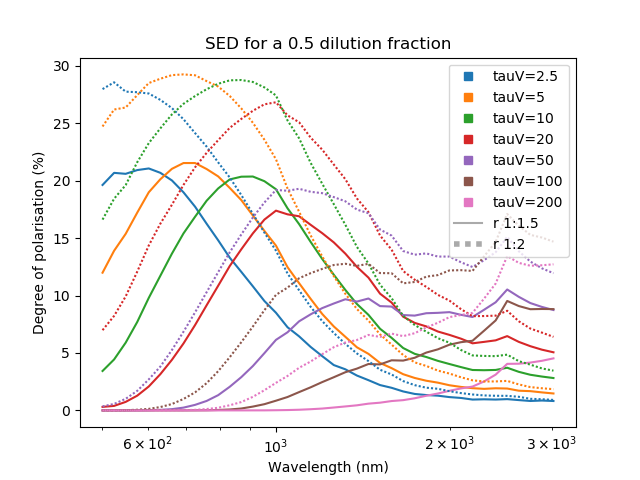}

 \includegraphics[width=0.48\textwidth,clip]{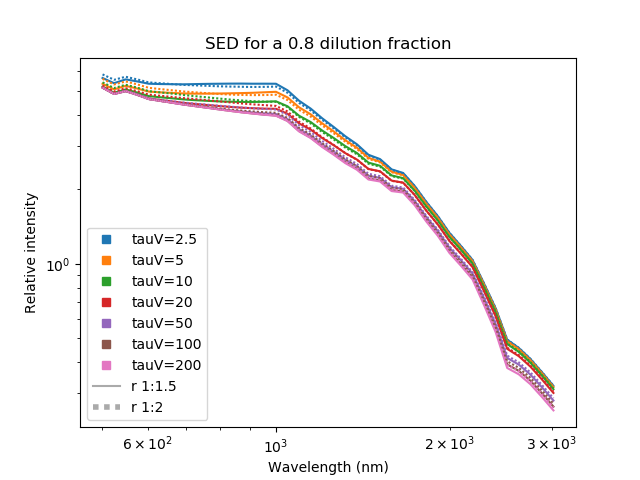}
 \includegraphics[width=0.48\textwidth,clip]{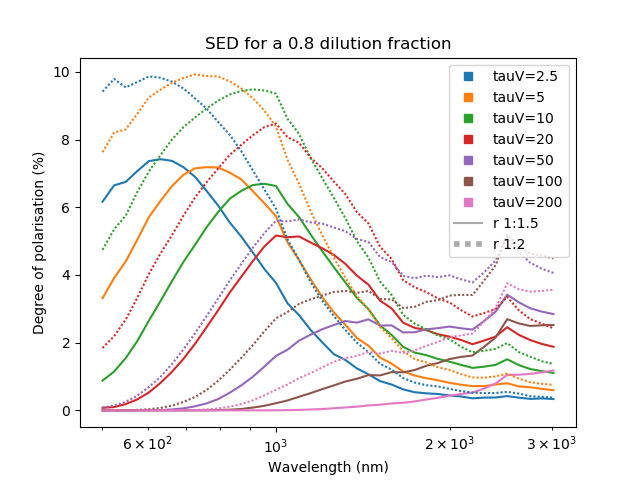}
  \caption{Spectra of relative intensity (first column) and degree of polarisation (second column) for an AGN with dilution fractions of 50~\% (first row) and 80~\% (second row) from MontAGN simulations. Intensity is normalised to the maximum of the dichroic flux as detailed in text.}
  \label{fig:dichro_dilu}
\end{figure*}

We used for dilution a library of spectra compiled by \cite{Marin2018dilu}. We used, similarly to them, the compiled spectrum of a Sbc galaxy (see their Figure~2), based on morphological classification of NGC~1068, such as \citealt{Balick1985}. A proper spectrum, has then be added to the computed intensity spectra, after scaling by dilution fraction, computed at 1~$\mu$m. Final SEDs and degree of polarisation are displayed in Figure~\ref{fig:dichro_dilu} for dilution fractions of 0.5 and 0.8.

Right column of Figure~\ref{fig:dichro_dilu} shows that dilution fraction and grain axis ratio do have a similar influence on the polarisation degree spectrum in mostly changing the amplitude but not the overall shape of the curve. This implies that these two parameters are to some extent degenerated. We also note that the optical depth is the main parameter driving the shape of the curve and thus will be constrained by comparison with observations.

\section{Comparison to observations}
\label{sec:simuvsobs}

\begin{figure*}[ht!]
 \centering
 \includegraphics[width=0.48\textwidth,clip]{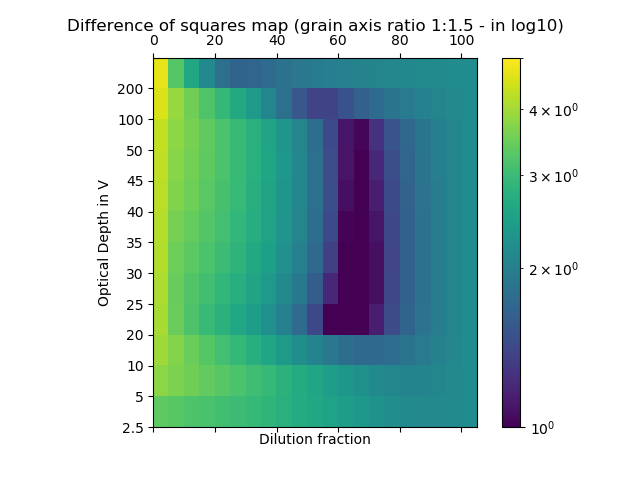}
 \includegraphics[width=0.48\textwidth,clip]{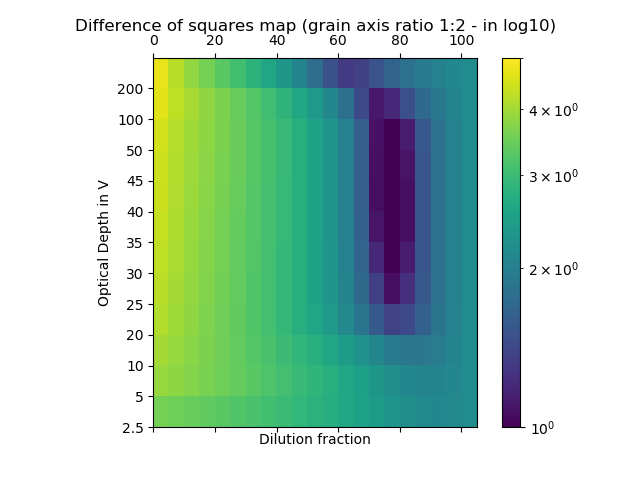}
  \caption{Least squares maps (in log scale) for ``r1:1.5'' (left panel) and ``r1:2'' (right panel) for dilution fraction varying between 0 and 100~\% (horizontal axis) and for optical depth $\tau_V$ between 2.5 and 200 (vertical axis).}
  \label{fig:errormap}
\end{figure*}

\begin{figure}[ht!]
 \centering
 \includegraphics[width=0.49\textwidth,clip]{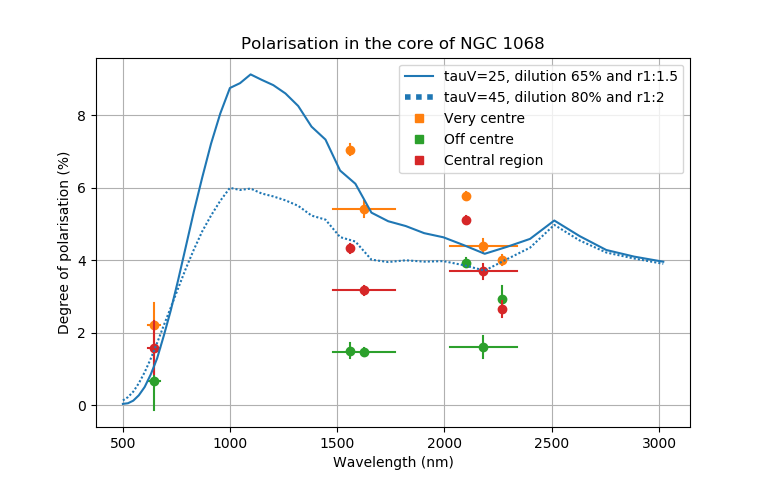}
  \caption{Degree of polarisation (in \%) as a function of wavelength (in nm) for both simulations and observations. Measurements are the same as on Figure~\ref{fig:centre} first panel and represent the ``very centre'' in orange, the ``off-centre'' region in green and the whole ``central region'' in red. Simulated spectra are shown in blue, continuous line for a grain ratio of 1:1.5 and dashed line for a ratio of 1:2, with corresponding optical depth and dilution fraction parameters to best fit the ``very-centre'' (orange) data points.}
  \label{fig:bestmodel}
\end{figure}

We computed fitting estimator maps as a function of optical depth and dilution fraction for both grain axis ratios. They were obtained by computing the squared difference between the observed and simulated -- through dichroism -- polarisation degree for each data point, on the very central aperture . Maps are displayed on Figure~\ref{fig:errormap}. Comparison of the observed data with the best model curve for each grain axis ratio is shown in Figure~\ref{fig:bestmodel}.

\subsection{Optical depth and dilution fraction}

As shown by Figure~\ref{fig:errormap}, the dilution fraction is constrained to a rather small range by the comparison between our observation and our simulations. Similarly, simulation also set a maximum and minimum limit values for the optical depth of the obscuring structure.

For the observed signal to be compatible with dichroic absorption, a dilution fraction of 60 to 85~\% is required, depending slightly on the optical depth and the grain axis ratio. Thus, the fraction of the total intensity received in the very central source of NGC~1068 through dichroism, and therefore directly coming from the CE, would be of 15 to 40~\% (at 1~$\mu$m), in rather good agreement with the $\approx$~5\% polarisation observed at the very centre between 1.5 and 2.3~$\mu$m.

The integrated optical depth of the structure is also constrained to values ranging between $\tau_V \approx$ {20 to 100}, slightly favouring values to the lower edge, with $\tau_V \leq$ 50. Such dust densities would allow enough light to propagate through the obscuring material in the NIR in one of the polarisation orientation to create the observed dichroic polarisation. This constraint does not depend on the geometry or on the clumpiness of the medium. Indeed, the obscuring material is very likely to be clumpy (see eg. \citealt{Marin2015}), with individual clouds' optical depth being currently estimated to $45 \leq \tau_V \leq 115$ for example by \cite{Lopez-Rodriguez2015,Audibert2017}. Thus, our constraint would be compatible with an estimate of 1 to 2 clouds in our line of sight to the CE.



\subsection{Role of grain axis ratio}


The other conclusion than can be driven from Figure~\ref{fig:errormap} is that even though the grain axis ratio does impact the optical depth difference between the two polarisation orientations, it does not translate into as large a difference in the measured polarisation degree. By comparing the two maps of Figure~\ref{fig:errormap}, the difference introduced by a change in the grain axis ratio only slightly shift both the required dilution fraction and structure's optical depth toward larger values. For instance, we measured for our second grain population a shift of $\approx$~10~\% in the dilution fraction and an increase of $\approx$~10 in the lower limit of the required optical depth in V.

We only tested here two grain ratio populations with our simple model, with an orientation identical for all of the grains. It is expected \citep{Efstathiou1997} that grains are not all perfectly aligned along the magnetic field and are precessing along the preferred orientation direction and that only a fraction of them are actually similarly orientated. \cite{Efstathiou1997} studies the impact of grain axis ratios and precession of the grains on the polarisation signal and concluded that the effect was small and comparable to a dilution of the signal, similarly to what can be seen in Figure~\ref{fig:errormap}. This is consistent since a misalignment of part of the grain population would be modelled by a slightly different shape of grain.

\subsection{Polarisation through scattering?}

\begin{figure*}[ht!]
 \centering
 \includegraphics[width=0.49\textwidth,clip]{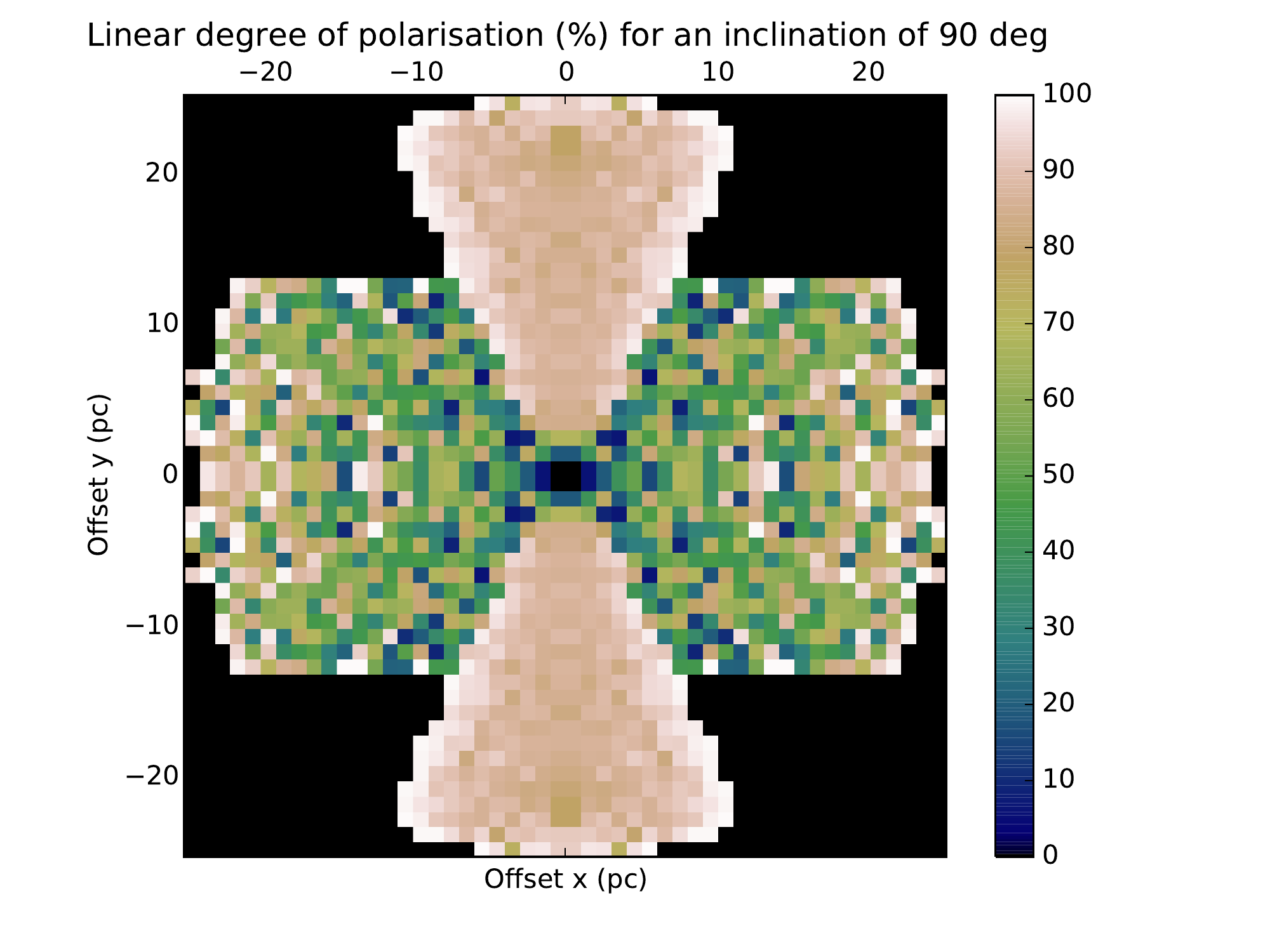}
 \includegraphics[width=0.49\textwidth,clip]{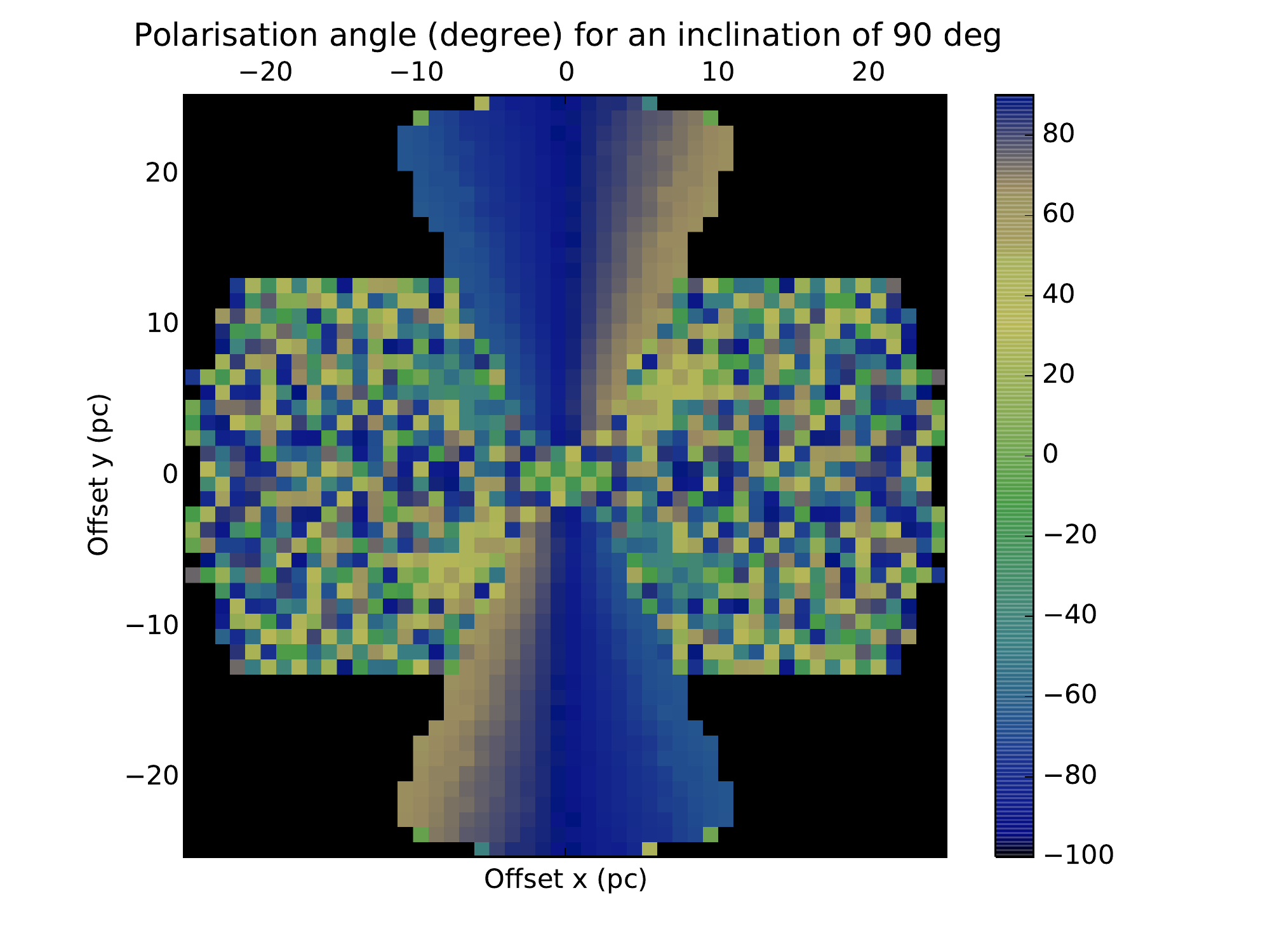}
  \caption{Simulated map of polarisation degree (left) and polarisation position angle (right) computed by MontAGN for an integrated optical depth of $\tau_V$~=~45 along the equatorial line of sight, with an inclination angle of 90$^\circ$ at 1.6 \textmu m (without dichroism and without dilution).}
  \label{fig:polar_scatt}
\end{figure*}

These results correspond to an optical depth lower than the one estimated by \cite{Grosset2018}, which were requiring $\tau_V \approx$~170, based on double scattering effect. However, this previous estimate used the polarimetric signal in all the central region (of 2'') of NGC~1068, thus including the very central emission. A high optical depth was therefore necessary to reproduce the polarimetric signal especially in this very central region, otherwise the unpolarised flux of the CE would mask the polarisation. However, if we consider that this central polarisation originates from dichroic absorption instead, we can lower the required optical depth and still produce double scattering polarisation signal out of the very centre. We verified this assumption and Figure~\ref{fig:polar_scatt} shows the polarimetric signal from scattering based on simulations developed in \cite{Grosset2018}, for an optical depth of 45 in V. Pattern on the cone and equatorial region of Figure~\ref{fig:polar_scatt} are similar to those of the $\tau_V$~=~170 map of \cite{Grosset2018} and differs only on the central $\approx$~3~pc. However, it is noteworthy that these two simulation works (\citealt{Grosset2018} ; this work) were conducted using different framework of only spherical dust grains, and only elongated dust grains respectively. In order to have a complete view of the mechanisms at the origin of the polarisation signal in the NIR, a full simulation framework would be required, and would be a very important extension of this work.

\subsection{Effect on large apertures}
\label{sec:large_ap}

Our polarisation measurements within the largest aperture (the red data points of Figure~\ref{fig:bestmodel}) are in fair agreement with the literature for similar or larger apertures. In particular, Figure~\ref{fig:centre} shows an increasing trends between 500 and 2000~nm that fits well within the compilation of polarimetric measurements in Figure~2 of \cite{Marin2018}. This tends to confirm that our small aperture measurements are reliable and that we are separating different polarisation mechanisms on these two regions that become combined in larger apertures. In particular, we can distinguish at least 3 regions with different polarisation mechanisms: the very inner central spot dominated by dichroic absorption; the equatorial region of the ``off-centre'' region, ranging between few pc to $\approx$~60~pc dominated by double scattering; and the bi-cone region between 60 and 200~pc in the polar directions dominated by single scattering.


\section{Discussion}

\subsection{Global view}
\label{sec:global_view}


Thanks to recent polarimetric studies \cite[this work]{Gratadour2015,Lopez-Rodriguez2015,Grosset2018,Marin2018}, we are proposing a global scheme of the polarisation mechanisms in AGNs and in particular in NGC~1068.

On the photo-centre, beyond 1~\textmu m we would detect light coming directly from hot dust in the very inner region of the torus, heated by the CE \citep{Lopez-Rodriguez2015}. Due to non-spherical grains more or less aligned by magnetic field, only one polarisation orientation preferentially propagates, the other being more absorbed, creating the observed polarisation in the NIR at the very centre. Dichroism is the most likely polarising mechanism in this region since it is the one able to produce such high polarisation degrees (up to 15~\%) and a constant polarisation orientation. Beyond 2~\textmu m, the optical depth become low enough for both the polarisation orientation to propagate through the material, leading to the observed decrease in the polarisation degree.

On the equatorial region, along the extension of the obscuring material, we detect a low polarisation likely to be produced by double scattering \citep{Grosset2018} on a region centred on the photo-centre and extending on both the equatorial direction to about 30~pc, with a 20~pc total thickness on the polar direction \citep{Gratadour2015}. This mechanism becomes negligible at the photo-centre because polarised flux becomes completely dominated by the effect of dichroic absorption at the photo-centre (the central PSF). The low polarisation signal is thus only detected at location offseted from the photo-centre. 

Finally, on the polar directions, we detect a signal well explainable by single scattering, on material located in the double ionisation cone between few parsecs to about 200~pc. The origin of this scattering signal is possibly due to multiple species, depending on the location.

\begin{figure}[ht!]
 \centering
 \includegraphics[width=0.49\textwidth,clip]{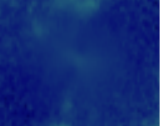}
  \caption{Zoom on the polar extension of the detected polarisation degree, close to the very centre, on the Cnt~K1 map (2.1~$\mu$m). Size of the image is $\approx$~1.5~$\times$~2'' and colour scale is the same as Figure~\ref{fig:NB2}.}
  \label{fig:ridge}
\end{figure}

In the innermost regions, the material responsible for the scattering (see maps of Figures~\ref{fig:NB1}, \ref{fig:NB2} and \ref{fig:NB3} until 0.5''), zoomed on Figure~\ref{fig:ridge}, is rather unclear. We observe an extension of the high polarisation degree along the polar direction, in H (1625~nm) and Ks (2182~nm) bands as already stated by \cite{Gratadour2015}, but also on Cnt H (1573~nm) and K1 (2091~nm) polarisation maps. This feature is remarkably perpendicular to the CO structure detected on the central 10~pc with ALMA by \cite{Garcia-Burillo2019}, and could be due to electron scattering on the first 10~pc of the bi-cone, as invoked by \cite{Antonucci1993} and observed by \cite{Antonucci1994} with the Hubble Space Telescope in the ultra-violet. But hot dust is known to be present in the polar outflow (\citealt{Ramos-Almeida2017} and references therein) and would also be a good candidate for scattering the light emitted by the CE. In both situations, it is expected that light emitted or scattered this close to the AGN centre would then cross through parts of the obscuring material, so that polarisation due to Thomson/Rayleigh/Mie scattering could be then combined to dichroic absorption by this material. 
As both these effects in these regions are expected to produce polarisation with a similar position angle, it would be very difficult to disentangle the exact contribution of each of these mechanisms. As Thomson scattering on electrons is wavelength-independent, the main polarising mechanism is also likely to change between visible and IR close to the very centre, depending on the optical depth of the obscuring material.

At larger scale, dust grains are likely to be dominant. Comparison with ALMA data reveals that the Southern arcs are very likely due to scattering on dusty material, present in the ionisation cone and directly illuminated by the CE \citep{Garcia-Burillo2016,Garcia-Burillo2019}. 



As discussed in Section~\ref{sec:large_ap}, depending on the size of the aperture used to measure the polarisation signal, different proportion of these -- at least -- three mechanisms must be used, thus leading to the differences between our measurements and the large aperture ones.

\subsection{Relative importance of the mechanisms}

Based on these measurements at HAR, we can attempt to estimate the relative contribution of these three mechanisms to the polarised intensity, as a function of their location.

As shown by Figure~\ref{fig:P_SA}, the degree of polarisation in the scattering region of the Southern arcs is increasing with wavelength. This variation is due both to evolution in the polarised intensity -- slightly increasing between H and Cnt K2 (from 1625 to 2266~nm)-- and to the global intensity which is strongly decreasing at large distance from the centre with increasing wavelength (
see upper panels of Figures~\ref{fig:NB1}, \ref{fig:NB2} and \ref{fig:NB3}). This is mostly due to the decrease of the stellar emission in the inner few arc-seconds between H and Ks (1625 and 2182~nm respectively), as detailed in \cite{Rouan2019}. A deeper study of the relation between the optical depth of the dust structures and the intensity and polarised intensity produced by this structure combined with data from ALMA, would be a very interesting extension of this work.

In the equatorial region, offset from the photo-centre, the degree of polarisation is low and slightly increases toward 2.2~\textmu m. This increase of the polarisation degree in Ks is most likely related to a decrease of the intensity as detailed in previous paragraph, a stronger effect than what is observed on the Southern arcs, the polarised intensity being quasi constant in the equatorial region. This is consistent with the results of \cite{Grosset2018} on double scattering that found a relatively constant low polarisation degree on the equatorial region (except on the very centre) regardless the optical depth of the structure, as long as it is optically thick (since we would detect the centrosymmetric single scattering pattern if it was optically thin).

Combining these measurements, we estimated the contribution of dichroic absorption (versus scattering) to the polarised flux in the central region for different aperture sizes. Results are detailed in Table~\ref{table:contrib}.

\begin{table}
\caption[Contribution of dichroism to Ip]{Contribution of dichroism to the polarised flux in different apertures for a given band (in \%)}
\label{table:contrib}
\centering
\begin{tabular}{|c|cccccc|}
\hline\hline
Aperture & & & Filter & name & & \\
size (in '')  & NR & Cnt H & H & Cnt K1 & Ks & Cnt K2 \\
\hline
Wvl (nm) & 646 & 1573 & 1625 & 2091 & 2182 & 2266 \\
\hline
0.2 & -$^{(1)}$ & 98.5 & 98.2 & 98.3 & 99.6 & 99.0 \\
0.5 & -$^{(1)}$ & 54 & 58 & 55 & 74 & 63 \\
1 & -$^{(1)}$ & 40 & 43 & 42 & 64 & 51 \\
3 & -$^{(1)}$ & 26 & 30 & 31 & 53 & 38 \\
\hline
\end{tabular}
\tablefoot{ 
Polarisation uncertainties range between 0.5 and 1~\% for all measurements except for NR and we thus took a conservative value for uncertainties of 1~\%. (1) The central polarisation feature that is expected to be due to dichroic absorption is weaker in NR band and estimation of different contributions is thus not reliable in this band (large incertitudes).}
\end{table}

Thus according to Table~\ref{table:contrib}, polarised intensity in the NIR, at the photo-centre, is expected to come at about 98~\%~$\pm$1~\% from dichroic absorption. But interestingly, this contribution decreases to 50~\% and less at larger apertures, when more and more scattered flux fell within the aperture. This contribution increases with the wavelength, with polarised flux in Ks band being still dominated by dichroism even on a 3'' aperture. This ratio might be decreasing on the Cnt~K2 band, around 2.3~\textmu m, as the fraction is lowered to 38~\%~$\pm$1~\%, however possible systematics do not allow to drive any firm conclusion.




\subsection{Geometry and optical depth of the dust structure}

As stated in Section~\ref{sec:simuvsobs}, our study tends to lower the estimation of the required optical depth along the LOS to the CE of the AGN derived by \cite{Grosset2018}. This lower values originates from a lower need of blocking the light directly coming from the centre, as polarisation is rather created by dichroic absorption (about 99~\%) than by double scattering (less than 2~\%) in the LOS. As discussed at the end of \cite{Grosset2018} and on Section~\ref{sec:simuvsobs} of the present work, dichroic absorption better reproduces the degree of polarisation at this location. We can thus set an upper value to the dust opacity at $\tau_V$~=~200, because if extinction is too high photons can barely escape the obscuring structure, which would be incompatible with a strong polarimetric signal produced by dichroic absorption (or any other mechanism). The range of optical depth of $\approx$~20~-~100 in $\tau_V$ constrained by our model is consistent with what is generally expected for the obscuring structure of AGN (as eg. \citealt{Packham1997,Lopez-Rodriguez2015,Audibert2017,Rouan2019}), lowering the previously estimation of \cite{Grosset2018}.

We only simulated here integrated absorption along the LOS to the CE, and there is therefore no assumption on the geometry or the nature (fragmentation) of the obscuring structure. In the $\approx$60~pc region of constant polarisation, double scattering process should still be involved and thus a dust density decreasing with the radius is still likely required to efficiently scatter photons \citep{Grosset2018}. Furthermore, this hypothesis is consistent with recent studies by \cite{Izumi2018}, who developed a multiphase dynamic torus model, with lower densities in the outer regions of the torus (R$>$10~pc), based on ALMA observations of the Circinus galaxy.

\vspace{0.5cm}
We also resolved an elongation along the polar axis with a high polarisation degree in most of the NIR maps, close to the very centre, on a $\approx$~10~pc scale. As discussed in Section~\ref{sec:global_view}, this could be due to scattering on electrons in the first parsecs of the ionisation cone or on hot dust in the outflow. What is surprising about this elongation is that both the polarisation degree and polarisation position angle are remarkably constant all along the structure, while the intensity undergo large variations (thus reflected in the polarised flux), and while different mechanisms would be involved. Would it be possible for two polarising effects (dichroism at the photo-centre and scattering on the polar extension encompassing the photo-centre) be able to produce such similar polarisation? It is likely that dichroism does also affect light scattered (or emitted) in the polar ionisation cone if close enough to the equatorial plane, since light would then have to cross through parts of the obscuring material. But where does the transition between this two effect take place? Estimating the exact contribution of both phenomena to polarisation in this polar extension is thus difficult and would be an interesting study.

\subsection{Magnetic field}

Our estimation of the 98 to 99~\% contribution of the dichroic absorption to the polarised flux observed in the NIR is consistent with the assumption of \cite{Lopez-Rodriguez2015}, who used a fraction of 100~\% to study the magnetic field in the central region of NGC~1068. Our measured polarisation degree in Ks (2182~nm) of 4.4$~\pm$~0.2~\% is also in very good agreement with these authors' measurement of 4.4$~\pm$~0.1 per cent. The optical depth range we estimated here, between 15 and 100 in the visible, corresponding to 1.7 to 12 in Ks, encompasses the value used by \cite{Lopez-Rodriguez2015} of $\tau_K\approx3.24$, based on \cite{Packham1997}. Having very similar values, we would derive identical strength estimation for the magnetic field of 4 to 139 mG, as estimated by \cite{Lopez-Rodriguez2015}. 
Our measurements also confirms that with HAR, the position angle for the polarisation orientation is very similar to those measured with larger apertures \citep{Lopez-Rodriguez2015} and therefore reinforces the hypothesis of a toroidal geometry for the magnetic field (parallel to the polarisation position angle for dichroic absorption).


\section{Conclusion and prospectives}

We presented in this paper new observations of the core of the archetypal Seyfert 2 galaxy NGC~1068 in NB filters with the extreme adaptive optics instrument SPHERE. Our polarisation measurements within large apertures are in good agreement with the literature and fits well the expected behaviour of the polarisation with wavelength. However, by separating the signal in smaller apertures, thanks to HAR imaging, we were able to trace rather different behaviours of polarisation regarding wavelength, depending on the location in the AGN environment, tracing as many different polarising mechanisms. Thanks to these polarimetric measurements, combined to the previous polarisation maps in H and Ks (1625 and 2182~nm respectively), we investigated the properties of three major regions within the central 200~pc of the AGN and their associated dominant polarising mechanism:
\begin{itemize}
\item the electrons and dust scattering in the ionisation cone, ranging between few tens of parsecs to $\approx$ 200~pc. In particular, the South-western arcs detected in the broad NIR bands by \cite{Gratadour2015} were re-observed and are compatible with single scattering, most likely on dust grains;
\item the dust double-scattering equatorial central region of the AGN (the first scattering would be in the polar region, and the second in the equatorial region) corresponding to constant polarisation orientation and degree on a 60~pc~$\times$~20~pc area, encompassing the photo-centre and tracing the outer envelop of the obscuring material (similarly to \citealt{Grosset2018});
\item the dichroic absorption of the light coming directly from the photo-centre of the AGN, from the CE.
\end{itemize}

Furthermore, we identified a fourth region, at the transition between the photo-centre and the ionisation cone up to $\approx$~10~pc in the polar direction, displaying similar polarisation properties to the photo-centre, with lower total intensity. We interpret this signal as a possible transition between dichroic absorption at the very centre to electron or dust scattering with increasing distance to the centre.

\vspace{0.5cm}

This study highlights the importance of the spatial resolution to study polarimetry and especially on regions where several polarising mechanisms are at work. We were able to disentangle the different mechanisms specific to each region, thanks to this combination of polarimetry and HAR, and by comparison with the numerical code MontAGN.

Invoking dichroic absorption and scattering simulations, we constrained the integrated optical depth of the dusty obscuring material and the dilution fraction by other emission. Integrated optical depth is constrained within a range of 20 to 100 in the visible for the obscuring material. For the scope of this work, we modelled both dichroism and scattering separately before combining the results. Combining these two effects in the same simulation will soon be implemented in MontAGN, to simulate more complex environments where both electron/dust scattering and dichroic absorption can be acting on the same photons, like in the very inner region of AGNs.

This work also argue in favour of an extension of the polarimetric investigations at HAR toward longer wavelengths and particularly to the 3 to 10~\textmu m domain where a switch in the polarisation position angle is expected due to a change in the dichroic absorption/emission at these wavelengths \citep{Efstathiou1997}.




\begin{acknowledgements}
Based on observations collected at the European Southern Observatory under ESO programmes 60.A-9361(A) and 097.B-0840(A). The authors would like to acknowledge financial support from the ``Programme National Hautes Energies'' (PNHE) and from ``Programme National de Cosmologie and Galaxies'' (PNCG) funded by CNRS/INSU-IN2P3-INP, CEA, and CNES, France. 
Authors thank the anonymous referee for useful comments helping to improve the clarity of the paper. The authors also thank E. Lopez-Rodriguez, T. Paumard and J. Milli for useful discussions that improved the manuscript. LG thanks N. T. Lam and P. Vermot for their contributions to the simulation code. This project has received funding from the European Union's Horizon 2020 research and innovation program under the Marie Sk\l{}odowska-Curie Grant agreement No. 665501 with the research Foundation Flanders (FWO) ([PEGASUS]$^2$ Marie Curie fellowship 12U2717N awarded to M.M.). This research has made use of the SIMBAD database, operated at CDS, Strasbourg, France. This research made use of Astropy,\footnote{http://www.astropy.org} a community-developed core Python package for Astronomy \citep{Astropy2013,Astropy2018} and Matplotlib, a 2D graphics package used for Python for application development, interactive scripting, and publication-quality image generation across user interfaces and operating systems \citep{Hunter2007}.

\end{acknowledgements}

%
%

\bibliographystyle{aa} 
\bibliography{grosset} 

\begin{thebibliography}{50}
\expandafter\ifx\csname natexlab\endcsname\relax\def\natexlab#1{#1}\fi

\bibitem[{{Alonso-Herrero} {et~al.}(2011){Alonso-Herrero}, {Ramos Almeida},
  {Mason}, {Asensio Ramos}, {Roche}, {Levenson}, {Elitzur}, {Packham},
  {Rodr{\'{\i}}guez Espinosa}, {Young}, {D{\'{\i}}az-Santos}, \&
  {P{\'e}rez-Garc{\'{\i}}a}}]{Alonso2011}
{Alonso-Herrero}, A., {Ramos Almeida}, C., {Mason}, R., {et~al.} 2011, \apj,
  736, 82

\bibitem[{{Antonucci}(1993)}]{Antonucci1993}
{Antonucci}, R. 1993, \araa, 31, 473

\bibitem[{{Antonucci} {et~al.}(1994){Antonucci}, {Hurt}, \&
  {Miller}}]{Antonucci1994}
{Antonucci}, R., {Hurt}, T., \& {Miller}, J. 1994, \apj, 430, 210

\bibitem[{{Antonucci} \& {Miller}(1985)}]{Antonucci1985}
{Antonucci}, R.~R.~J. \& {Miller}, J.~S. 1985, \apj, 297, 621

\bibitem[{{Astropy Collaboration} {et~al.}(2018){Astropy Collaboration},
  {Price-Whelan}, {Sip{\H{o}}cz}, {G{\"u}nther}, {Lim}, {Crawford}, {Conseil},
  {Shupe}, {Craig}, {Dencheva}, {Ginsburg}, {Vand erPlas}, {Bradley},
  {P{\'e}rez-Su{\'a}rez}, {de Val-Borro}, {Aldcroft}, {Cruz}, {Robitaille},
  {Tollerud}, {Ardelean}, {Babej}, {Bach}, {Bachetti}, {Bakanov}, {Bamford},
  {Barentsen}, {Barmby}, {Baumbach}, {Berry}, {Biscani}, {Boquien}, {Bostroem},
  {Bouma}, {Brammer}, {Bray}, {Breytenbach}, {Buddelmeijer}, {Burke},
  {Calderone}, {Cano Rodr{\'\i}guez}, {Cara}, {Cardoso}, {Cheedella}, {Copin},
  {Corrales}, {Crichton}, {D'Avella}, {Deil}, {Depagne}, {Dietrich}, {Donath},
  {Droettboom}, {Earl}, {Erben}, {Fabbro}, {Ferreira}, {Finethy}, {Fox},
  {Garrison}, {Gibbons}, {Goldstein}, {Gommers}, {Greco}, {Greenfield},
  {Groener}, {Grollier}, {Hagen}, {Hirst}, {Homeier}, {Horton}, {Hosseinzadeh},
  {Hu}, {Hunkeler}, {Ivezi{\'c}}, {Jain}, {Jenness}, {Kanarek}, {Kendrew},
  {Kern}, {Kerzendorf}, {Khvalko}, {King}, {Kirkby}, {Kulkarni}, {Kumar},
  {Lee}, {Lenz}, {Littlefair}, {Ma}, {Macleod}, {Mastropietro}, {McCully},
  {Montagnac}, {Morris}, {Mueller}, {Mumford}, {Muna}, {Murphy}, {Nelson},
  {Nguyen}, {Ninan}, {N{\"o}the}, {Ogaz}, {Oh}, {Parejko}, {Parley}, {Pascual},
  {Patil}, {Patil}, {Plunkett}, {Prochaska}, {Rastogi}, {Reddy Janga},
  {Sabater}, {Sakurikar}, {Seifert}, {Sherbert}, {Sherwood-Taylor}, {Shih},
  {Sick}, {Silbiger}, {Singanamalla}, {Singer}, {Sladen}, {Sooley},
  {Sornarajah}, {Streicher}, {Teuben}, {Thomas}, {Tremblay}, {Turner},
  {Terr{\'o}n}, {van Kerkwijk}, {de la Vega}, {Watkins}, {Weaver}, {Whitmore},
  {Woillez}, {Zabalza}, \& {Astropy Contributors}}]{Astropy2018}
{Astropy Collaboration}, {Price-Whelan}, A.~M., {Sip{\H{o}}cz}, B.~M., {et~al.}
  2018, \aj, 156, 123

\bibitem[{{Astropy Collaboration} {et~al.}(2013){Astropy Collaboration},
  {Robitaille}, {Tollerud}, {Greenfield}, {Droettboom}, {Bray}, {Aldcroft},
  {Davis}, {Ginsburg}, {Price-Whelan}, {Kerzendorf}, {Conley}, {Crighton},
  {Barbary}, {Muna}, {Ferguson}, {Grollier}, {Parikh}, {Nair}, {Unther},
  {Deil}, {Woillez}, {Conseil}, {Kramer}, {Turner}, {Singer}, {Fox}, {Weaver},
  {Zabalza}, {Edwards}, {Azalee Bostroem}, {Burke}, {Casey}, {Crawford},
  {Dencheva}, {Ely}, {Jenness}, {Labrie}, {Lim}, {Pierfederici}, {Pontzen},
  {Ptak}, {Refsdal}, {Servillat}, \& {Streicher}}]{Astropy2013}
{Astropy Collaboration}, {Robitaille}, T.~P., {Tollerud}, E.~J., {et~al.} 2013,
  \aap, 558, A33

\bibitem[{{Audibert} {et~al.}(2017){Audibert}, {Riffel}, {Sales}, {Pastoriza},
  \& {Ruschel-Dutra}}]{Audibert2017}
{Audibert}, A., {Riffel}, R., {Sales}, D.~A., {Pastoriza}, M.~G., \&
  {Ruschel-Dutra}, D. 2017, \mnras, 464, 2139

\bibitem[{{Balick} \& {Heckman}(1985)}]{Balick1985}
{Balick}, B. \& {Heckman}, T. 1985, \aj, 90, 197

\bibitem[{{Bastien} \& {Menard}(1990)}]{Bastien1990}
{Bastien}, P. \& {Menard}, F. 1990, \apj, 364, 232

\bibitem[{{Beuzit} {et~al.}(2008){Beuzit}, {Feldt}, {Dohlen}, {Mouillet},
  {Puget}, {Wildi}, {Abe}, {Antichi}, {Baruffolo}, {Baudoz}, {Boccaletti},
  {Carbillet}, {Charton}, {Claudi}, {Downing}, {Fabron}, {Feautrier},
  {Fedrigo}, {Fusco}, {Gach}, {Gratton}, {Henning}, {Hubin}, {Joos}, {Kasper},
  {Langlois}, {Lenzen}, {Moutou}, {Pavlov}, {Petit}, {Pragt}, {Rabou}, {Rigal},
  {Roelfsema}, {Rousset}, {Saisse}, {Schmid}, {Stadler}, {Thalmann}, {Turatto},
  {Udry}, {Vakili}, \& {Waters}}]{Beuzit2008}
{Beuzit}, J.-L., {Feldt}, M., {Dohlen}, K., {et~al.} 2008, in \procspie, Vol.
  7014, Ground-based and Airborne Instrumentation for Astronomy II, 701418

\bibitem[{{Clarke} \& {Stewart}(1986)}]{Clarke1986}
{Clarke}, D. \& {Stewart}, B.~G. 1986, Vistas in Astronomy, 29, 27

\bibitem[{{Dohlen} {et~al.}(2008){Dohlen}, {Langlois}, {Saisse}, {Hill},
  {Origne}, {Jacquet}, {Fabron}, {Blanc}, {Llored}, {Carle}, {Moutou}, {Vigan},
  {Boccaletti}, {Carbillet}, {Mouillet}, \& {Beuzit}}]{Dohlen2008}
{Dohlen}, K., {Langlois}, M., {Saisse}, M., {et~al.} 2008, in \procspie, Vol.
  7014, Ground-based and Airborne Instrumentation for Astronomy II, 70143L

\bibitem[{{Efstathiou} {et~al.}(1997){Efstathiou}, {McCall}, \&
  {Hough}}]{Efstathiou1997}
{Efstathiou}, A., {McCall}, A., \& {Hough}, J.~H. 1997, \mnras, 285, 102

\bibitem[{{Everett} \& {Weisberg}(2001)}]{Everett2001}
{Everett}, J.~E. \& {Weisberg}, J.~M. 2001, \apj, 553, 341

\bibitem[{{Fusco} {et~al.}(2006){Fusco}, {Rousset}, {Sauvage}, {Petit},
  {Beuzit}, {Dohlen}, {Mouillet}, {Charton}, {Nicolle}, {Kasper}, {Baudoz}, \&
  {Puget}}]{Fusco2006}
{Fusco}, T., {Rousset}, G., {Sauvage}, J.-F., {et~al.} 2006, Optics Express,
  14, 7515

\bibitem[{{Garc{\'{\i}}a-Burillo} {et~al.}(2019){Garc{\'{\i}}a-Burillo},
  {Combes}, {Ramos Almeida}, {Usero}, {Alonso-Herrero}, {Hunt}, {Rouan},
  {Aalto}, {Querejeta}, {Viti}, {van der Werf}, {Vives-Arias}, {Fuente},
  {Colina}, {Mart{\'\i}n-Pintado}, {Henkel}, {Mart{\'\i}n}, {Krips},
  {Gratadour}, {Neri}, \& {Tacconi}}]{Garcia-Burillo2019}
{Garc{\'{\i}}a-Burillo}, S., {Combes}, F., {Ramos Almeida}, C., {et~al.} 2019,
  \aap, 632, A61

\bibitem[{{Garc{\'{\i}}a-Burillo} {et~al.}(2016){Garc{\'{\i}}a-Burillo},
  {Combes}, {Ramos Almeida}, {Usero}, {Krips}, {Alonso-Herrero}, {Aalto},
  {Casasola}, {Hunt}, {Mart{\'{\i}}n}, {Viti}, {Colina}, {Costagliola},
  {Eckart}, {Fuente}, {Henkel}, {M{\'a}rquez}, {Neri}, {Schinnerer}, {Tacconi},
  \& {van der Werf}}]{Garcia-Burillo2016}
{Garc{\'{\i}}a-Burillo}, S., {Combes}, F., {Ramos Almeida}, C., {et~al.} 2016,
  \apjl, 823, L12

\bibitem[{{Gratadour} {et~al.}(2015){Gratadour}, {Rouan}, {Grosset},
  {Boccaletti}, \& {Cl{\'e}net}}]{Gratadour2015}
{Gratadour}, D., {Rouan}, D., {Grosset}, L., {Boccaletti}, A., \& {Cl{\'e}net},
  Y. 2015, \aap, 581, L8

\bibitem[{{Gravity Collaboration} {et~al.}(2020){Gravity Collaboration},
  {Pfuhl}, {Davies}, {Dexter}, {Netzer}, {H{\"o}nig}, {Lutz}, {Schartmann},
  {Sturm}, {Amorim}, {Brandner}, {Cl{\'e}net}, {de Zeeuw}, {Eckart},
  {Eisenhauer}, {F{\"o}rster Schreiber}, {Gao}, {Garcia}, {Genzel},
  {Gillessen}, {Gratadour}, {Kishimoto}, {Lacour}, {Millour}, {Ott}, {Paumard},
  {Perraut}, {Perrin}, {Peterson}, {Petrucci}, {Prieto}, {Rouan}, {Shangguan},
  {Shimizu}, {Sternberg}, {Straub}, {Straubmeier}, {Tacconi}, {Tristram},
  {Vermot}, {Waisberg}, {Widmann}, \& {Woillez}}]{Pfuhl2020}
{Gravity Collaboration}, {Pfuhl}, O., {Davies}, R., {et~al.} 2020, \aap, 634,
  A1

\bibitem[{{Gravity Collaboration} {et~al.}(2018){Gravity Collaboration},
  {Sturm}, {Dexter}, {Pfuhl}, {Stock}, {Davies}, {Lutz}, {Cl{\'e}net},
  {Eckart}, {Eisenhauer}, {Genzel}, {Gratadour}, {H{\"o}nig}, {Kishimoto},
  {Lacour}, {Millour}, {Netzer}, {Perrin}, {Peterson}, {Petrucci}, {Rouan},
  {Waisberg}, {Woillez}, {Amorim}, {Brandner}, {F{\"o}rster Schreiber},
  {Garcia}, {Gillessen}, {Ott}, {Paumard}, {Perraut}, {Scheithauer},
  {Straubmeier}, {Tacconi}, \& {Widmann}}]{Sturm2018}
{Gravity Collaboration}, {Sturm}, E., {Dexter}, J., {et~al.} 2018, \nat, 563,
  657

\bibitem[{{Grosset} {et~al.}(2018){Grosset}, {Rouan}, {Gratadour}, {Pelat},
  {Orkisz}, {Marin}, \& {Goosmann}}]{Grosset2018}
{Grosset}, L., {Rouan}, D., {Gratadour}, D., {et~al.} 2018, \aap, 612, A69

\bibitem[{{Hunter}(2007)}]{Hunter2007}
{Hunter}, J.~D. 2007, Computing in Science and Engineering, 9, 90

\bibitem[{{I.A.U.}(1973)}]{IAU1973}
{I.A.U.} 1973, {XVth General Assembly, Sydney, Australia.}

\bibitem[{{Imanishi} {et~al.}(2018){Imanishi}, {Nakanishi}, {Izumi}, \&
  {Wada}}]{Imanishi2018}
{Imanishi}, M., {Nakanishi}, K., {Izumi}, T., \& {Wada}, K. 2018, \apjl, 853,
  L25

\bibitem[{{Imanishi} {et~al.}(2020){Imanishi}, {Nguyen}, {Wada}, {Hagiwara},
  {Iguchi}, {Izumi}, {Kawakatu}, {Nakanishi}, \& {Onishi}}]{Imanishi2020}
{Imanishi}, M., {Nguyen}, D.~D., {Wada}, K., {et~al.} 2020, \apj, 902, 99

\bibitem[{{Impellizzeri} {et~al.}(2019){Impellizzeri}, {Gallimore}, {Baum},
  {Elitzur}, {Davies}, {Lutz}, {Maiolino}, {Marconi}, {Nikutta}, {O'Dea}, \&
  {Sani}}]{Impellizzeri2019}
{Impellizzeri}, C.~M.~V., {Gallimore}, J.~F., {Baum}, S.~A., {et~al.} 2019,
  \apjl, 884, L28

\bibitem[{{Izumi} {et~al.}(2018){Izumi}, {Wada}, {Fukushige}, {Hamamura}, \&
  {Kohno}}]{Izumi2018}
{Izumi}, T., {Wada}, K., {Fukushige}, R., {Hamamura}, S., \& {Kohno}, K. 2018,
  \apj, 867, 48

\bibitem[{{Kervella} {et~al.}(2015){Kervella}, {Montarg{\`e}s}, {Lagadec},
  {Ridgway}, {Haubois}, {Girard}, {Ohnaka}, {Perrin}, \&
  {Gallenne}}]{Kervella2015}
{Kervella}, P., {Montarg{\`e}s}, M., {Lagadec}, E., {et~al.} 2015, \aap, 578,
  A77

\bibitem[{{Langlois} {et~al.}(2014){Langlois}, {Dohlen}, {Vigan}, {Zurlo},
  {Moutou}, {Schmid}, {Mili}, {Beuzit}, {Boccaletti}, {Carle}, {Costille},
  {Dorn}, {Gluck}, {Hubin}, {Feldt}, {Kasper}, {Lizon}, {Madec}, {Le Mignant},
  {Mouillet}, {Puget}, {Sauvage}, \& {Wildi}}]{Langlois2014}
{Langlois}, M., {Dohlen}, K., {Vigan}, A., {et~al.} 2014, in \procspie, Vol.
  9147, Ground-based and Airborne Instrumentation for Astronomy V, 91471R

\bibitem[{{Lopez-Rodriguez} {et~al.}(2019){Lopez-Rodriguez}, {Alonso-Herrero},
  {Garc{\'\i}a-Burillo}, {Gordon}, {Ichikawa}, {Imanishi}, {Kameno},
  {Levenson}, {Nikutta}, \& {Packham}}]{Lopez-Rodriguez2019}
{Lopez-Rodriguez}, E., {Alonso-Herrero}, A., {Garc{\'\i}a-Burillo}, S.,
  {et~al.} 2019, arXiv e-prints, arXiv:1905.08802

\bibitem[{{Lopez-Rodriguez} {et~al.}(2015){Lopez-Rodriguez}, {Packham},
  {Jones}, {Nikutta}, {McMaster}, {Mason}, {Elvis}, {Shenoy}, {Alonso-Herrero},
  {Ram{\'{\i}}rez}, {Gonz{\'a}lez Mart{\'{\i}}n}, {H{\"o}nig}, {Levenson},
  {Ramos Almeida}, \& {Perlman}}]{Lopez-Rodriguez2015}
{Lopez-Rodriguez}, E., {Packham}, C., {Jones}, T.~J., {et~al.} 2015, \mnras,
  452, 1902

\bibitem[{{Lopez-Rodriguez} {et~al.}(2016){Lopez-Rodriguez}, {Packham},
  {Roche}, {Alonso-Herrero}, {D{\'\i}az-Santos}, {Nikutta},
  {Gonz{\'a}lez-Mart{\'\i}n}, {{\'A}lvarez}, {Esquej}, {Espinosa}, {Perlman},
  {Ramos Almeida}, \& {Telesco}}]{Lopez-Rodriguez2016}
{Lopez-Rodriguez}, E., {Packham}, C., {Roche}, P.~F., {et~al.} 2016, \mnras,
  458, 3851

\bibitem[{{Maire} {et~al.}(2016){Maire}, {Langlois}, {Dohlen}, {Lagrange},
  {Gratton}, {Chauvin}, {Desidera}, {Girard}, {Milli}, {Vigan}, {Zins},
  {Delorme}, {Beuzit}, {Claudi}, {Feldt}, {Mouillet}, {Puget}, {Turatto}, \&
  {Wildi}}]{Maire2016}
{Maire}, A.-L., {Langlois}, M., {Dohlen}, K., {et~al.} 2016, in \procspie, Vol.
  9908, Ground-based and Airborne Instrumentation for Astronomy VI, 990834

\bibitem[{{Marin}(2018{\natexlab{a}})}]{Marin2018}
{Marin}, F. 2018{\natexlab{a}}, \mnras, 479, 3142

\bibitem[{{Marin}(2018{\natexlab{b}})}]{Marin2018dilu}
{Marin}, F. 2018{\natexlab{b}}, \aap, 615, A171

\bibitem[{{Marin} {et~al.}(2015){Marin}, {Goosmann}, \& {Gaskell}}]{Marin2015}
{Marin}, F., {Goosmann}, R.~W., \& {Gaskell}, C.~M. 2015, \aap, 577, A66

\bibitem[{{Mathis} {et~al.}(1977){Mathis}, {Rumpl}, \&
  {Nordsieck}}]{Mathis1977}
{Mathis}, J.~S., {Rumpl}, W., \& {Nordsieck}, K.~H. 1977, \apj, 217, 425

\bibitem[{{Milli} {et~al.}(2017){Milli}, {Mouillet}, {Fusco}, {Girard},
  {Masciadri}, {Pena}, {Sauvage}, {Reyes}, {Dohlen}, {Beuzit}, {Kasper},
  {Sarazin}, \& {Cantalloube}}]{Milli2017arXiv}
{Milli}, J., {Mouillet}, D., {Fusco}, T., {et~al.} 2017, arXiv e-prints,
  arXiv:1710.05417

\bibitem[{{Murakawa}(2010)}]{Murakawa2010}
{Murakawa}, K. 2010, \aap, 518, A63

\bibitem[{{Nenkova} {et~al.}(2002){Nenkova}, {Ivezi{\'c}}, \&
  {Elitzur}}]{Nenkova2002}
{Nenkova}, M., {Ivezi{\'c}}, {\v{Z}}., \& {Elitzur}, M. 2002, \apj, 570, L9

\bibitem[{{Packham} {et~al.}(2007){Packham}, {Young}, {Fisher}, {Volk},
  {Mason}, {Hough}, {Roche}, {Elitzur}, {Radomski}, \& {Perlman}}]{Packham2007}
{Packham}, C., {Young}, S., {Fisher}, S., {et~al.} 2007, \apjl, 661, L29

\bibitem[{{Packham} {et~al.}(1997){Packham}, {Young}, {Hough}, {Axon}, \&
  {Bailey}}]{Packham1997}
{Packham}, C., {Young}, S., {Hough}, J.~H., {Axon}, D.~J., \& {Bailey}, J.~A.
  1997, \mnras, 288, 375

\bibitem[{{Ramos Almeida} \& {Ricci}(2017)}]{Ramos-Almeida2017}
{Ramos Almeida}, C. \& {Ricci}, C. 2017, Nature Astronomy, 1, 679

\bibitem[{{Rouan} {et~al.}(2019){Rouan}, {Grosset}, \& {Gratadour}}]{Rouan2019}
{Rouan}, D., {Grosset}, L., \& {Gratadour}, D. 2019, \aap, 625, A100

\bibitem[{{Rowan-Robinson}(1995)}]{Rowan-Robinson1995}
{Rowan-Robinson}, M. 1995, \mnras, 272, 737

\bibitem[{{Schmid} {et~al.}(2018){Schmid}, {Bazzon}, {Roelfsema}, {Mouillet},
  {Milli}, {Menard}, {Gisler}, {Hunziker}, {Pragt}, {Dominik}, {Boccaletti},
  {Ginski}, {Abe}, {Antoniucci}, {Avenhaus}, {Baruffolo}, {Baudoz}, {Beuzit},
  {Carbillet}, {Chauvin}, {Claudi}, {Costille}, {Daban}, {de Haan}, {Desidera},
  {Dohlen}, {Downing}, {Elswijk}, {Engler}, {Feldt}, {Fusco}, {Girard},
  {Gratton}, {Hanenburg}, {Henning}, {Hubin}, {Joos}, {Kasper}, {Keller},
  {Langlois}, {Lagadec}, {Martinez}, {Mulder}, {Pavlov}, {Podio}, {Puget},
  {Quanz}, {Rigal}, {Salasnich}, {Sauvage}, {Schuil}, {Siebenmorgen}, {Sissa},
  {Snik}, {Suarez}, {Thalmann}, {Turatto}, {Udry}, {van Duin}, {van Holstein},
  {Vigan}, \& {Wildi}}]{Schmid2018}
{Schmid}, H.~M., {Bazzon}, A., {Roelfsema}, R., {et~al.} 2018, \aap, 619, A9

\bibitem[{{Siebenmorgen} {et~al.}(2015){Siebenmorgen}, {Heymann}, \&
  {Efstathiou}}]{Siebenmorgen2015}
{Siebenmorgen}, R., {Heymann}, F., \& {Efstathiou}, A. 2015, \aap, 583, A120

\bibitem[{{Simmons} \& {Stewart}(1985)}]{Simmons1985}
{Simmons}, J.~F.~L. \& {Stewart}, B.~G. 1985, \aap, 142, 100

\bibitem[{{Tinbergen}(1996)}]{Tinbergen1996}
{Tinbergen}, J. 1996, {Astronomical Polarimetry}, 174

\bibitem[{{Zallat} \& {Heinrich}(2007)}]{Zallat2007}
{Zallat}, J. \& {Heinrich}, C. 2007, Optics Express, 15, 83

\end{thebibliography}

\begin{appendix} 
\section{Matrix inversion method}
\label{App:matrinvers}

The matrix inversion method was inspired by polarisation state analysers methodologies, see \cite{Zallat2007} for instance. Indeed, by observing with a polariser, we apply to the initial Stokes parameters on the incoming light the following transformation matrix W to get the measured intensities:

\begin{equation}
I_{meas} = W \times S,
\end{equation}
with
\begin{equation}
S =
\begin{bmatrix}
I \\
Q \\
U \\
\end{bmatrix}
\end{equation}
and
\begin{equation}
I_{meas} =
\begin{bmatrix}
I_1 \\
I_2 \\
... \\
I_n \\
\end{bmatrix}.
\end{equation}

W depends on the angles $\theta_n$ of the polariser for each image recorded:
\begin{equation}
W =
\begin{pmatrix}
\cos^2(\theta_1) & \cos(\theta_1) & \sin(\theta_1)\\
\cos^2(\theta_2) & \cos(\theta_2) & \sin(\theta_2)\\
& ... &\\
\cos^2(\theta_n) & \cos(\theta_n) & \sin(\theta_n)\\
\end{pmatrix}.
\end{equation}

Because W may not be a square matrix, it is not always invertible. In practice, it will never be the case (having only three polarimetric measurements/images is rare because of the symmetry of measurements) and we apply the pseudo-inverse $(W^T W)$. Therefore, by applying $(W^T W)^{-1} W^T$ on both sides, we can compute S directly:
\begin{equation}
S= (W^T W)^{-1} W^T I_{meas}.
\label{eq:S}
\end{equation}

In our case, we have eight measurements with the four following positions of the polariser:
\begin{equation}
I_{meas}=
\begin{bmatrix}
Q_+ \\
Q_- \\
U_+ \\
U_- \\
Q_+ \\
Q_- \\
U_+ \\
U_- \\
\end{bmatrix}.
\end{equation}

We then get the W matrix as following
\begin{equation}
W =\frac{1}{2}
\begin{pmatrix}
1 & 1 & 0\\
1 & -1 & 0\\
0 & 0 & 1\\
0 & 0 & -1\\
1 & 1 & 0\\
1 & -1 & 0\\
0 & 0 & 1\\
0 & 0 & -1\\
\end{pmatrix}.
\end{equation}

\section{Polarisation uncertainty}
\label{App:polar_sig}

We first verified the significance of the measured polarisation according to the received intensity. Using formula developed on \cite{Simmons1985} and \cite{Everett2001}, we derived maps of the ratio of the polarised intensity over the standard deviation of the corresponding intensity $Ip/\sigma_I$. Assuming that photon noise is dominant in the NIR bands, the derived values close to the photo-centre are high (above 30) and we get $Ip/\sigma_\gamma \approx 4 - 6$ for the rest of the selected regions (the arcs, the equatorial region). As this is a lower limit for the standard deviation ($\sigma_I>\sigma_\gamma$), we estimate the true $Ip/\sigma_I$ value to be about 3 to 4. Thus, the lowering correction to be applied to the polarisation estimation, based on equation 11 of \cite{Everett2001} is of $\approx$~5\% of its value (thus about 0.5\% in polarisation degree). The value of $Ip/\sigma_I$ is lower in the case of the NR band and our measurements should therefore be considered as maximum values for polarisation in this band.

In order to evaluate the uncertainty due to polarisation variation within the apertures and compare the methods, we also made an analysis of the local variations of the degree and angle of polarisation. As detailed in \cite{Clarke1986}, we subtracted to each pixel a mean of the four closest pixels, creating a ``pseudo-noise'' map $m_\sigma$ and then look at the dispersion of these values on all the maps, as follows:

\begin{equation}
\label{eq:sigma_pol}
\sigma_{pol} = 1.4826 \times \mathrm{med}\left(\left|\sqrt{0.8}~m_\sigma-\mathrm{med}\left(\sqrt{0.8}~m_\sigma\right)\right|\right)
\end{equation}

Thus we ensure to compare the variations in polarisation on regions where the SNR of the intensity maps is almost identical, which is required since the SNR will affect the determination of the degree of polarisation.

In our apertures, this variation is generally ranging between 15 and 20~\% (see error bars of Figure~\ref{fig:SA} and Figure~\ref{fig:centre}) and thus larger than the correction to be applied to the measured polarisation of 5~\% determined previously. We thus used in this study as polarisation estimator the measured one, $P_{\mathrm{TRUE}} \approx P_{\mathrm{MES}}$. As highlighted by \cite{Simmons1985}, this estimator is not very efficient for low signal to noise ratio but is converging towards the other polarisation estimators at $Ip/\sigma_I \geq 2$, a value well within our error bars.

\section{SPHERE Derotator}
\label{App:derot}

\begin{figure}[ht!]
 \centering
 \includegraphics[width=0.35\textwidth,clip]{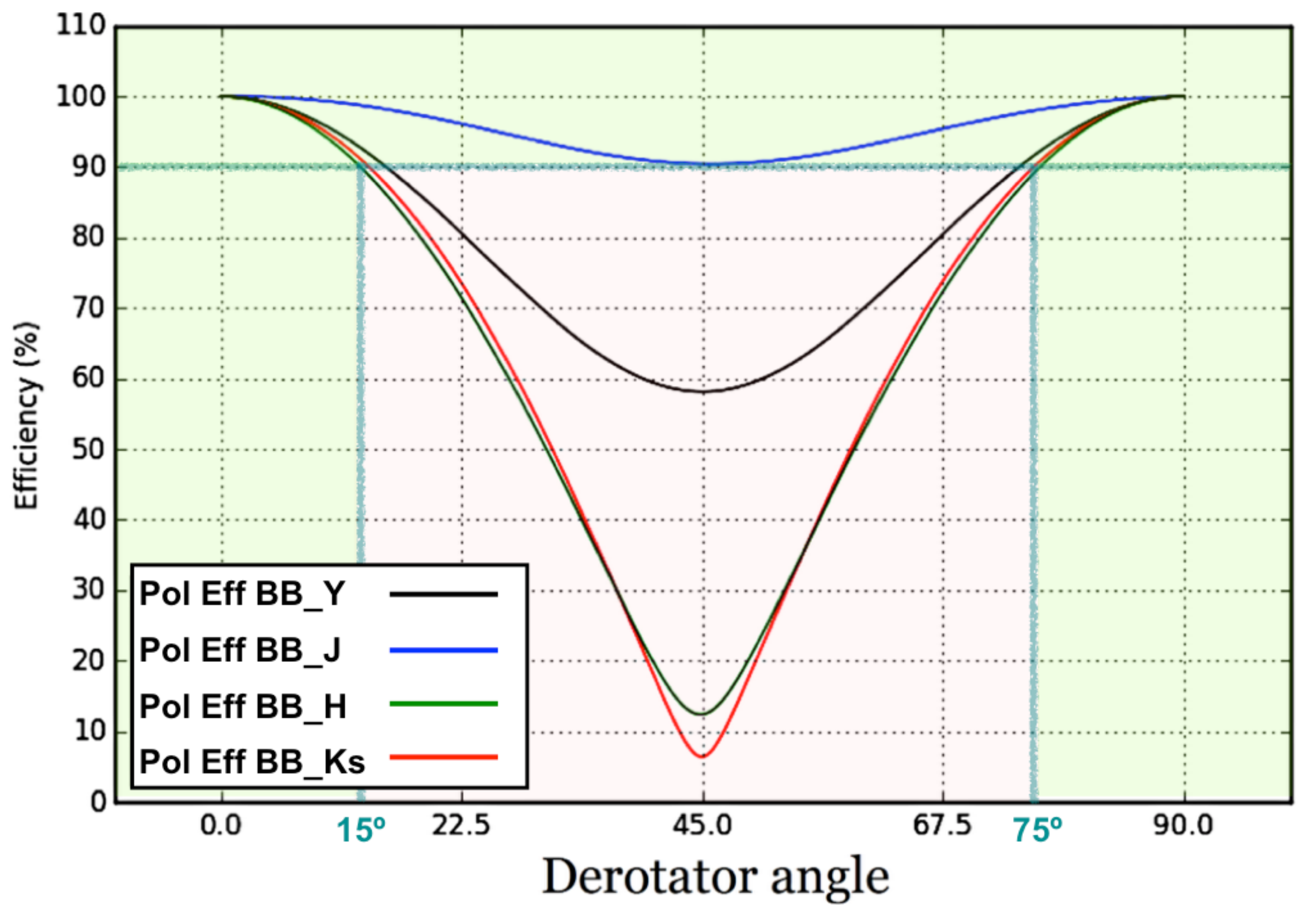}
  \caption[SPHERE-IRDIS instrumental polarisation efficiency.]{Measurements of the instrumental polarisation efficiency (not accounting the telescope) for four BB filters. For best use of the DPI mode, one should avoid the pink zone where the efficiency drops below 90$^\circ$ (> 10$^\circ$ loss due to cross-talks). For that, one should make sure the derotator angle stays < 15$^\circ$ or > 75$^\circ$. From SPHERE User Manual.}
  \label{fig:derotator_eff}
 \end{figure}

\begin{figure*}[ht!]
 \centering
 \includegraphics[width=0.30\textwidth,clip]{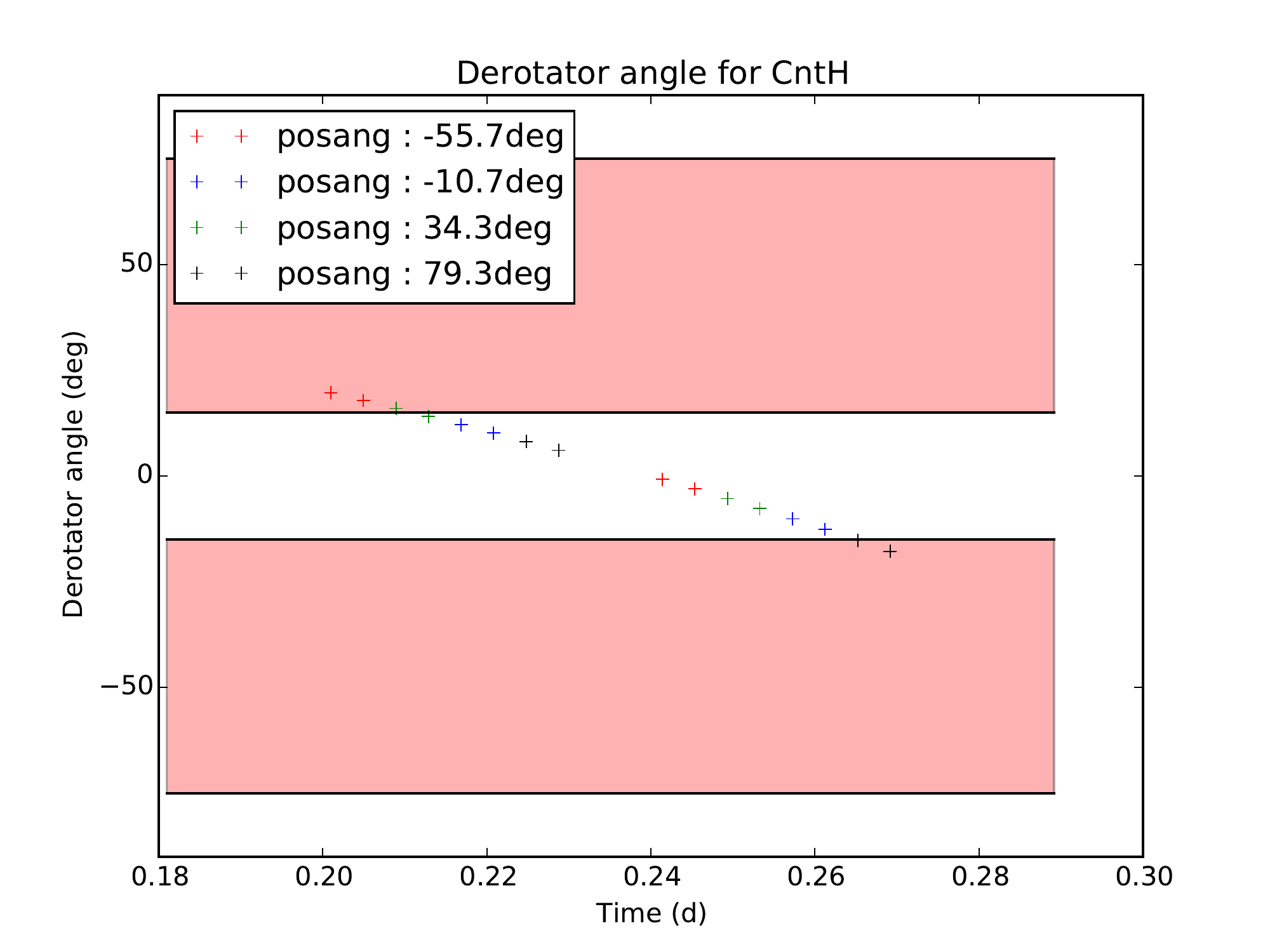}
 \includegraphics[width=0.30\textwidth,clip]{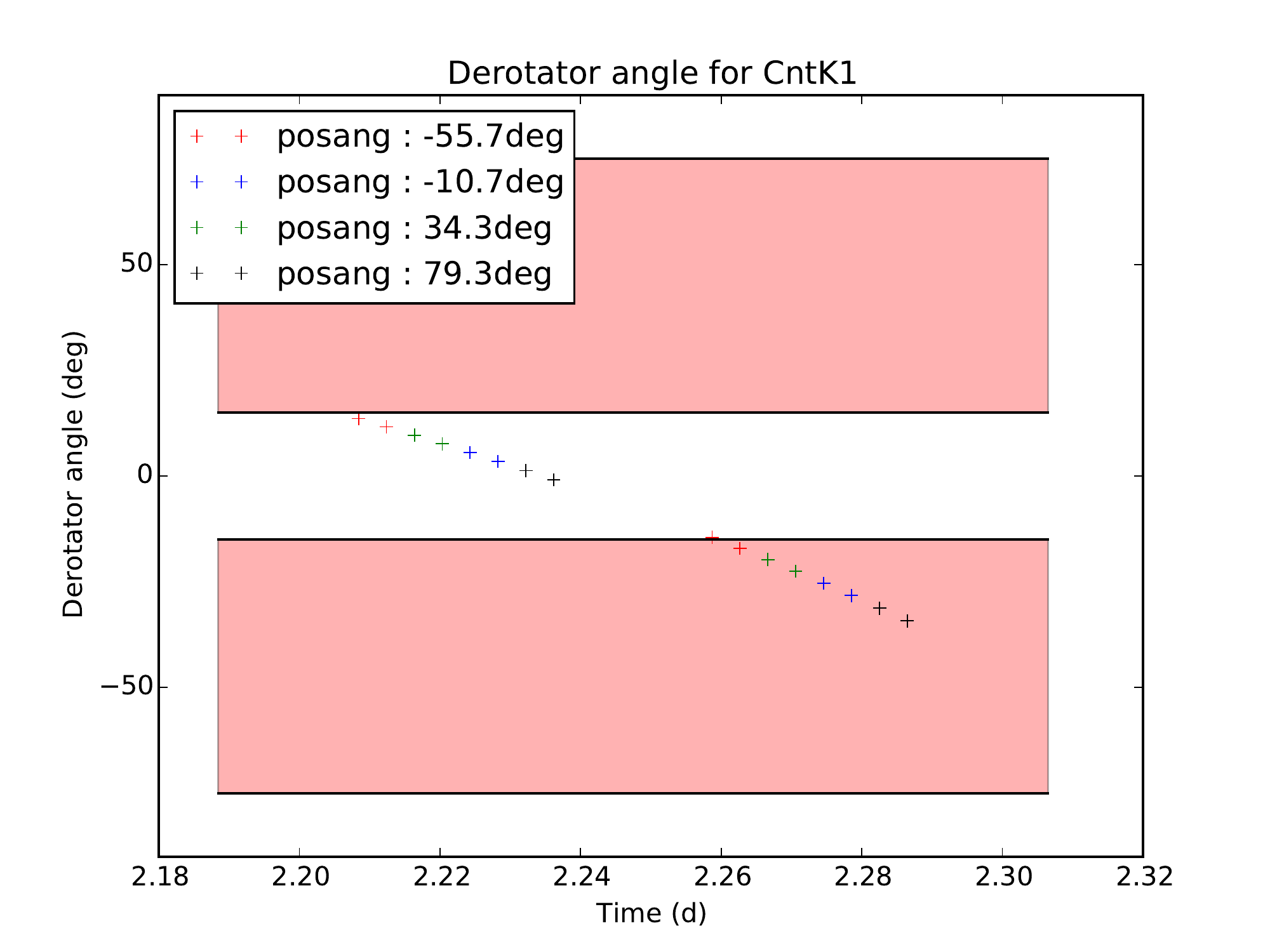}
 \includegraphics[width=0.30\textwidth,clip]{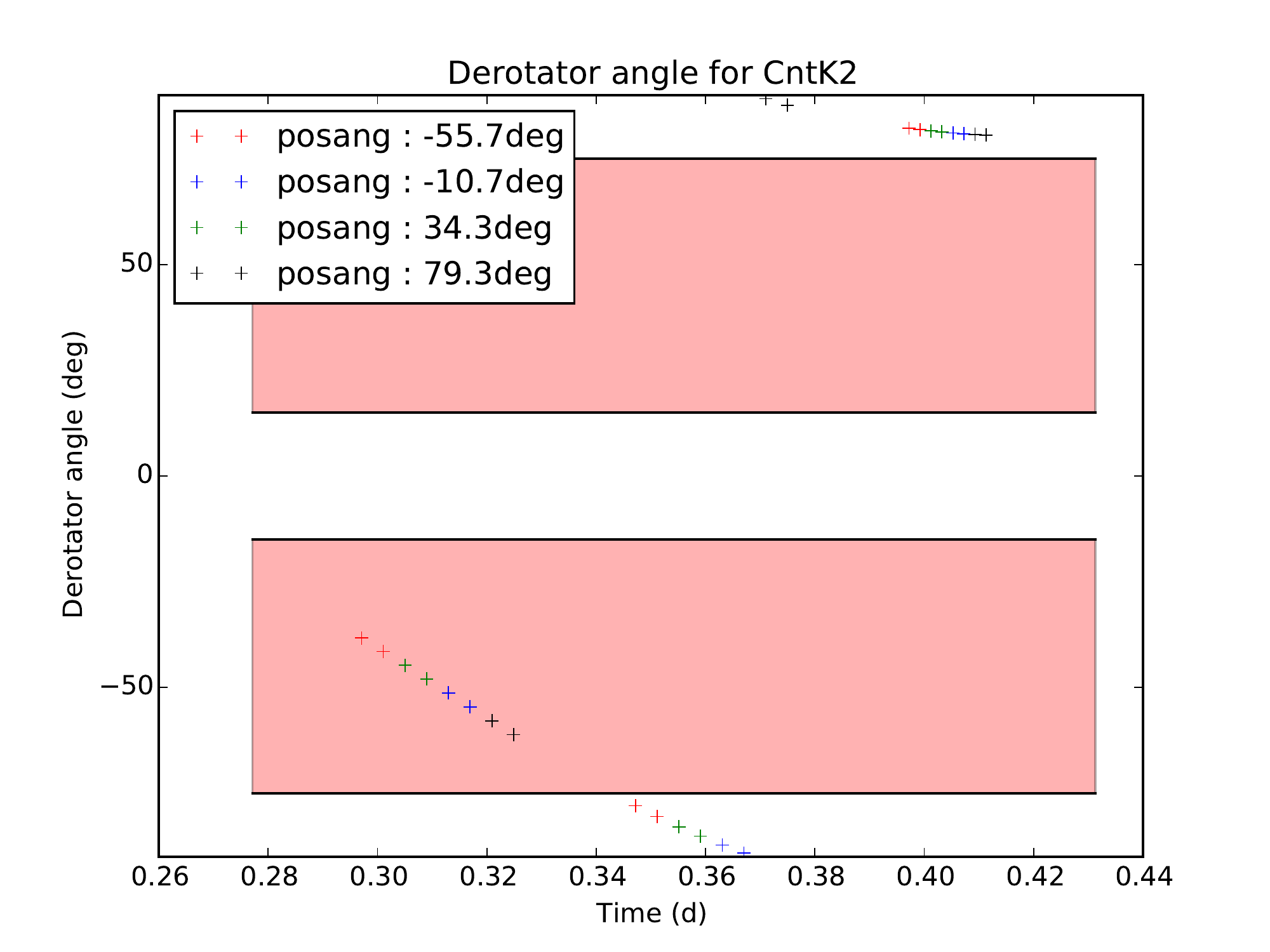}
  \caption[Derotator position for NB observations.]{Derotator position for NB observations, CntH, CntK1 and CntK2. The red bands indicates the zones that one should avoid when using BBs.}
  \label{fig:derotator_pos}
 \end{figure*}

The derotator angle is an important parameter for polarimetric observations with SPHERE. It affects the polarisation measurements and should therefore be carefully planned or taken into account. Figure~\ref{fig:derotator_eff} shows the polarimetric efficiency depending on the broad filter and derotator angle, extracted from the SPHERE instrument User Manual. Contrary to broad filters, NB ones have not been tested and their polarimetric efficiency is therefore unknown. We expect their efficiency to follow the general trends of the other filters, but we have no proper way to verify this, nor to constrain exactly what is the efficiency for a given angle. We placed our observations on a graph indicating for each of them the derotator angle (figure~\ref{fig:derotator_pos}).

\begin{figure*}[ht!]
 \centering
  \includegraphics[width=0.3\textwidth,clip]{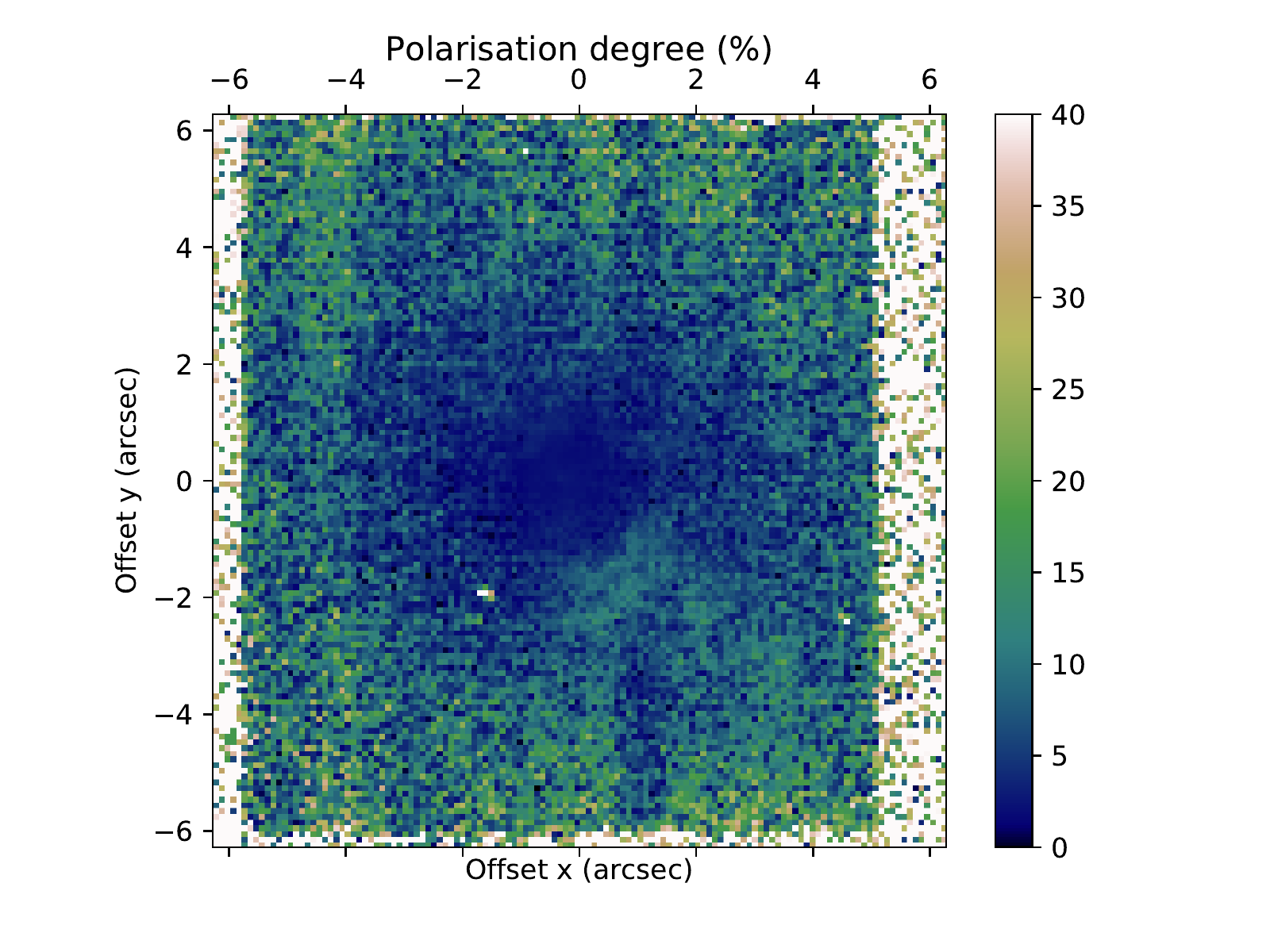}
  \includegraphics[width=0.3\textwidth,clip]{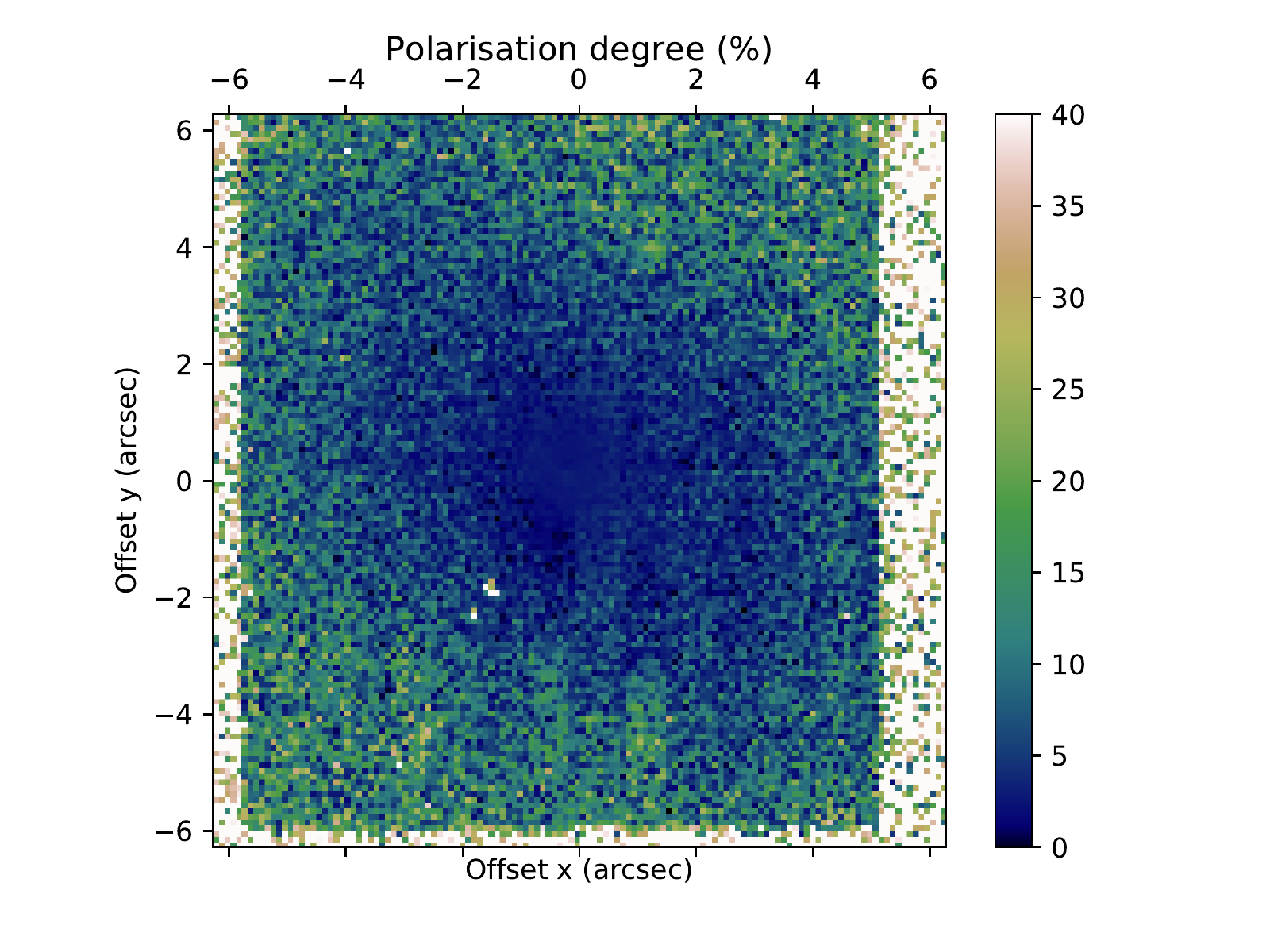}
  \includegraphics[width=0.3\textwidth,clip]{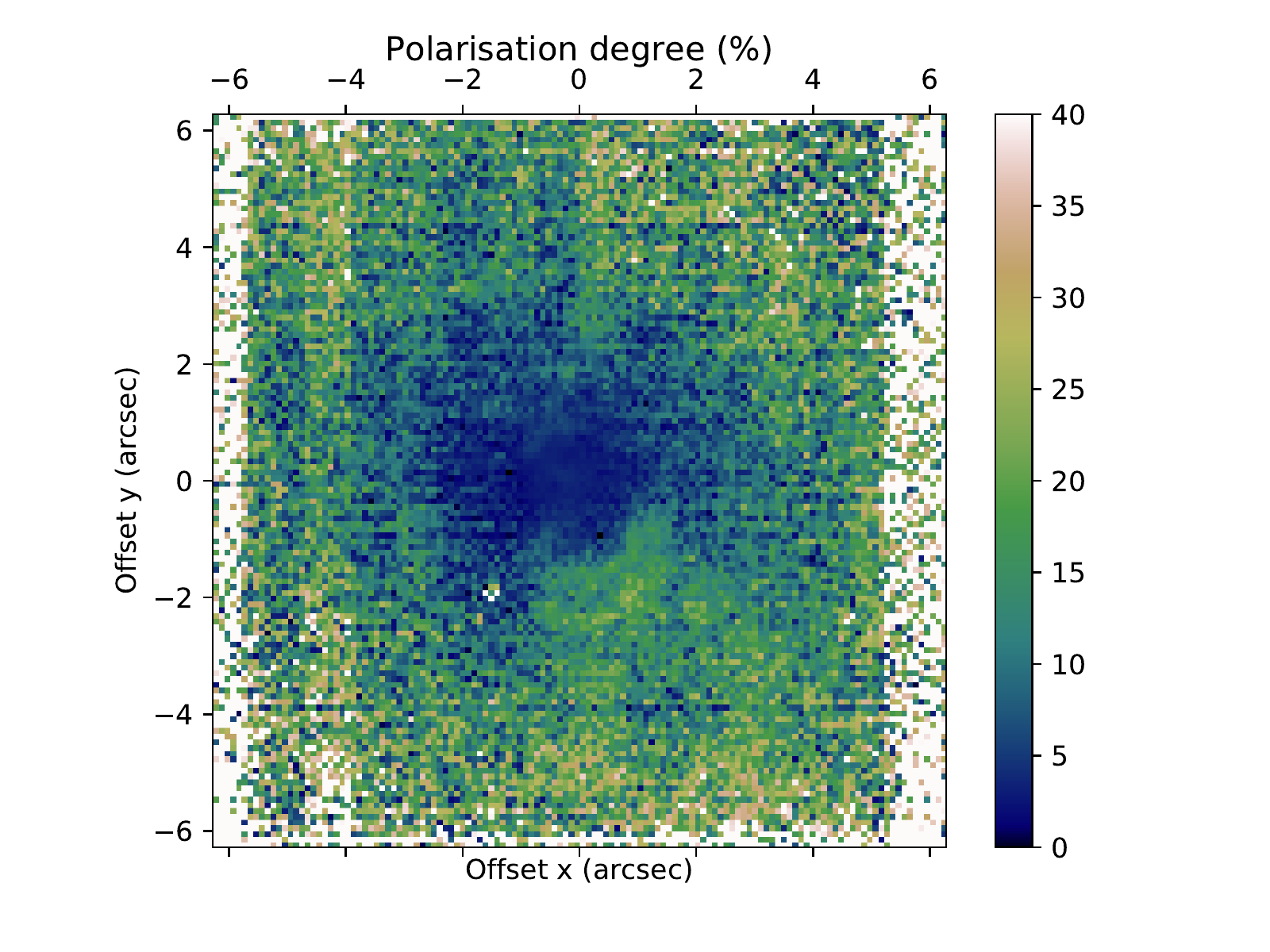}
   \caption{Binned maps of linear degree of polarisation (in \%) in NGC~1068 with Cnt~K2. These maps were created using all available raw images for first panel; only the 60~\% of the frames outside the 15-75$^\circ$ red zone of Fig.~\ref{fig:derotator_pos} (last two series of points) for second panel; and only the first series of points (40~\%), within the red zone for the third panel. Third panel corresponds to the binned version of maps displayed in Fig.~\ref{fig:NB3}.} 
  \label{fig:sel_test_P}
 \end{figure*}

\begin{figure*}[ht!]
 \centering
  \includegraphics[width=0.3\textwidth,clip]{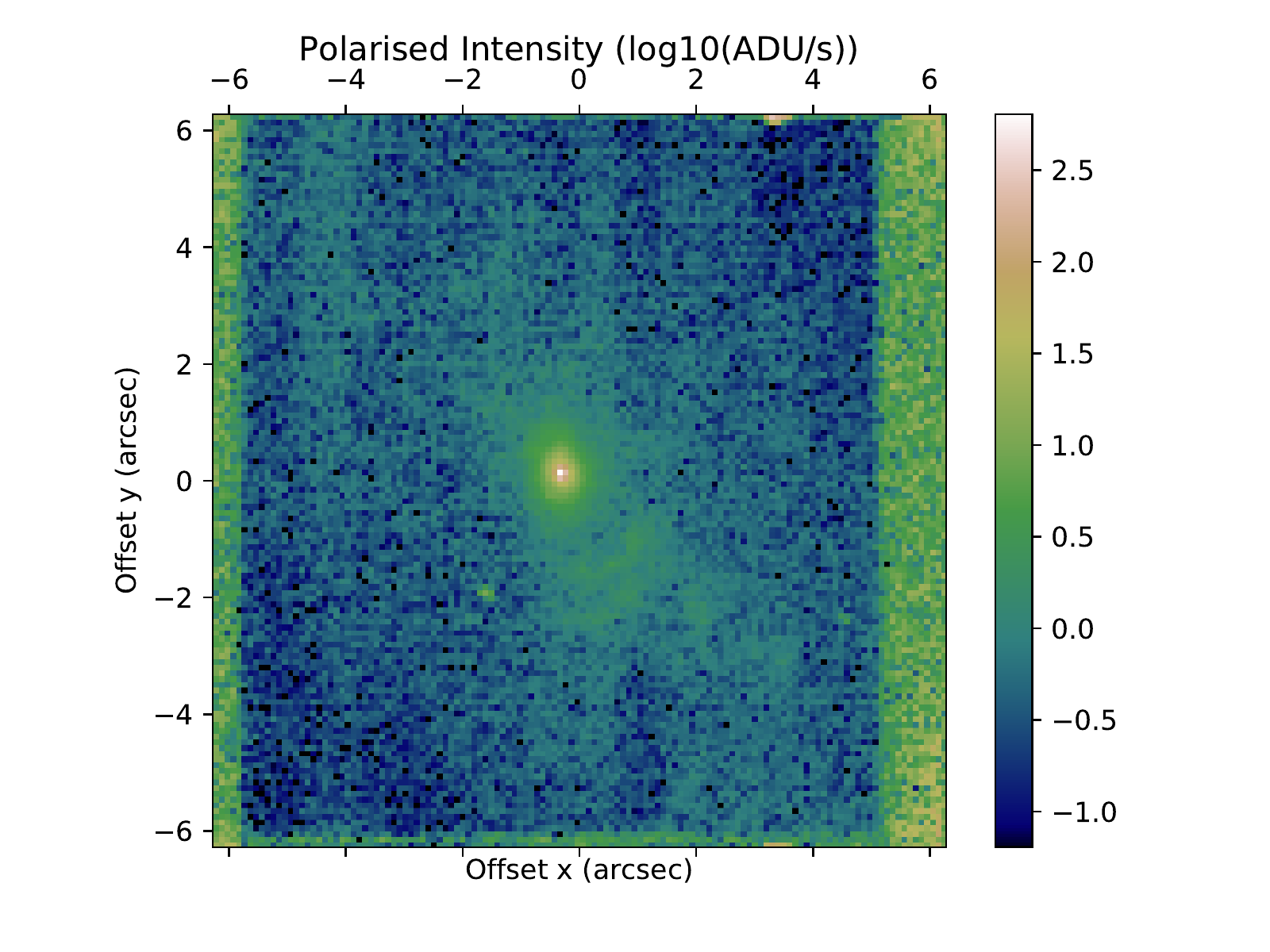}
  \includegraphics[width=0.3\textwidth,clip]{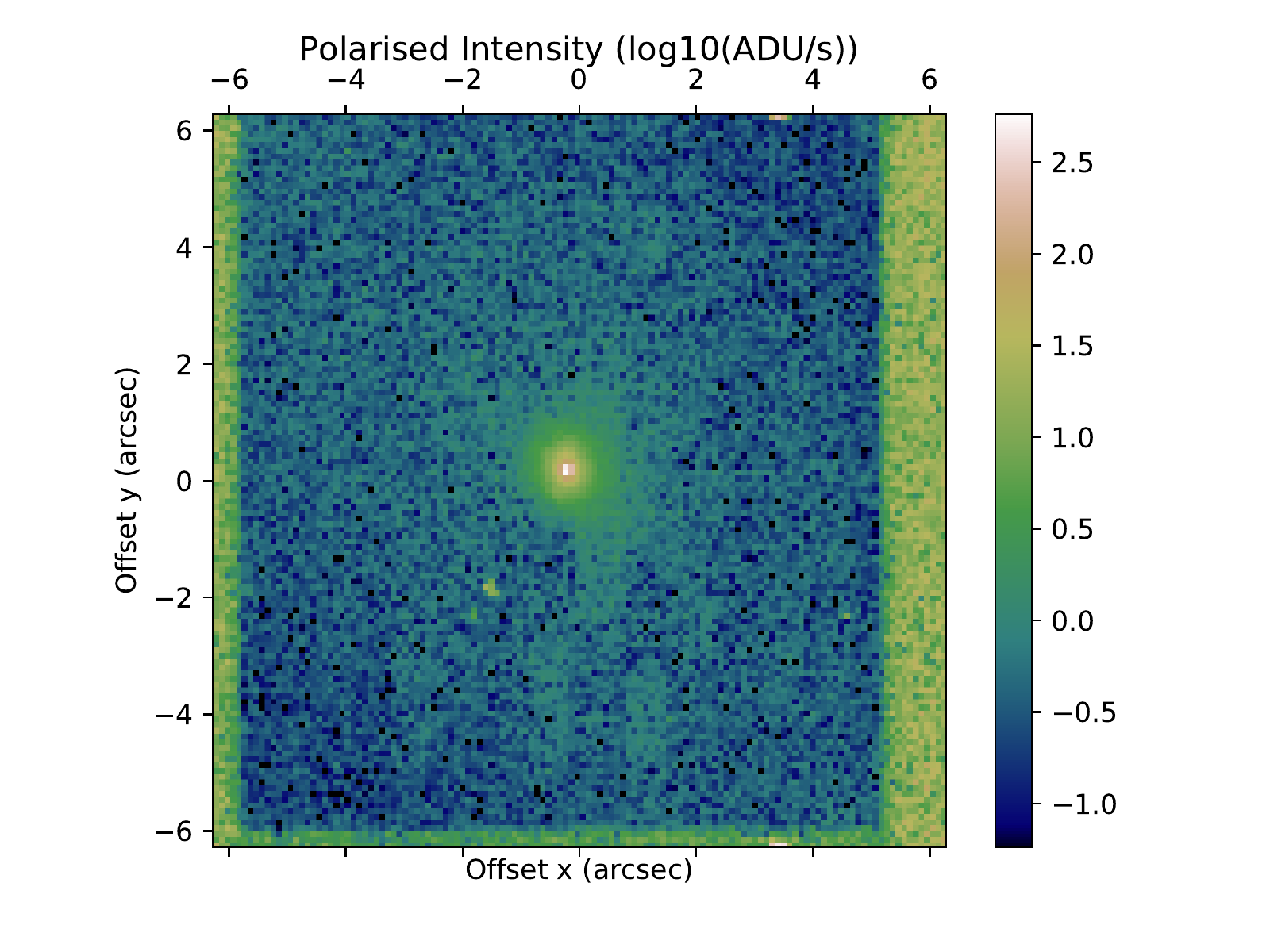}
  \includegraphics[width=0.3\textwidth,clip]{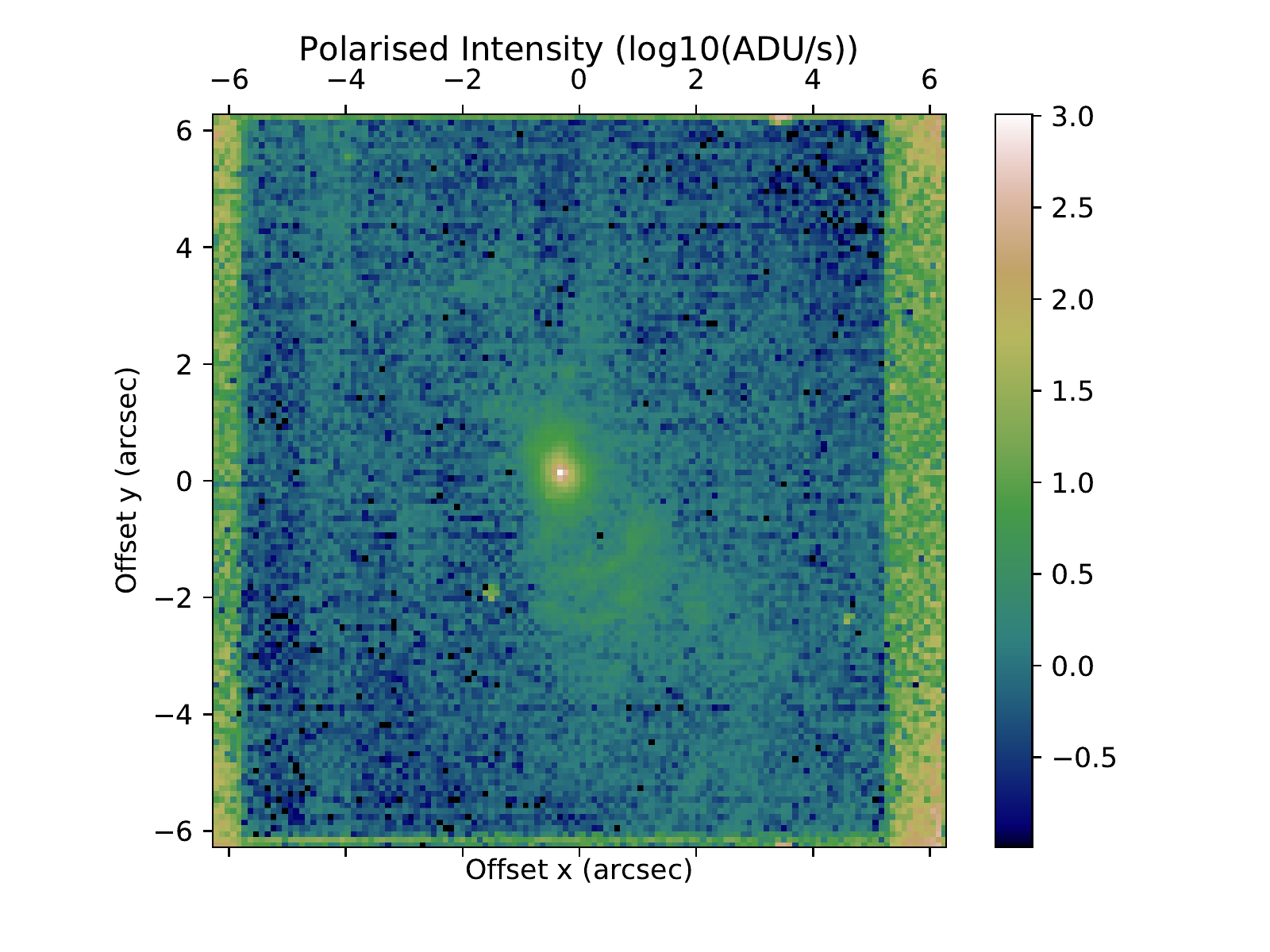}
   \caption{Same as Fig.~\ref{fig:sel_test_P} but for linearly polarised flux (in ADU/s).}
  \label{fig:sel_test_Ip}
 \end{figure*}

\begin{figure*}[ht!]
 \centering
  \includegraphics[width=0.3\textwidth,clip]{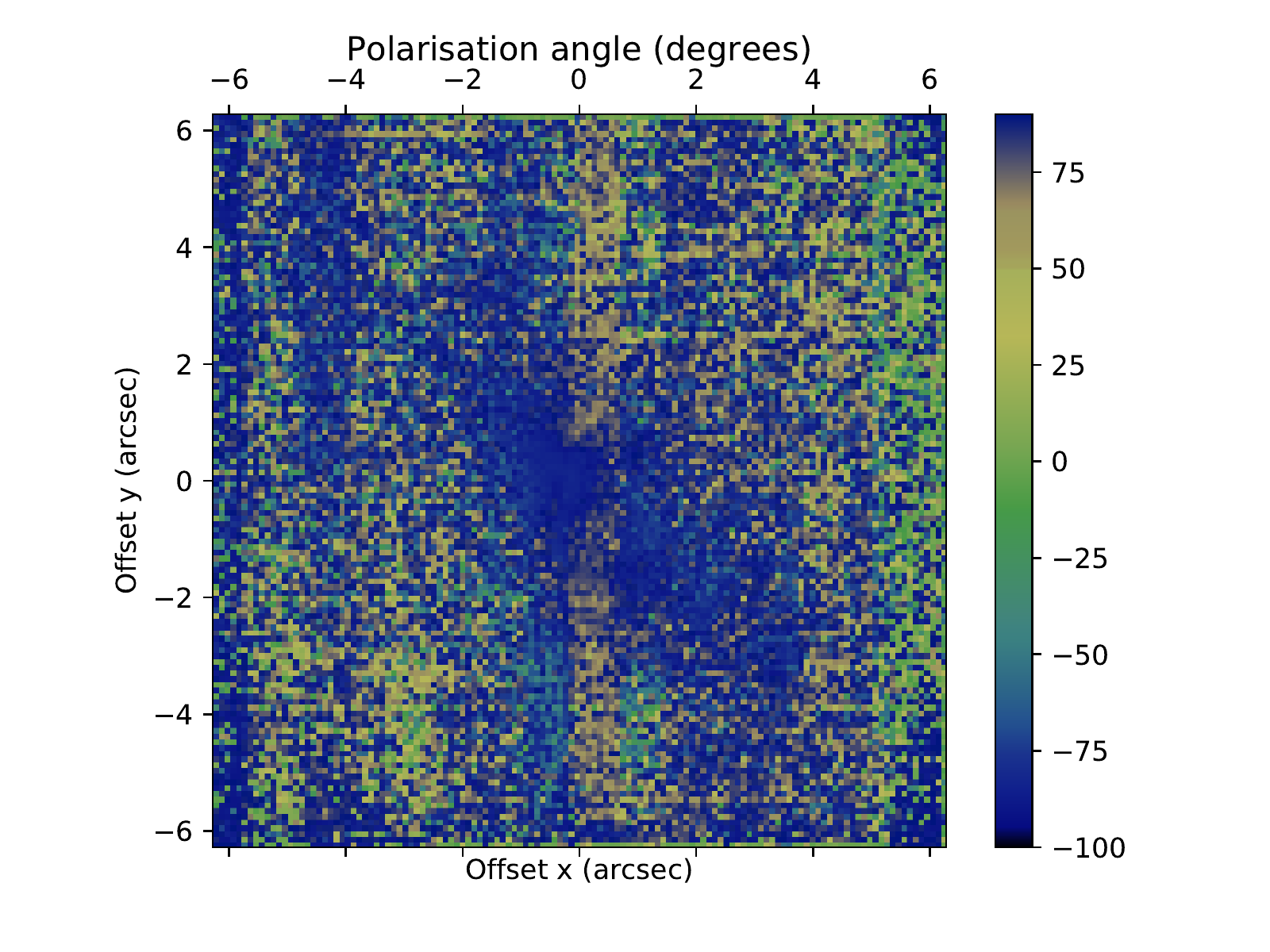}
  \includegraphics[width=0.3\textwidth,clip]{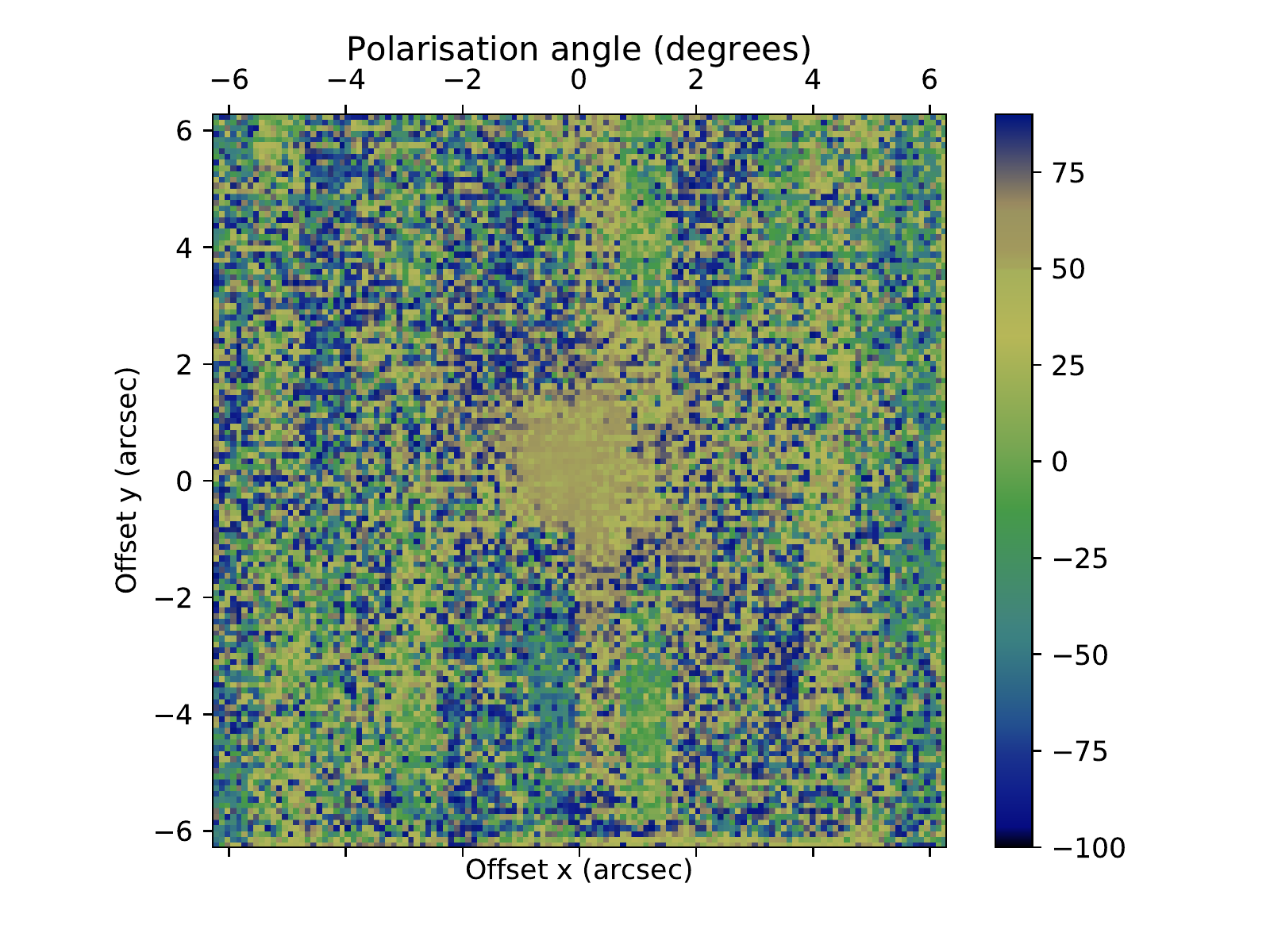}
  \includegraphics[width=0.3\textwidth,clip]{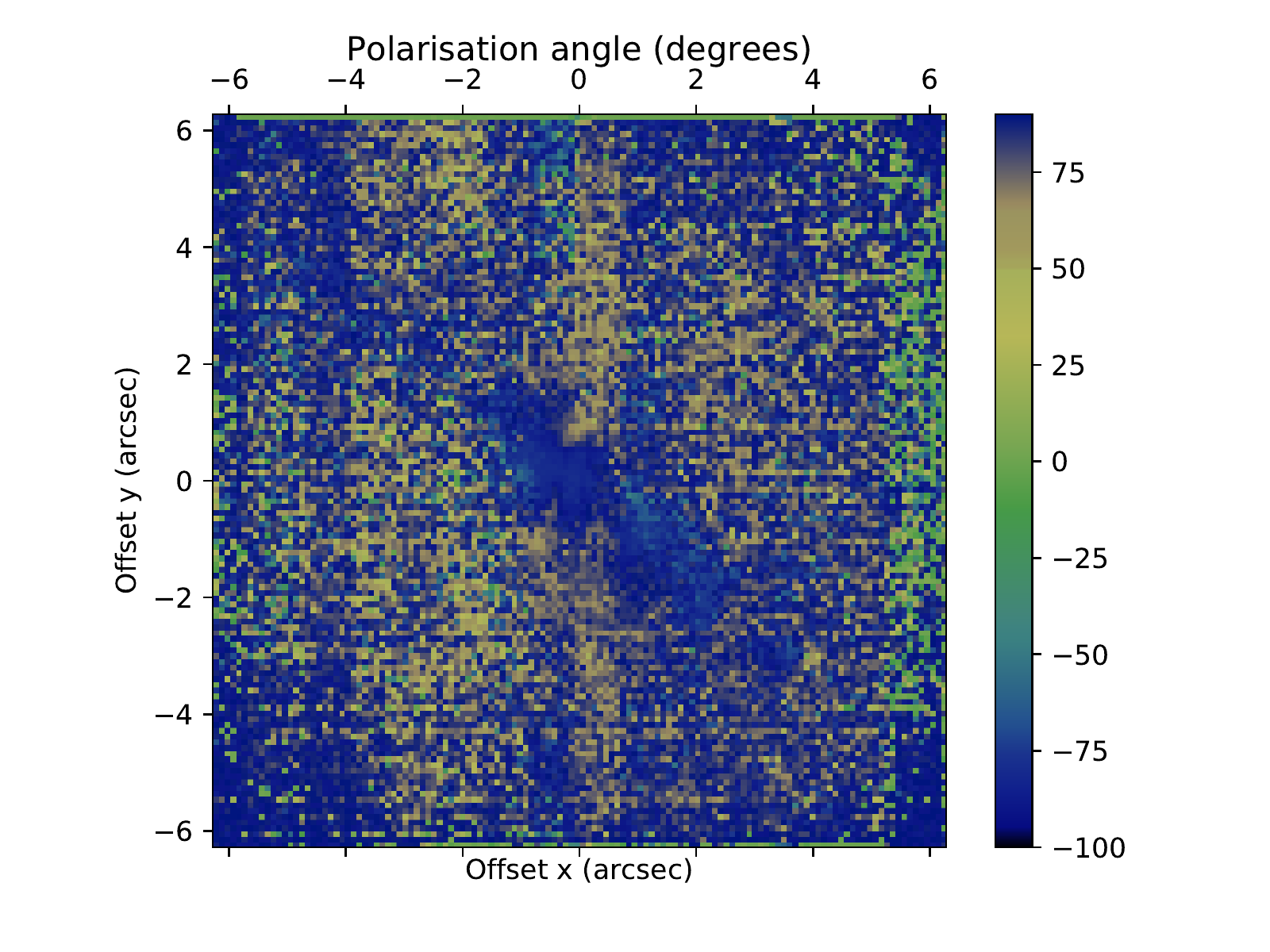}
   \caption{Same as Fig.~\ref{fig:sel_test_P} but for linear polarisation position angle (in degrees).}
  \label{fig:sel_test_theta}
 \end{figure*}

A decent fraction of our observations have been conducted within the $-15^\circ \rightarrow$ 15$^\circ$ range, i.e. on optimal position for BB filters. However this is notably not the case for some images taken with the Cnt~K1 and Cnt~K2 filters. Contrary to Cnt~K1 whose results look very compatible to what was observe in BBs, Cnt~K2 maps, when creating without selection (First panel of Figs.~\ref{fig:sel_test_P}, \ref{fig:sel_test_Ip} and \ref{fig:sel_test_theta}), shows low polarimetric signal. It also exhibits not realistic polarisation position angle (First panel of Figure~\ref{fig:sel_test_theta}), as discussed in Section~\ref{sec:K2}. We therefore investigated the possible relation between this lack of signal and the derotator position and thus compared the final maps computed with all raw images, and those obtained with a selection of raw images with an optimised derotator position. All final images are shown in Figure~\ref{fig:sel_test_P},\ref{fig:sel_test_Ip} and \ref{fig:sel_test_theta} and reveal that selection does affect significantly the measured polarisation as the South-western arcs displays very different polarisation (flux and degree). 

As revealed by Figs.~\ref{fig:sel_test_P}, \ref{fig:sel_test_Ip} and \ref{fig:sel_test_theta}, derotator position angle affects the measured polarisation degree and position angle, but also when outside the 15 -- 75$^\circ$ angle range, for Cnt~K2. Maps created using data taken with angles $>75^\circ$ do show very low levels of polarisation. For this reason, we did selected the first 40~\% of frames, within the 15-75$^\circ$ region, but with lower instrumental depolarisation for Cnt~K2 data reduction (Fig.~\ref{fig:NB3}).

\vspace{0.5cm}
We also conducted for the same reasons, despite having some higher polarimetric signal, the same experiment on Cnt K1 filter. Selection processes do not displays such differences between final products and we thus keep the complete set of data for the data reduction for this NB.

\section{The case of Cnt~K2}
\label{App:CntK2}

It is already known (see eg. the SPHERE user manual) that the SPHERE derotator position has an impact on the measured polarisation, at least in the BBs. This has been studied and depicted in the SPHERE instrument documentation. However, investigations by the SPHERE team have been conducted only on the BB filters, while the NB filters, including Cnt~K2, have not been characterised yet. Therefore, we can not derive strong conclusions about the exact impact of the depolarisation on the images in the NB filters. As the NB measurements in Cnt~H and Cnt~K1 filters are rather consistent with BB measurements, we are however more confident in the derived parameters, while Cnt~K2 output should be considered carefully, especially the polarisation position angle. An analysis of the derotator positions in our data sets is detailed in Appendix~\ref{App:derot}, and shows a clear variation of the polarisation as a function of the derotator for Cnt~K2 (no clear evidences were found for other bands). We thus selected accordingly to this study the image set that introduced the lower possible depolarisation.

\begin{figure*}[ht!]
 \centering
 \includegraphics[width=0.48\textwidth,clip]{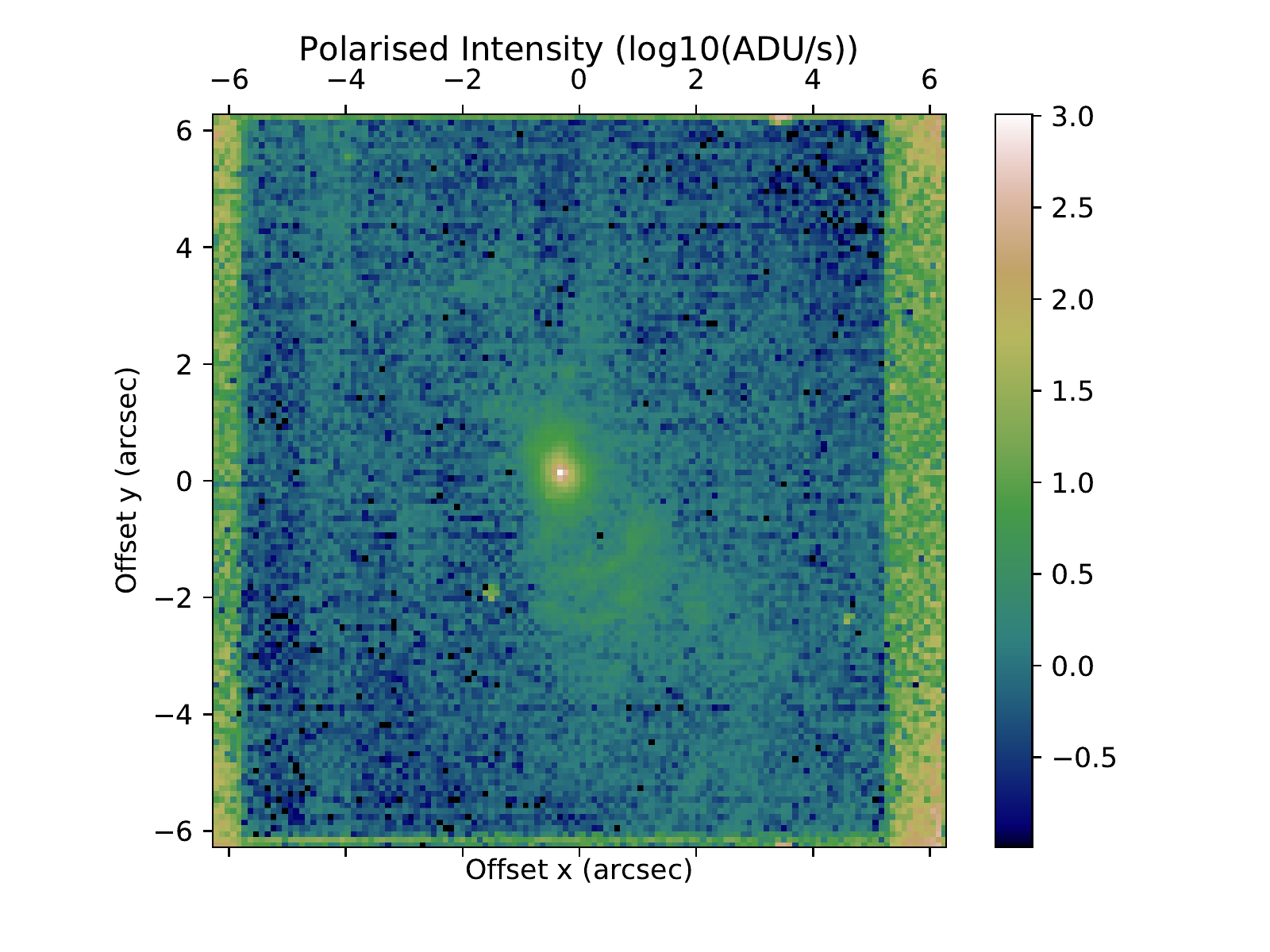}
 \includegraphics[width=0.48\textwidth,clip]{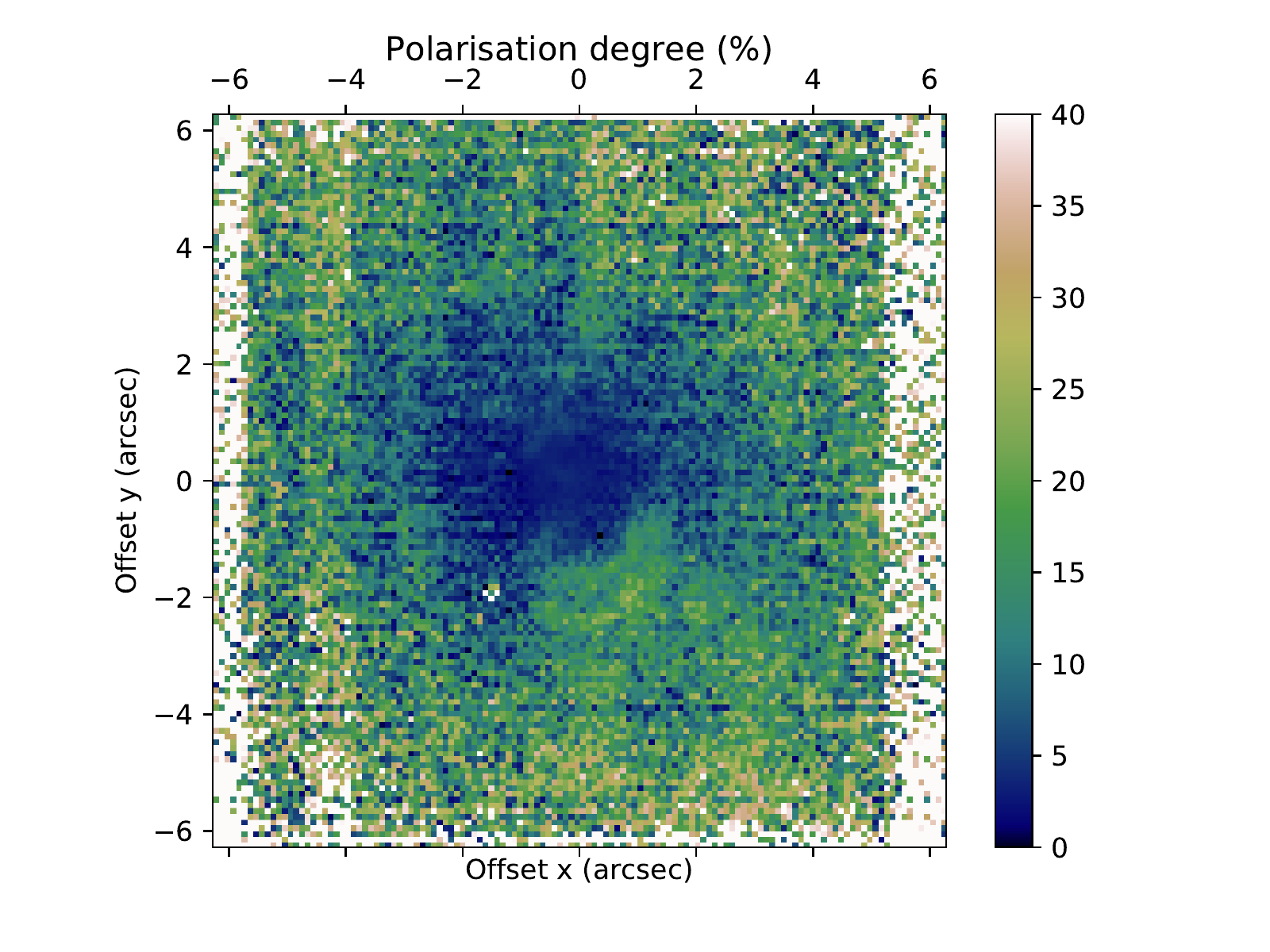}
 \includegraphics[width=0.48\textwidth,clip]{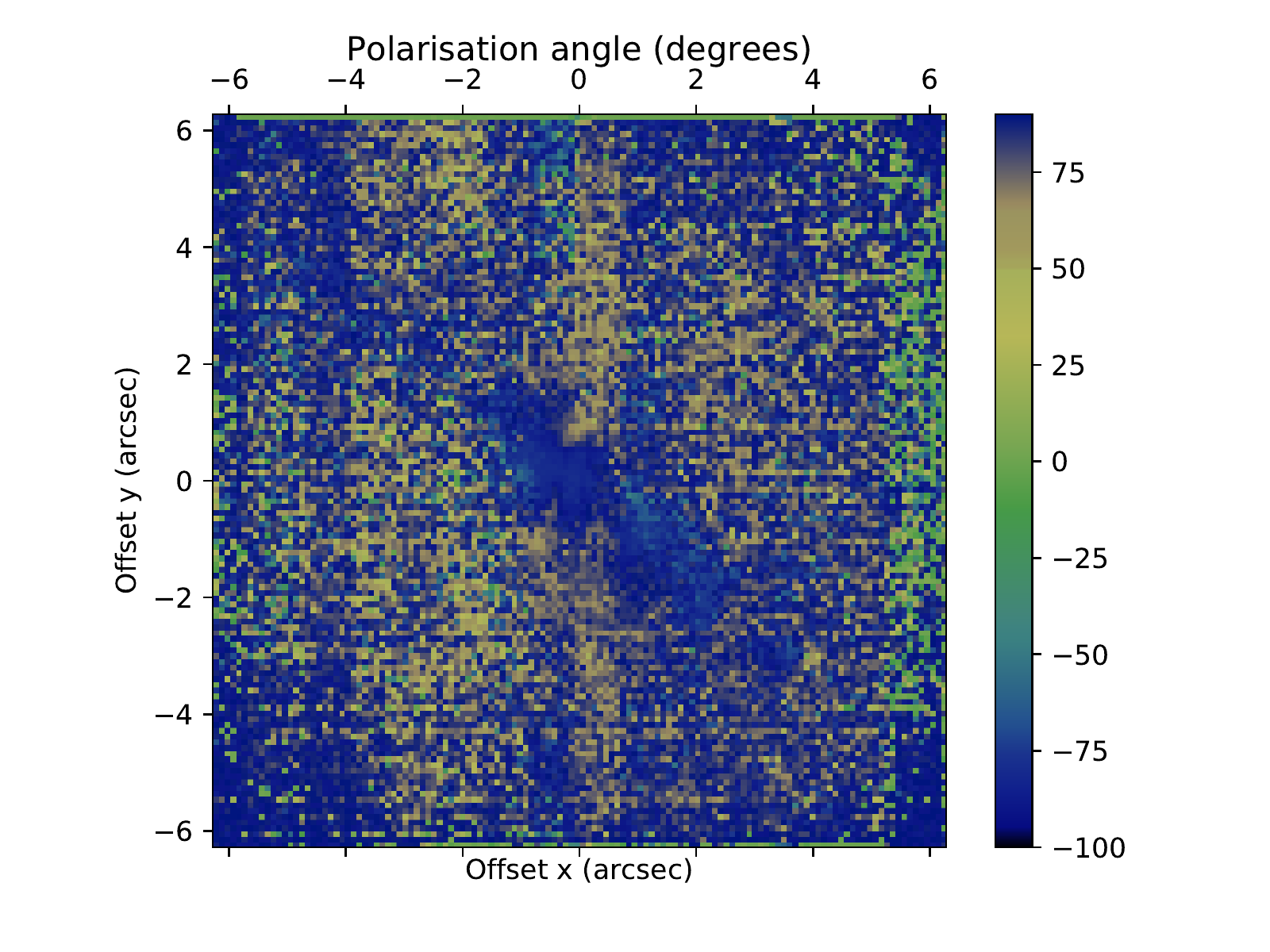}
  \caption{Binned $8 \times 8$ maps of NGC~1068 in Cnt~K2. These maps displays the binned version of the Cnt~K2 maps of Figure~\ref{fig:NB3} for polarised intensity (in $\log_{10}$(ADU/s), first row, left panel) with polarisation vectors over-plotted, linear degree of polarisation (in \%, first row, right panel) and linear angle of polarisation (in degrees, second row).}
  \label{fig:cntK2}
\end{figure*}

With this set, it is clear that polarised signal does exist in Cnt~K2 band as shown by the polarised intensity map of Figure~\ref{fig:NB3} 
(upper right and bottom left panels), harbouring the two South-western arcs. In order to better display the polarimetric signal in the maps in this NB, we display in Figure~\ref{fig:cntK2} polarimetric maps using binned I, Q and U maps. The South-western arcs are clearly identified on the polarisation degree map (binning reduces the effect of the noise onto the polarimetric signal by averaging the intensity of a group of pixels). However, this binning does not seem to improve significantly the polarisation position angle map.

\end{appendix}
\end{document}